\documentclass[twocolumn,floatfix,altaffilletter,superscriptaddress,preprintnumbers,
               tightenlines,showpacs,showkeys,nofootinbib]{revtex4-1}

\usepackage[utf8]{inputenc}
\usepackage[colorlinks=true,citecolor=blue,linkcolor=blue]{hyperref}
\usepackage[normalem]{ulem}
\usepackage{amsmath,amssymb, mathrsfs}
\usepackage{epsfig}
\usepackage{graphicx}               
\usepackage{url}
\usepackage{color}
\usepackage[dvipsnames]{xcolor}
\usepackage{soul}
\usepackage[capitalise, english]{cleveref}
\usepackage{siunitx}
\usepackage{xspace}
\usepackage{enumitem}
\usepackage{fontawesome}
\usepackage{mathtools}
\usepackage{float}
\usepackage[format=plain,justification=centerlast]{caption}
\usepackage[format=plain,justification=centerlast]{subcaption}

\newcommand\myshade{80}
\colorlet{mylinkcolor}{ForestGreen}
\colorlet{mycitecolor}{Red}
\colorlet{myurlcolor}{violet}
\definecolor{babypink}{rgb}{0.96, 0.76, 0.76}

\hypersetup{
  linkcolor  = mylinkcolor!\myshade!black,
  citecolor  = mycitecolor!\myshade!black,
  urlcolor   = myurlcolor!\myshade!black,
  colorlinks = true
}

\DeclareSIUnit\parsec{pc}

\newcommand{\lam}{\lambda}
\newcommand{\kap}{\kappa}

\newcommand{\nbicon}{{\color{tobycolour}\faFileCodeO}\xspace}
\newcommand{\nblink}[1]{\href{https://github.com/github_url/notebooks/#1.ipynb}{\nbicon}}

\begin{document}







\newcommand{\AddrBonn}{%
Bethe Center for Theoretical Physics\\ \& Physikalisches Institut der 
Universit\"at Bonn\\  Nu{\ss}allee 12, 
 53115 Bonn, Germany
}

\title{A $\nu$ Approach to Analyzing Neutrino Data in the $\mathbf{R}$-Parity-Violating MSSM}

\author{Herbi K. Dreiner}\email{dreiner@uni-bonn.de}
\affiliation{\AddrBonn}

\author{Dominik K\"{o}hler}\email{koehler@physik.uni-bonn.de}
\affiliation{\AddrBonn}

\author{Saurabh Nangia}\email{nangia@physik.uni-bonn.de}
\affiliation{\AddrBonn}

\preprint{BONN-TH-2022-21}

\begin{abstract}
\noindent
The $R$-parity-violating Minimal Supersymmetric Standard Model (RPV-MSSM) can
naturally accommodate massive neutrinos as required by the oscillation data. However, 
studying the phenomenology is complicated due to the large number of undetermined
parameters involved. Thus, studies are usually restricted to specific submodels. In this work, we develop an approach that allows us to be less restrictive. Working in (almost) the completely general RPV-MSSM setting, we analyze the structure of the neutrino mass matrix, and identify -- for the case of two massive neutrinos -- only four minimal classes of structures that can solve the neutrino data; we call these Minimal Oscillation Models (MOMs). We study the general features of each MOM class, and present numerical fits to the oscillation data. Our approach allows us to study all RPV models satisfying the neutrino data in a unified manner, as long as they satisfy the MOM criteria. Through several examples, we show that this indeed holds for many interesting scenarios.
\end{abstract}

\maketitle

\section{Introduction}
\label{sec:1}
The Standard Model of particle physics (SM) is incomplete. The nature of gravity, dark matter, dark energy, the baryon asymmetry, etc. are important unresolved issues. However, the most conclusive sign of physics beyond the SM comes from the precise neutrino oscillation data. It is now established that at least two of the neutrino species are massive. One way to give neutrinos mass is to add right-handed neutrinos to the SM spectrum. Via the see-saw mechanism, one then `naturally' obtains very light neutrinos, as required 
indirectly by cosmology $\left(\sum m_{\nu_i} <\SI{0.12} {\electronvolt}\right) 
$~\cite{Workman:2022ynf} or, directly, for example, by the KATRIN experiment 
$\left(m_{\nu}<\SI{0.8}{\electronvolt}\right)$~\cite{KATRIN:2021uub}. However, this 
requires the right-handed neutrinos to be very heavy.

Supersymmetry (SUSY), a well-motivated extension of the 
SM~\cite{Nilles:1983ge, Martin:1997ns}, is an attractive alternative. The simplest 
phenomenological realization, the Minimal Supersymmetric Standard Model (MSSM), has
been studied extensively. An equally well-motivated~\cite{Dreiner:1997uz,Barbier:2004ez} setting is provided by adding 
$R$-parity-violating (RPV) terms to the MSSM Lagrangian, giving the 
RPV-MSSM~\cite{Allanach:2003eb}. This framework leads to a starkly different 
phenomenology compared to the MSSM, allows for lepton- and baryon-number violation, as well as flavor 
violation. Most importantly for this paper: Neutrino masses arise for free, without the
need for any heavy right-handed partners~\cite{Hall:1983id,Hirsch:2000ef}.

Neutrino-mass generation in the RPV-MSSM framework has been studied extensively in the literature. Early work on the tree-level calculation can be found in 
Refs.~\cite{Hall:1983id,Joshipura:1994ib,Nowakowski:1995dx,Banks:1995by}, and on the 
loop-level one in 
Refs.~\cite{Hall:1983id,Hempfling:1995wj,Chun:1999bq,Kaplan:1999ds,Grossman:1997is,Grossman:1998py,Grossman:2003gq,Hirsch:2000ef,Diaz:2003as,Davidson:2000uc,Abada:2001zh}. 
Ref.~\cite{Davidson:2000ne} gives a (nearly) complete list of one-loop contributions, 
presented in a basis-independent formalism. Detailed accounts of the one-loop 
calculation can be found in Refs.~\cite{Dedes:2006ni,Allanach:2007qc}. 

There has also been a lot of work to fit the theory calculations to the neutrino data;
see the above references as well as 
Refs.~\cite{Borzumati:1996hd,Drees:1997id,Chun:1998gp,Joshipura:1999hr,Choi:1998wc,Kong:1998bs,Rakshit:1998kd,Adhikari:1999pa,Abada:2000xr,Rakshit:2004rj,Dreiner:2011ft}. 
The main obstacle to a systematic study is the unmanageably large number of contributions to the neutrino mass matrix in the most general RPV-MSSM. Thus, 
all numerical studies are performed within specific submodels; for instance 
bilinear-only RPV models~\cite{Hirsch:2000ef,Diaz:2003as,Romao:1999up}, trilinear-only 
RPV models~\cite{Drees:1997id}, 
mixed models~\cite{Borzumati:1996hd,Cheung:1999az,Kong:1998bs,Chun:1998gp}, and
constrained MSSM (cMSSM) models extended by one (or two) RPV 
couplings~\cite{Allanach:2007qc,Dreiner:2011ft}. For an overview of the various 
types of models that have been considered, see Ref.~\cite{Barbier:2004ez}.

The above studies allow an interpretation of the neutrino data within a predictive framework, but are limited in their scope. In this work, we approach the problem from a different 
perspective. Working in the general RPV-MSSM setting, allowing for \textit{all} terms, 
we analyze the possible resulting structures (textures) of the neutrino mass matrix. To this end, 
we first argue that the most general neutrino mass matrix in 
the RPV-MSSM, to a good approximation, can be written as a sum of just two types of terms. This expression is 
general and simple but still has far-too-many free variables to be predictive. However,
appealing to minimality, we identify just four structures of the mass matrix that are relevant for
the case of two massive neutrinos. We refer to these as Minimal Oscillation Models 
(MOMs). The advantage of this approach is its simplicity and generality. By analyzing 
just four cases, it allows us to study qualitative and quantitative features of all RPV 
models in a unified, model-independent way, as long as they satisfy the MOM criteria;
we demonstrate through examples that many interesting scenarios do indeed 
fulfill this condition. If, in turn, new neutrino measurements arise, then this data can be 
systematically analyzed in terms of the MOMs we present here, instead of in terms of 
the many, many different RPV-MSSM neutrino-mass models.

This paper is organized as follows. In~\cref{sec:2}, we introduce the RPV-MSSM
(and our notation). In~\cref{sec:3}, we discuss neutrino masses in the 
$R$-parity-violating context. In~\cref{sec:4}, we define the MOM framework and classify the four relevant structures of the neutrino mass matrix that arise in the RPV-MSSM. In~\cref{sec:5}, we summarize the current status of the neutrino data. We then analyze the four classes of MOMs, studying their general features in~\cref{sec:6}. We solve each class by numerically fitting to the neutrino data in~\cref{sec:7}. Finally, in~\cref{sec:8}, we consider example applications to show how results from the MOM framework can be directly translated to specific neutrino-mass models in the RPV-MSSM. We conclude in~\cref{sec:9}.


\section{\texorpdfstring{$\mathbf{R}$}{}-Parity Violation: Theoretical Framework}
\label{sec:2}
Assuming the $N=1$ SUSY algebra, and the MSSM particle spectrum, the most general renormalizable superpotential invariant under the SM gauge group is,
\begin{equation}
W = W_{\mathrm{MSSM}} + W_{\mathrm{LNV}} + W_{\mathrm{BNV}}\,,
\label{eq:TFeq1}
\end{equation}
with,
\begin{eqnarray}
W_{\mathrm{MSSM}} &=& h_e^{ij}H_dL_i\Bar{E}_j + 
h_d^{ij}H_dQ_i\Bar{D}_j + h_u^{ij}Q_iH_u\Bar{U}_j\nonumber \\
&&+ {\mu}H_uH_d\,, \nonumber \\
W_{\mathrm{LNV}} &=& \frac{1}{2}\lam^{ijk}L_iL_j\Bar{E}_k + 
\lam'^{ijk}L_iQ_j\Bar{D}_k + \kappa^{i}H_uL_i\,, \nonumber\\
W_{\mathrm{BNV}} &=& 
\frac{1}{2}\lam''^{ijk}\Bar{U}_i\Bar{D}_j\Bar{D}_k\,.
\label{eq:TFeq1a}
\end{eqnarray}
In the notation we employ, $L$ $(Q)$ and $\Bar{E}$ $(\Bar{U},\Bar{D})$
label the lepton (quark) $\mathrm{SU}(2)_{L}$-doublet and -singlet 
chiral superfields, respectively, while $H_u, H_d$ refer to the 
$\mathrm{SU}(2)_{L}$-doublet Higgs chiral superfields. All
gauge indices are suppressed while the generational ones have been retained explicitly: $i, j, k =1,
2, 3$, with a summation implied over repeated indices. The $\lam$'s and 
the $h$'s are dimensionless Yukawa couplings, while $\mu$ and the 
$\kap$'s are dimension-one mass parameters.

In~\cref{eq:TFeq1}, the $W_{\mathrm{MSSM}}$ terms conserve both lepton-
$\left(L\right)$ and baryon-number $\left(B\right)$, the $W_{\mathrm 
{LNV}}$ terms violate only $L$, and the $W_{\mathrm{BNV}}$ terms 
violate only $B$. A disconcerting consequence of allowing unsuppressed $L$- 
and $B$-violating terms simultaneously is proton decay at a rate that 
is disallowed by experimental constraints on the proton lifetime, $\tau
_p >\SI{3.6e29}{yrs}$~\cite{Workman:2022ynf}. The usual approach
in the MSSM is to invoke $R$-parity~\cite{Farrar:1978xj}, a 
$\mathbb{Z}_2$ symmetry that allows $W_{\mathrm{MSSM}}$, while 
disallowing the $R$-parity-violating terms, $W_{\mathrm{RPV}} 
\equiv W_{\mathrm{LNV}} + W_{\mathrm{BNV}}$. However, to stabilize
the proton, $R$-parity is sufficient, but not necessary. For 
instance, forbidding either the $W_{\mathrm{BNV}}$ or the $W_
{\mathrm{LNV}}$ terms alone results in a stable proton.\footnote{If
the lightest neutralino is lighter than the proton~\cite{Dreiner:2009ic}, then the proton can also decay with just 
$W=W_{\mathrm{MSSM}}  + W_{\mathrm{BNV}}$, \textit{e.g.}, $p\to 
K^+\tilde\chi^0_1$~\cite{Chamoun:2020aft}.} Baryon triality, $B_3$, 
is such a symmetry that forbids the former and leaves the latter~\cite{Dreiner:2006xw}. In fact, unlike $R$-parity, $B_3$ even 
forbids potentially dangerous proton-decay operators of dimension five. We note that $R$-parity and $B_3$ are the only $\mathbb{Z}_2$ or $\mathbb{Z}_3$ symmetries possible with the MSSM 
low-energy particle content free from gauge anomalies~\cite{Ibanez:1991hv,Ibanez:1991pr}; the 
higher symmetries have been classified in Ref.~\cite{Dreiner:2005rd}.
  
$R$-parity-violating phenomenology differs strongly from the 
$R$-parity-conserving case~\cite{Dreiner:1991pe,Dreiner:1997uz,Barbier:2004ez,Dercks:2017lfq}. 
Collider signals are no longer dominated by missing transverse 
momentum, the lightest neutralino is no longer a dark 
matter candidate, and baryogenesis, lepton-flavor violation and 
neutrino masses arise naturally. We summarize the last point, 
central to the further discussion.

\section{Neutrino Masses and \texorpdfstring{$\mathbf{R}$}{}-parity Violation}
\label{sec:3}

For neutrino masses at next-to-leading order, without loss of 
generality, we specialize to the $B_3$-MSSM, and abusively call it the 
RPV-MSSM. Our superpotential is,
\begin{align}
W_{B_3} &= W_{\mathrm{MSSM}} + W_{\mathrm{LNV}}.
\label{eq:TFeq2a}
\end{align}
There is no quantum number
distinguishing $H_d$ from $L_i$ and hence, we define the following 
vectors and matrix: 
\begin{align}
L_\alpha &\equiv \left(H_d, L_1, L_2, L_3\right)\,,\label{eq:4-A}\\
\kappa^\alpha &\equiv \left(\mu, \kappa^1, \kappa^2, \kappa^3\right) 
\,,\\
\lambda'^{\alpha j k} &\equiv \left(h_d^{j k}, \lambda'^{1 j k}, 
\lambda'^{2 j k}, \lambda'^{3 j k}\right)\,,\\
\lambda^{\alpha \beta k} &\equiv \begin{pmatrix} 0 & h_e^{1 k} 
& h_e^{2 k} & h_e^{3 k}\\
-h_e^{1 k} & 0 & \lambda^{1 2 k} & \lambda^{1 3 k}\\
-h_e^{2 k} & \lambda^{2 1 k} & 0 &  \lambda^{2 3 k}\\
-h_e^{3 k} & \lambda^{3 1 k} & \lambda^{3 2 k} & 0\\
\end{pmatrix}\,. \label{eq:4-D}
\end{align}
$\alpha, \beta = 0, 1, 2, 3$ label the vector and matrix components, 
\textit{e.g.}, $L_0\equiv H_d$, and $\lam^{\prime0jk}\equiv h_d
^{jk}$.  $j,k=1,2,3$ are as before. We can thus write the 
superpotential as,
\begin{align}
W_{B_3} &=  \frac{1}{2}\lambda^{\alpha \beta k}L_\alpha L_\beta 
\Bar{E}_k + \lambda'^{\alpha j k}L_\alpha Q_j \Bar{D}_k + h_u^{ij} Q_i 
H_u \Bar{U}_j\notag\\
&\quad + \kappa^\alpha H_u L_\alpha \,.
\label{eq:TFeq2}
\end{align}
In addition, there are the soft-breaking terms,
\begin{align}
\mathcal{L}_{soft} &= \text{mass terms} + \frac{1}{2}A^{\alpha \beta k}\tilde{L}_\alpha \tilde{L}_\beta \tilde{\Bar{E}}_k + A'^{\alpha j k}\tilde{L}_\alpha \tilde{Q}_j \tilde{\Bar{D}}_k\notag\\
&\qquad + A_u^{ij} \tilde{Q}_i H_u \tilde{\Bar{U}}_j + B^\alpha H_u \tilde{L}_\alpha + h.c.\,,
\label{eq:TFeq3}
\end{align}
where the fields appearing in the above equation are the scalar 
components of the corresponding chiral superfields. 
The definitions of the parameters with one $\left(B^\alpha,
A'^{\alpha j k}\right)$, and two $\left(A^{\alpha \beta k}\right)$ 
Greek indices are obvious generalizations of the MSSM
soft-breaking parameters, \textit{cf.}
Eqs.~(\ref{eq:4-A})-(\ref{eq:4-D}).

After spontaneous symmetry breaking,  the neutrinos, the neutral 
gauginos, and the higgsinos mix, leading to a 7$\times$7 mass matrix. 
At tree level in the gauge eigenbasis, $\left(-i\tilde{B}; -i\tilde{W}^0; 
\tilde{H}^0_u;\nu_\alpha\right)$, with $\nu_\alpha\equiv \left(\tilde{H}^0_d, 
\nu_i\right)$, we have the overall mass matrix,
\begin{equation}
\mathbf{M}_N = \begin{pmatrix} \mathbf{M}_{4\times4} & 
\mathbf{m}_{4\times3}\\[2mm]
\mathbf{m}^T_{3\times4} & \mathbf{0}_{3\times3}
\end{pmatrix}\,,
\label{eq:TFeq4}
\end{equation}
with $\mathbf{M}_{4\times4}$ corresponding to the 
MSSM neutralino mass matrix,
\begin{equation}
\mathbf{M}_{4\times4} = \begin{pmatrix} M_1 & 0 & \frac{g_1v_u}{2} 
& \frac{-g_1v_d}{2}\\[1.7mm]
0 & M_2 & \frac{-g_2v_u}{2} & \frac{g_2v_d}{2}\\[1.6mm]
\frac{g_1v_u}{2} & \frac{-g_2v_u}{2} & 0 & -\kappa^0\\[1.6mm]
\frac{-g_1v_d}{2} &  \frac{g_2v_d}{2} & -\kappa^0 & 0
\end{pmatrix}\,,
\end{equation}
and the sub-block $\mathbf{m}_{4\times3}$ containing the RPV terms,
\begin{equation}
\mathbf{m}_{4\times3} = \begin{pmatrix} \frac{-g_1v_1}{2} & 
\frac{-g_1v_2}{2} & \frac{-g_1v_3}{2}\\[1.6mm]
\frac{g_2v_1}{2} & \frac{g_2v_2}{2} & \frac{g_2v_3}{2}\\[1.6mm]
-\kappa^1 & -\kappa^2 & -\kappa^3\\
0 & 0 & 0\\
\end{pmatrix}\,.
\end{equation}
$\tilde{B}$ and $\tilde{W}^0$ denote the neutral gauginos,
$\tilde{H}^0_u$, $\tilde{H}^0_d$ the neutral higgsinos, and $\nu_
i$ the neutrinos. $M_1, M_2$, and $g_1, g_2$ are the electroweakino
soft-breaking masses and gauge couplings, respectively.
$\frac{v_u}{\sqrt{2}},\frac{v_d}{\sqrt{2}},\frac
{v_i}{\sqrt{2}}$, with $i=1,2,3$, are the vacuum expectation 
values (vevs) of the two neutral Higgs fields and the three 
sneutrinos, respectively.

The mass matrix of~\cref{eq:TFeq4} has been discussed abundantly
in the literature. The details of the diagonalization procedure can be 
found, for instance, in Ref.~\cite{Dedes:2006ni}. The scales in the various blocks are expected to have a hierarchy. Given the lower mass 
bounds on sparticles from the LHC, one expects the 
lepton-number-conserving SUSY scales of $\mathbf{M}_ 
{4\times4}$ to be at least $\sim\mathcal{O}\left( \SI{1} {\tera
\electronvolt}\right)$, while the lepton-number-violating scales 
of $\mathbf{m}_{4\times3} $ are constrained by various stringent
low-energy bounds to be much smaller~\cite{Barbier:2004ez}. For example, the cosmological limit on neutrino masses implies
$v_i,\,\kappa^i \lesssim \mathcal{O}\left(\SI{1}{
\mega\electronvolt}\right)$~\cite{Allanach:2003eb}. One can then proceed \`{a} la see-saw, and end up with an approximately 
block-diagonal matrix,
\begin{equation}
\mathbf{M}_N^{\text{diag}} \approx \begin{pmatrix} 
\mathbf{M}_{4\times4} & 0\\[2mm]
0 & \mathbf{M}_{\nu}\\
\end{pmatrix}\,,
\end{equation}
where,
\begin{equation}
\mathbf{M}_{\nu}^{ij} \equiv
\frac{\left(M_1g^2_2+M_2g_1^2\right)}{4\det\left(\mathbf{M}_{4
\times4}\right)}\left(v_i\kap^0-v_d\kap^i\right)\!\left(v_j
\kap^0-v_d\kap^j\right)\,.
\label{eq:TFeq5}
\end{equation}
The neglected contributions to $\mathbf{M}_{4\times4}$ and $\mathbf 
{M}_{\nu}$ in the above approximation are of order $\frac{\| 
\mathbf{m}^2_{4\times3}\|}{\|\mathbf{M}_{4\times4}\|}\lesssim
\mathcal{O}\left(\SI{1}{\electronvolt}\right)$, and $\frac{\| 
\mathbf{m}^3_{4\times3}\|}{\|\mathbf{M}^2_{4\times4}\|}\lesssim
\mathcal{O}\left(\SI{e-6}{\electronvolt}\right)$, respectively~\cite{Dedes:2006ni}. 
To this order, $\mathbf{M}_{4
\times4}$ is unaffected compared to the MSSM and we identify it 
as the neutralino mass matrix, and the corresponding mass  
eigenstates as the neutralinos. The high-scale-suppressed 
$\mathbf{M}_\nu$ can then be identified as the effective neutrino 
mass matrix.

Importantly, $\mathbf{M}_\nu$ is rank one at tree level, 
\textit{i.e.}, it has only one non-zero eigenvalue. However, at
least two neutrinos must be massive to explain the oscillation 
data. This can be achieved by including one-loop corrections~\cite{Hall:1983id}, which have been computed several times in the
literature. We shall use the results of Ref.~\cite{Davidson:2000ne}.
There, an almost complete list of the one-loop contributions to 
$\mathbf{M}_{\nu}$ is presented; certain contributions whose effects
are expected to be negligible have been dropped. 

The advantage of using the formalism of Ref.~\cite{Davidson:2000ne} is that the contributions have been written in terms of basis invariants. It is common practice in the literature to 
use the $\mathrm{U}(4)$ `flavor' freedom to rotate $L_{\alpha}$ to 
a specific basis. Various useful choices have been identified -- 
the most common being the vanishing-$\kappa^i$ 
basis~\cite{Hall:1983id,Dreiner:2003hw}, and the vanishing-sneutrino-vev basis~\cite{Grossman:1998py}. The notation of
Ref.~\cite{Davidson:2000ne} is invariant under this $\mathrm{U}
(4)$ and is useful to compare results across works using 
different bases. 

We present an adapted version of all the contributions calculated in
Ref.~\cite{Davidson:2000ne} in~\cref{tab:TFtab1}. Each entry can have 
multiple diagrams contributing. Further, the expressions are not exact 
but are meant to indicate the resulting form. For brevity, we have
set all the SUSY mass scales to $m_\text{SUSY}$, all gauge couplings to
$g$,  dropped some factors involving the ratio of vevs -- $\tan\beta$, and taken some scalar-sector flavor matrices as diagonal. We discuss 
the implications of this point in more detail shortly. The $\delta$'s 
appearing in the table are the basis invariants. Throughout, the 
constraints we derive apply to them but the results can always be 
translated into a specific basis using the general expressions~\cite{Davidson:2000ne}. For instance, in the vanishing-sneutrino-vev 
basis, we have,
\begin{align}
    \delta^i_{\kappa} &= \frac{\kappa^i}{|\kappa|}\,,  &\delta^i_{B} &= \frac{B^i}{|B|}\,,\notag\\
    \delta^{ijk}_{\lambda} &= \lambda^{ijk}\,,  &\delta^{ijk}_{\lambda'} &= \lambda'^{ijk}\,,
    \label{eq:TFeq5a}
\end{align}
with, 
\begin{eqnarray}
|\kappa|^2 \equiv {\sum\limits_{\alpha=0}^3 \left|\kappa^{\alpha}\right|^2}, \qquad
|B|^2\equiv {\sum\limits_{\alpha=0}^3 \left|B^{\alpha}\right|^2}\,.
\end{eqnarray}
Motivated by the above expressions, we often loosely refer to the $\delta$'s as `RPV couplings'.
\begin{table}[t]
\captionsetup{justification=raggedright,
}
\begin{center}
\scalebox{0.8}{
\begin{tabular}{ |c|c| } 
\hline\hline
Contribution & $16 \pi^2 m_\text{SUSY} \mathbf{M}_{\nu}^{ij}$\\
\hline 
Tree-Level & $16\pi^2m_0m_{\text{SUSY}}\delta_{\kappa}^{i}\delta_
{\kappa}^{j}$\\[1.2mm]
$1$ & $\delta_{\lambda}^{ink}\delta_{\lambda}^{jkn}m_{e_n}m_{e_k} 
+ \left(i \leftrightarrow j\right)$\\[1.2mm]
$2$ & $3\delta_{\lambda'}^{ink}\delta_{\lambda'}^{jkn}m_{d_n}m_
{d_k} + \left(i \leftrightarrow j\right)$\\[1.2mm]
$3$ & $g^2 \delta_\text{B}^i \delta_\text{B}^j 
m^2_\text{SUSY}/4$\\[1.2mm]
$4$ & $3 \left( \delta_\kappa^i \delta_{\lambda'}^{jkk} + 
\delta_\kappa^j \delta_{\lambda'}^{ikk}\right) m^2_{d_k} 
h^k_d$\\[1.2mm]
$5$  & $\delta_{\lambda}^{ijk}\delta_{\text{B}}^{k}m_{e_k} 
\left(m_{e_j} h^j_e - m_{e_i} h^i_e \right)$\\[1.2mm]
$6$ &  $\left( \delta_\kappa^i \delta_{\lambda}^{jkk} + \delta_\kappa^j \delta_{\lambda}^{ikk}\right) m^2_{e_k} h^k_e$\\[1.2mm]
$7$ & $\delta^i_\kappa \delta^j_\kappa m_{e_i} m_{e_j} h^i_e h^j_e + \left(i \leftrightarrow j\right)$\\[1.2mm]
$8$ & $\delta^i_\kappa \delta^j_\kappa \left[ \left(m_{e_i} h^i_e \right)^2+ \left(m_{e_j} h^j_e\right)^2 \right]$\\[1.2mm]
$9$ & $\delta_\text{B}^i \delta^j_\kappa \left(m_{e_i}h^i_e \right)^2 +\delta^j_\text{B} \delta_\kappa^i  \left(m_{e_j}h^j_e \right)^2$\\[1.2mm]
$10$ & $\delta^{ijk}_\lambda \delta^{k}_\kappa m_{e_k} \left(m_{e_i} h^i_e - m_{e_j} h^j_e \right)$\\[1.2mm]
$11$ & $\left(\delta_\text{B}^i \delta_\kappa^j + \delta_\text{B}^j  \delta_\kappa^i\right) h^i_e h_e^j m_{e_i} m_{e_j}$\\[1.2mm]
$12$ & $g \left(\delta^i_\kappa \delta^j_\text{B} m^2_{e_i} +\delta^j_\kappa \delta^i_\text{B} m^2_{e_j} \right)$\\[1.2mm]
$13$ & $g\, \delta^i_\kappa \delta^j_\kappa \left(m^2_{e_i}+ m^2_{e_j} \right)$\\[1.2mm]
$14$ & $g\, m_\text{SUSY} \left( \delta_\kappa^i \delta_{\lambda}^{jkk} + \delta_\kappa^j \delta_{\lambda}^{ikk}\right) m_{e_k}$\\[1.2mm]
$15$ & $3\,g\, m_\text{SUSY} \left( \delta_\kappa^i \delta_{\lambda'}^{jkk} + \delta_\kappa^j \delta_{\lambda'}^{ikk}\right) m_{d_k}$\\[1.2mm]
$16$ & $g^2 m_\text{SUSY}^2 \left(\delta^i_\text{B}\delta^j_\kappa +\delta^j_\text{B}\delta^i_\kappa \right)/4$\\[1.2mm]
$17$ & $g \left(\delta^i_\kappa \delta^j_\text{B} m^2_{e_j} +\delta^j_\kappa \delta^i_\text{B} m^2_{e_i} \right)$\\
[1ex]
\hline\hline
\end{tabular}}
\end{center}
\caption{$\mathbf{M}_{\nu}$ contributions as calculated in Ref.~\cite{Davidson:2000ne}. 
The numbered entries are due to one-loop diagrams. Summation is implied over all 
repeated indices other than $i, j$. The $\delta$'s are the RPV basis invariants. $m_0$ is the tree-level mass scale of~\cref{eq:TFeq5}, the remaining $m$'s are
the SM fermion masses, and the $h$'s are the Yukawas.}
\label{tab:TFtab1}
\end{table}

Even though the contributions in~\cref{tab:TFtab1} are in terms of 
basis invariants, they have been written in a specific basis which 
corresponds approximately to the charged lepton mass basis. 
Analogous to the neutral case, the uncolored $5\times5$ 
charged fermion mass matrix mixes the charged gaugino, 
charged Higgsino, and the three charged leptons. However, it also has a 
hierarchical structure and can be approximately block-diagonalized to 
obtain separate $3\times3$ and $2\times2$ mass matrices, corresponding to the charged leptons and  charginos, respectively. The charged lepton matrix, subject to small neglected terms, 
can then be diagonalized as usual.

With a diagonal charged lepton mass matrix, one can then diagonalize 
the effective neutrino mass matrix $\mathbf{M}_{\nu}$:
\begin{equation}
\mathbf{M}_{\nu} = 
U^*_{\text{PMNS}}\mathbf{M}^{\text{diag}}_{\nu}U^{\dag}_
{\text{PMNS}}\,,
\label{eq:TFeq6}
\end{equation}
where $\mathbf{M}_{\nu}^{\text{diag}}$ is the diagonalized neutrino mass matrix, and $U_{\text{\text{PMNS}}}$ is the PMNS matrix that appears in the charged-current interactions of the neutrinos. It 
should be clear that the PMNS matrix, as defined here, is a $3\times3$ sub-matrix 
inside the larger $5\times7$ matrix describing the mixing between all the $5$ 
charged fermions and  $7$ neutral fermions. Thus $U_{\text{PMNS}}$
is not exactly unitary, here. However, these effects are suppressed by 
the high-energy scales and we ignore them~\cite{Dedes:2006ni}.


\section{Minimal Oscillation Models}
\label{sec:4}
The matrix equation to be solved is,
\begin{equation}
{\color{white}.}\!\!\!\!    \mathbf{M}_{\nu} \overset{!}{=} 
\mathbf{M}^{\text{exp}}_{\nu} = 
U^*_{\text{PMNS}}\text{diag}\left(m_{\nu_1}, 
    m_{\nu_2}, m_{\nu_3}\right)U^{\dag}_{\text{PMNS}}\,,
\label{eq:MOMeq1}
\end{equation}
where $\mathbf{M}_{\nu}$ is the one-loop effective neutrino mass matrix computed from 
the RPV Lagrangian and the right-hand side is to be determined through fits to the 
neutrino oscillation data. The difficulty of numerically analyzing the 
\textit{most general} RPV neutrino-mass model should be evident from the large number of contributions in~\cref{tab:TFtab1}. The goal of this paper 
is to show that -- despite this -- due to the structure of the entries, only a small set of truly `distinct models' is possible. These, in turn, can be systematically analyzed.

~\cref{eq:MOMeq1} is a set of six complex equations, or $12$ real constraints. 
Nine of these are physical, corresponding to the three neutrino masses, the three mixing angles, and the three $CP$-violating phases in the PMNS matrix (see the parameterization of the 
PMNS matrix below). The remaining three are not physical 
constraints. They correspond to arbitrary phases in the PMNS matrix that can be rotated 
away~\cite{Dreiner:2007yz}. 

Looking at~\cref{tab:TFtab1}, it is clear that the most general one-loop mass matrix 
arising in RPV models, entering~\cref{eq:MOMeq1} on the left-hand side, has too many 
parameters; the system is very much underdetermined. Just the RPV 
superpotential has 
$\left(\kappa^i, \lam^{ijk}, \lam'^{ijk} 
\right)$  $3+9+27 =$ 39 free complex (or $78$ real) parameters. 
As mentioned,  the usual approach of numerical studies has been to assume specific 
models. For instance, bilinear-only models $\left(\lam^{ijk}=\lam'^{ijk} =0\right)$~\cite{Hirsch:2000ef}, or unification approaches that begin with a small number of
non-zero $\lam$'s at $M_X$, which then generate other non-zero couplings at the low 
scale through renormalization-group effects~\cite{Dreiner:2011fp}, etc. Our aim in this 
work is to remain as general as possible.

In a first step, we observe that all the contributions of~\cref{tab:TFtab1} (except 
entries 5 and 10 -- we return to this point) can be reduced to combinations of 
just two types of structures:
\begin{itemize}
    \item[1.] $x^ix^j$
    \item[2.] $x^iy^j + y^ix^j$
\end{itemize}
Here, the $x^i$ and $y^i$ are place-holding variables with mass-dimension $\left[M\right]^{1/2}$ that are directly proportional to the $\delta$'s of~\cref{tab:TFtab1}. For instance, 
when the first one-loop entry of the table is expanded out, we get,
\begin{align}
    \mathbf{M}_{\nu}^{ij} &= \frac{1}{8\pi^2 m_\text{SUSY}}\left(\delta_{\lam}^{i33}\delta_{\lam}^{j33}m^2_{\tau} + \delta_{\lam}
    ^{i23}\delta_{\lam}^{j32}m_{\tau}m_{\mu}\right. \notag\\
    &+ \left. \delta_{\lam}^{i32}\delta_{\lam}^{j23}m_{\tau}m_{\mu} + \delta_{\lam}^{i22}\delta_{\lam}^{j22}m^2_{\mu} + \ldots \right)\notag\\
    &= x_1^ix_1^j + \left(x_3^ix_4^j + x_4^ix_3^j\right) + x_2^ix_2^j +\ldots\,,
\end{align}
with
\begin{align}
x_1^i&\equiv \frac{\delta^{i33}_\lam m_\tau}{2\pi\sqrt{2\, m_\text{SUSY}}} \,, \quad x_2^i\equiv \frac{\delta^{i22}_\lam m_\mu}{2\pi\sqrt{2\, m_\text{SUSY}}}\,,\notag\\
x_3^i&\equiv \frac{\delta^{i23}_\lam m_\mu}{2\pi\sqrt{2 m_\text{SUSY}}} \,,\quad
x_4^i\equiv \frac{\delta^{i32}_\lam m_\tau}{2\pi\sqrt{2 m_\text{SUSY}}} \,.
\end{align}
We see that the first and fourth terms correspond to an $x^ix^j$ structure while the 
second and third terms together form an $x^iy^j + y^ix^j$ structure. The choice of 
the variables is non-unique. For instance, one can multiply $x_3^i$ by a 
constant and divide $x_4^i$ by the same constant without changing the total contribution. Similarly, $x^i_1, x^i_2$ are defined only up to a sign. The important point is that the variables are chosen to be directly proportional to the $\delta$'s.

One can similarly check the other entries. So, (ignoring the two exceptions) the most
general one-loop effective neutrino mass matrix in RPV models can symbolically be written as,
\begin{align}
    \mathbf{M}_{\nu}^{ij} &= \sum_\alpha x^i x^j + \sum_\beta \left(x^iy^j + y^ix^j\right)\,,
    \label{eq:MOMeq2}
\end{align}
where the sum over $\alpha \left(\beta\right)$ is such that all the contributions of the first (second) type in~\cref{tab:TFtab1} are included.  Given~\cref{eq:MOMeq2}, the simplest
neutrino mass matrix that one can construct in the RPV-MSSM is with only one set, $x^i$:
\begin{align}
    \mathbf{M}_{\nu}^{ij} = x^i x^j\,.
\end{align}
The rank of this matrix is one, leading to two massless neutrinos which is inconsistent 
with oscillation data. The next simplest case involves two sets $x^i, x'^i$. Consider, 
for instance,
\begin{align}
    \mathbf{M}_{\nu}^{ij} = x^ix^j + x'^i x'^j\,.
\label{eq:MOMeq3}
\end{align}
This is, in general, a rank two structure and could possibly explain neutrino data if the
lightest neutrino is massless. However, it does not work if the two sets are linearly 
dependent. To see this, let $x'^i = k x^i$; we get,
\begin{align}
    \mathbf{M}_{\nu}^{ij} &= x^ix^j + k^2 x^i x^j
    = (1+k^2)x^i x^j\notag\\
    &= \tilde{x}^i\tilde{x}^j\,,
\end{align}
where $\tilde{x}^i \equiv \sqrt{1+k^2}\,x^i$. The structure reduces to the rank one case.
Thus, we must have two linearly independent sets.

We emphasize that the number of linearly independent $x^i$ sets is not the same as the 
number of RPV-coupling sets that give rise to them. For instance, one can check that 
reducing the tree-level contribution and entry 7 of~\cref{tab:TFtab1} to the form of~\cref{eq:MOMeq2} requires two linearly independent sets, $x^i$ and $x'^i$, even if both 
contributions arise from just a single RPV-coupling set, $\delta^i_{\kappa}$. The 
inverse is also possible: Several RPV parameters can be written in terms of just one set $x^i$, \textit{cf.}~\cref{sec:B}. 

With the above in mind, all possible structures that can be written with two linearly independent sets, $x^i$ and $x'^i$, are:
\begin{itemize}
    \item Class~1: $\mathbf{M}_{\nu}^{ij} = x^ix'^j + x'^ix^j$
    \item Class~2: $\mathbf{M}_{\nu}^{ij} = x^ix^j + (x^ix'^j + x'^ix^j)$
    \item Class~3: $\mathbf{M}_{\nu}^{ij} = x^ix^j + x'^ix'^j$
    \item Class~4: $\mathbf{M}_{\nu}^{ij} = x^ix^j + x'^ix'^j + A\left(x^ix'^j + x'^ix^j\right)$
\end{itemize}
These four structures\footnote{The Class~4 structure follows by using~\cref{eq:MOMeq2} to write the most general 
expression involving only $x^i, x'^i$, or couplings that are a linear combination of the two; and then suitably redefining the variables such that all the proportionality constants appear only in $A$. The detailed steps are given in~\cref{sec:B}.} are all rank two -- the minimum required, and are the only possible solutions to the neutrino data, as long as one is interested in a minimal setup. This is a crucial observation of this paper. We analyze these structures in the following. 

Let us now discuss the exceptions mentioned above -- entries 5 and 10 in~\cref{tab:TFtab1}. Before proceeding, we note that the various contributions to the neutrino mass matrix in~\cref{tab:TFtab1} have a natural hierarchy. For instance, consider a scenario with only the $\delta^i 
_{\kappa}\not=0$, leading to four contributions: The tree-level term, and entries  7, 8 
and 13. Contributions 7 and 8 are suppressed by at least two extra powers of the small 
lepton-Yukawas compared to the other two. Thus, to a first approximation, we can neglect 
them.\footnote{One should make sure that the $\tan\beta$ factors, not shown in~\cref{tab:TFtab1}, cannot undo the hierarchies. As discussed in~\cref{sec:8}, 
this is indeed not the case here.} The remaining two contributions can be reduced to a 
Class~2 MOM structure, \textit{cf.}~\cref{sec:8}. This is a general trend, not specific to this example; we explore several examples later. 

Indeed, the exceptions 5, 10 are not too worrisome for the same reason. They are Yukawa suppressed compared to the 
other terms involving the same sets of couplings. Let us see this explicitly for 
entry 5. The RPV parameters involved are $\delta_B^i$ and $\delta^{ijk}_\lam$. Assuming other couplings vanish, this
entry would be competing with entries 1 and 3. We can estimate the magnitudes of the 
three contributions as:
\begin{align}
    \text{Entry 1} &\sim |\delta_\lam|^2 m^2_{\tau}\,,\notag \\
    \text{Entry 3} &\sim \frac{g^2|\delta_B|^2m^2_{\text{SUSY}}}{4\cos^2\beta}\,,\notag \\
    \text{Entry 5} &\sim |\delta_\lam||\delta_B| m^2_{\tau} h_{\tau}\tan\beta\,,
\end{align}
where we have assumed a common magnitude for all generations of a 
particular coupling and hence dropped the latin indices. Further, we 
have only retained the terms proportional to the dominant $\tau$ lepton
Yukawas for entries 1 and 5. The $\tan\beta$ and $\cos\beta$ factors are 
read off from the expressions found in Ref.~\cite{Davidson:2000ne}. Substituting the known values, and taking $m_{\text{SUSY}} \sim 
\mathcal{O\left(\SI{1}{\tera\electronvolt}\right)}$, one can easily prove that there is no configuration of parameters for which Entry 5 becomes important relative to the other two contributions.
A similar argument can be made for entry 10. 

Going beyond rank two, it is possible that all three neutrinos are massive, requiring a rank three structure and a third linearly independent set, $x''^i$.  Three linearly independent sets is the most general case and hence this approach would capture all RPV-MSSM neutrino-mass models.
However, the number of classes to be
considered is large 
making them 
less conducive for systematic numerical studies.
In this paper we focus only on 
the rank two case.

We should note that the MOM approach does not cover the most general rank two structure possible in an RPV 
model. In~\cref{tab:TFtab1}, we assume some scalar-sector mixing matrices are diagonal in 
the charged lepton mass basis we are working in. This includes matrices that diagonalize the charged doublet and singlet sleptons and down-type squarks, and matrices that 
describe the left-right sparticle mixings; 
that is, we assume the sparticle and particle flavors are aligned
with no inter-generational mixing. The fact that all contributions can be reduced to one 
of just two types of structures relies on this assumption. 
Further, by setting all SUSY scales in~\cref{tab:TFtab1} common, we have neglected the 
possibility that strong hierarchies in the scalar sector may undo some of the hierarchies that we saw above. Finally, it is possible that three linearly independent sets --$x^i, x'^i, x''^i$ -- lead to a rank two structure through specific cancellations (see~\cref{sec:B} 
for an illustration of this point). The four structures listed above with only two sets would 
not capture such models. Hence, we shall refer to these as Minimal 
Oscillation Models (MOMs). MOMs are not minimal in the sense of having the fewest number 
of RPV parameters. They are, rather, minimal in the sense that the mass matrix has the 
minimal structure demanded by the data.

In the absence of any experimental information about the scalar sector, we believe the 
MOM framework provides a minimal setting that is widely applicable for the interpretation
of neutrino data. It is simple and predictive. After briefly reviewing the neutrino data,
we analyze qualitative and quantitative features of the models in the subsequent  
sections.

\section{Neutrino Data}
\label{sec:5}
The PMNS matrix can be parameterized~\cite{Workman:2022ynf} by the three mixing angles 
$\left(\theta_{12}, \theta_{13}, \theta_{23}\right)$, one $CP$-violating Dirac phase $\left(\delta_ 
{CP}\right)$, and two $CP$-violating Majorana phases $\left(\eta_1, \eta_2\right)$:
\begin{widetext}
\begin{align}
U_\text{PMNS}
=&
\begin{pmatrix} 
 c_{12}c_{13} & s_{12}c_{13} & s_{13} e^{-i \delta_{{CP}}}\\
 -s_{12}c_{23} - c_{12}s_{23}s_{13} e^{i \delta_{{CP}}} & c_{12}c_{23} - 
 s_{12}s_{23} s_{13} e^{i \delta_{{CP}}} & s_{23}c_{13} \\
 s_{12}s_{23} - c_{12} c_{23}s_{13} e^{i \delta_{{CP}}}  & -c_{12}s_{23} 
 -s_{12}c_{23} s_{13} e^{i \delta_{{CP}}}  & c_{23}c_{13}
\end{pmatrix} 
\begin{pmatrix} 
e^{i\eta_1} & 0 & 0 \\
0 & e^{i\eta_2} & 0 \\
0 & 0 & 1
\end{pmatrix}\,,
\label{eq:ND1}
\end{align} 
\end{widetext}
where $\sin \theta_{ij}$ and $\cos \theta_{ij}$ are written as $s_{ij}$ and $c_{ij}$ 
respectively. Without loss of generality, the angles $\theta_{ij}$ can be taken to lie in 
the first quadrant, \textit{i.e.}, $\theta_{ij} \in [0, \pi/2]$, and the phases $\delta 
_{CP}, \eta_i \in [0,2\pi]$.

We summarize neutrino oscillation data from Ref.~\cite{Esteban:2020cvm} in~\cref{tab:neudata}. We follow their assumption of three active oscillating neutrinos. 
They present the best-fit values of the combined global analysis of atmospheric, solar, 
reactor, and accelerator neutrinos. Here, we specifically choose their fit including the SK atmospheric data~\cite{Hosaka:2005um,Ashie:2004mr}. The data still allows one neutrino to be
massless; we work in this limit. For Normal Ordering (NO) $\left(m_1 < m_2 < m_3\right)$ this 
means $m_1 \approx 0$, and for Inverted Ordering (IO) $\left(m_3 < m_1 < m_2\right)$ it means $m_3 
\approx0$. In the global neutrino fit, the Normal Ordering is preferred over the inverted
ordering, however this has become less pronounced with more recent data~\cite{Esteban:2020cvm,Kelly:2020fkv}. 
\begin{table}[h]
\centering
\begin{tabular}{|c| c || c|}
\hline\hline
 & \;Normal Ordering\; & \;Inverted Ordering\; \\ [0.5ex] 
\hline && \\[-2ex]
$\theta_{12} /^\circ$& $33.44^{+0.77}_{-0.74}$ & $33.45^{+0.78}_{-0.75}$\\[0.5ex]
$\theta_{23} /^\circ$ & $49.2^{+0.9}_{-1.2}$ &$49.3^{+0.9}_{-1.1}$\\[0.5ex]
$\theta_{13} /^\circ$ & $8.57^{+0.12}_{-0.12}$ & $8.60^{+0.12}_{-0.12}$\\[0.5ex]
$\delta_{{CP}} /^\circ$ & $197^{+27}_{-24}$ & $282^{+26}_{-30}$\\
$\frac{\Delta m_{21}^2}{\SI{e-5}{\electronvolt}}$ & $7.42^{+0.21}_{-0.20}$ & $7.42^{+0.21}_{-0.20}$\\
$\frac{\Delta m_{3l}^2}{\SI{e-3}{\electronvolt}}$ & $+2.517^{+0.026}_{-0.028}$ & $-2.498^{+0.028}_{-0.028}$\\
[1ex]
\hline \hline
\end{tabular}
\caption{Neutrino oscillation parameters from a global fit to data. The first (second) column depicts the best fit assuming NO (IO). Note 
that $\Delta^2_{3l} \equiv \Delta_{31}^2 >0$ for NO and $\Delta^2_{3l} \equiv \Delta_{32}^2 <0$ for IO.}
\label{tab:neudata}
\end{table}

We use the data as presented in~\cref{tab:neudata} for our numerical fits, except we set $\delta_{CP}=0$. Further, we also set the as-yet-undetermined Majorana phases to be zero. That is, we work in the $CP$-conserving scenario. We do this merely for convenience; the solution space is more symmetric. Nevertheless, to show our analysis can accommodate $CP$ violation, we show a sample plot in~\cref{app:CP} for $\delta_{CP}\not=0$. 

We will also find it convenient, at times, to use the so-called tri-bi-maximal (TBM) approximation\footnote{See 
Ref.~\cite{Dreiner:2011fp} for relating the TBM to RPV neutrino-mass models.}
for the angles instead of the values in~\cref{tab:neudata}~\cite{Harrison:2002er}:
\begin{align}
\sin^2(\theta_{12}) = \frac{1}{3}\,,\;\; \sin^2(\theta_{23}) = \frac{1}{2}\,, \;\;
\sin^2(\theta_{13})=0\,,\;\;\delta_{CP}=0\,.
\end{align}
Even though this scenario is ruled out by the $\sin\theta_{13}$ measurement, it gives 
convenient analytical expressions, provides initialization for numerical fits, and allows 
studying qualitative features that carry through to the experimentally viable scenarios.



\section{General Features of our Results}
\label{sec:6}
In the following, we present solutions to~\cref{eq:MOMeq1} for each of
the four classes of MOMs. As we explain below, the solution space is an infinite set. Furthermore, since the neutrino data are quite precise,
we shall ignore the experimental errors in the graphical presentation of our 
results below; technically each line in the plot should be understood to have a finite width.

There are two subtle points applying to all MOM classes worth mentioning before we solve 
them. The first concerns the basis choice. Even with our basis fixed to the (approximate) charged 
lepton mass basis, there is remnant freedom in the $U_{\text{PMNS}}$ 
matrix. This corresponds to the freedom to multiply $U_{\text{PMNS}}$ by three arbitrary 
phases~\cite{Dreiner:2007yz}:
\begin{equation}
U_{\text{PMNS}} \mapsto \text{diag}\left(e^{i\alpha_1}, 
    e^{i\alpha_2}, e^{i\alpha_3}\right)U_{\text{PMNS}}\,.
\end{equation}
Using~\cref{eq:TFeq6}, this corresponds to shifting $\mathbf{M}_\nu$:
\begin{eqnarray}
\mathbf{M}_\nu &\mapsto& \text{diag}\left(e^{-i\alpha_1}, 
    e^{-i\alpha_2}, e^{-i\alpha_3}\right)\times\mathbf{M}_\nu\times \nonumber \\ 
    &&\times \text{diag}\left(e^{-i\alpha_1}, 
    e^{-i\alpha_2}, e^{-i\alpha_3}\right)\,.
\end{eqnarray}
This, in turn, can be interpreted as shifts in the phases of the $x^i, x'^i$
variables. For instance, if $\mathbf{M}_\nu$ has a Class~1 MOM structure,
the above equation becomes:
\begin{eqnarray}
\left(x^ix'^j + x'^ix^j\right) &\mapsto& \sum_{a,b}e^{-i\alpha_i} 
    \delta^{ia}\left(x^ax'^b + x'^ax^b\right)\delta^{bj}e^{-i\alpha_j}\notag\\
    &=&\left(e^{-i\alpha_i} 
    x^i\right)\left(e^{-i\alpha_j}x'^j\right) \notag  \\
    &&+ \left(e^{-i\alpha_i}x'^i\right)\left(e^{-i\alpha_j}x^j\right)\,,
\end{eqnarray}
which is equivalent to the simultaneous transformations:
\begin{align}
x^i &\mapsto \tilde{x}^i \equiv \left(e^{-i\alpha_i} 
    x^i\right)\,, 
    \notag\\
x'^i &\mapsto \tilde{x}'^i \equiv \left(e^{-i\alpha_i} 
    x'^i\right)\,.
\label{eq:GFeq1}
\end{align}
A change of basis induces simultaneous phase rotations on the RPV couplings. This
holds for all MOM classes.

The second subtlety is the issue of degrees of freedom. MOM classes 1-3 have six free 
(complex) parameters while the fourth has seven. One might expect the six (complex) 
equations in~\cref{eq:MOMeq1} are enough to determine the system of variables for at 
least the first three classes. However, for the case at hand, the experimental matrix [right-hand side of~\cref{eq:MOMeq1}] is rank two. Hence, its last row can be written as a linear combination of 
the first two rows; the sixth constraint is redundant. We, thus, have an infinite set of 
solutions characterized by one unconstrained variable. Correspondingly, for Class~4 MOMs, we have two unconstrained variables.

To summarize, our solution space is an infinite set parameterized by one (or two) free 
variables. Further, the phases of the variables are only meaningful once the basis is 
completely specified. Our results are presented in the basis $\alpha_1,\alpha_2,\alpha_3
= 0$ with $U_{\text{PMNS}}$ given by~\cref{eq:ND1}. 

We now study the solution spaces for MOMs in detail. The analytical expressions are
presented in~\cref{sec:A}; our emphasis here is on a qualitative discussion of the 
general features. We exclude a study of Class~4 models. They are straightforward to solve numerically (see~\cref{sec:7} for the discussion on numerical fits), but the analytical expressions are rather long and 
awkward. Furthermore, a visual representation would require non-intuitive 
three-dimensional plots.

\subsection{Class~1: \texorpdfstring{$x^ix'^j + x'^ix^j$}{}}
\label{subsec:VIA}
The equations we solve are quadratic in $x^i, x'^i$. Thus, there are
multiple distinct solution sets for each MOM class. For instance, from~\cref{eq:App1} in~\cref{sec:A}, we see that Class~1 
MOMs have four solution sets. However, using the symmetries of the 
equations, we can relate these to each other. Let us assume we know one 
solution set. Taking $x^1$ to be our free variable and expressing the other 
variables as a function of it, this set has the form:
\begin{align}
S_{\mathrm{I}}:  \left[x_{\mathrm{I}}^2\left(x^1\right), x_{\mathrm{I}}^3\left(x^1\right), x_{\mathrm{I}}'^1\left(x^1\right),  x_{\mathrm{I}}'^2\left(x^1\right),x_{\mathrm{I}}'^3\left(x^1\right)\right].
\end{align}
The subscript I labels the solution set. More explicitly, let 
us choose the constraints corresponding to the elements $ij=11, 12, 13,
22, 33$ of $\mathbf{M}_{\nu}^{ij}$ as our five independent conditions. 
Then, the Class~1 equations are invariant under the simultaneous transformations,
\begin{align}
    x^2\left(x^1\right) &\mapsto -x^2\left(-x^1\right)\,, \qquad  x'^2 \mapsto -x'^2\left(-x^1\right)\,.
    \label{eq:transf}
\end{align}
To see this, consider the constraint corresponding to $ij=12$; for the others, the check is trivial. We have,
\begin{align}
x^1x'^2\left(x^1\right)+x^2\left(x^1\right)x'^1\left(x^1\right)\,.
\label{eq:eq12}
\end{align}
Making the transformations of~\cref{eq:transf}, we get,
\begin{align}
&x^1\left[-x'^2\left(-x^1\right)\right]+\left[-x^2\left(-x^1\right)\right]x'^1\left(x^1\right)\,, \notag \\
= &\left[-x^1\right]x'^2\left(-x^1\right)+x^2\left(-x^1\right)\left[-x'^1\left(x^1\right)\right]\,, \notag \\
= &\left[-x^1\right]x'^2\left(-x^1\right)+x^2\left(-x^1\right)x'^1\left(-x^1\right)\,,
\end{align}
where, in the last line, we have used $x'^1\left(x^1\right) = -x'^1\left(-x^1\right)$ which follows straightforwardly from the $ij=11$ constraint. Finally, replacing the dummy variable $-x^1\mapsto x^1$, we see that we recover the left-hand side of~\cref{eq:eq12}.

Thus, given set $S_{\mathrm{I}}$, we can obtain a new solution set: 
\begin{align}
S_{\mathrm{II}}:\,  \left[x_{\mathrm{II}}^2\left(x^1\right), x_{\mathrm{II}}^3\left(x^1\right), x_{\mathrm{II}}'^1\left(x^1\right),  x_{\mathrm{II}}'^2\left(x^1\right),x_{\mathrm{II}}'^3\left(x^1\right)\right],
\end{align}
with, 
\begin{align}
    x_{\mathrm{II}}^2\left(x^1\right) &= -x_{\mathrm{I}}^2\left(-x^1\right)\,,\notag\\
        x_{\mathrm{II}}'^1\left(x^1\right) &= x_{\mathrm{I}}'^1\left(x^1\right)\,,\;\notag\\
    x_{\mathrm{II}}'^2\left(x^1\right) &= -x_{\mathrm{I}}'^2\left(-x^1\right)\,,\notag \\
    x_{\mathrm{II}}^{(')3}\left(x^1\right) &= x_{\mathrm{I}}^{(')3}\left(x^1\right)\,.
\end{align}
The third set can be obtained by transforming the $x^3, x'^3$ variables
instead of the $x^2, x'^2$ variables in an analogous manner, and the last one can be obtained by making the transformations on both sets 
simultaneously. 

Consulting the analytical expressions in~\cref{sec:A}, we see that, as 
long as $\mathbf{M}_{\nu}^{ii} \neq 0$ for any $i$, the solution implies 
that the magnitudes of the $x'^i$ couplings are inversely proportional 
to the magnitude of $x^1$ while those of $x^2, x^3$ are directly
proportional to it. Thus, a solution point where any of the $|x'^i|$ 
are small comes at the price of bigger $|x^i|$, and vice-versa. Knowing
which RPV coupling can be made smaller by trading for another is useful 
from a model-building perspective, since the low-energy bounds on RPV 
couplings are non-democratic, varying over orders of magnitude~\cite{Allanach:1999ic}. We draw upon this point further when we study 
applications to specific models. 

\begin{figure}[t]
\centering
\includegraphics[width=0.4\textwidth]{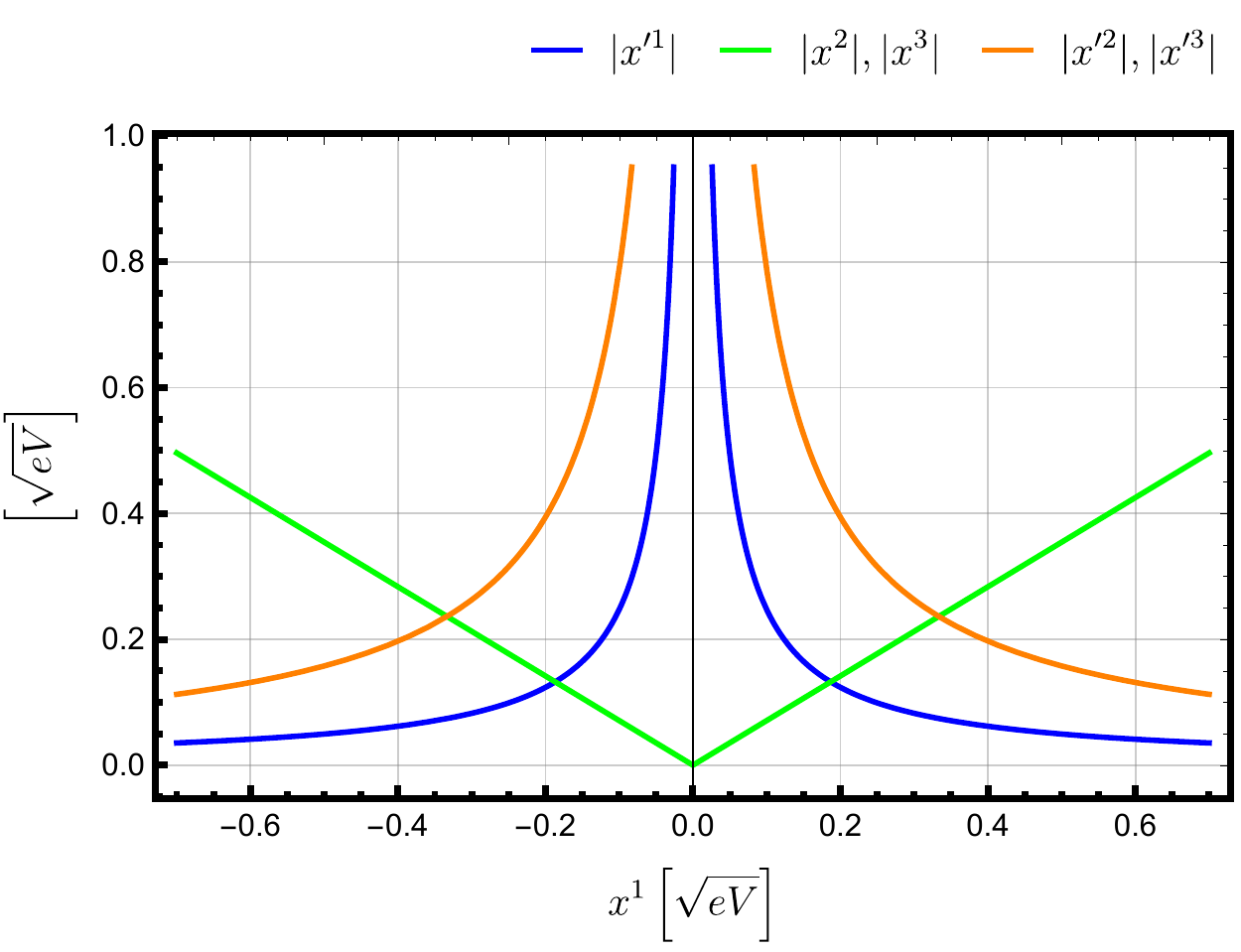}%
\caption{Absolute values of the couplings required to fit the IO limit of the TBM 
scenario in models with Class~1 structure.}
\label{fig:GFplot1}%
\end{figure}

As an illustration, we plot one solution-set for the IO 
limit ($m_3 \approx 0$), assuming TBM values for the angles\footnote{Even though we use the TBM limit for
illustration in this section, all features we discuss are general.} in~\cref{fig:GFplot1}. For visualization, we restrict ourselves to real 
$x^1$ values. The solution then constrains $x'^1$ to be real, while the other 
couplings are complex, in general. The behavior of the couplings is as 
described above. We observe a symmetry under $x^1 \leftrightarrow -x^1$; 
this is an 
intrinsic feature of the model structure. More generally, for a complex
$x^1$, the magnitude of the couplings is unchanged if $|x^1|$ 
is unchanged. The relation between the magnitudes of $x^2({x'}^
2)$ and $x^3 ({x'}^3)$ in~\cref{fig:GFplot1} 
is a peculiarity of the numbers involved in the TBM 
case;\footnote{This arises due to the fact that the TBM-IO mass matrix is antisymmetric under an interchange of the second 
and third columns.} it is not present when using experimental data.

Another point of interest is the ``total amount of RPV" a particular model requires to 
explain the neutrino data. As an illustration, consider how the $x^i, x'^i$ variables 
relate to the RPV parameters, \textit{i.e}, the $\delta's$ of~\cref{tab:TFtab1}:
\begin{equation}
    x^i = A^i\delta^i\,,\qquad  x'^i = A'^i\delta'^i\,,
\label{eq:GFeq2}
\end{equation}
where no summation is implied. In the above, $\delta$ and $\delta'$ are general symbols corresponding to any of the invariants in 
~\cref{tab:TFtab1}; they can both also correspond to the same invariant. One 
measure of the ``total amount of RPV" we can define in the model is the sum,
\begin{align}
    \sigma \equiv \sum_i\left|\delta^i\right| + \sum_i\left|\delta'^i\right|\,.
\label{eq:GFeq3}
\end{align}
The two terms represent the amount of RPV arising due to each individual set. Substituting~\cref{eq:GFeq2}, 
\begin{align}
    \sigma = \sum_i\left(\frac{\left|x^i\right|}{\left|A^i\right|} + \frac{\left|x'^i\right|}{\left|A'^i\right|}\right)\,.
\label{eq:GFeq4}
\end{align}
This will be a function of $x^1$. One could use the analytical expressions in~\cref{sec:A} to study how the RPV-amount demanded by each point varies with $x^1$ and find the point where it is minimal or maximal. In general, this requires that we first fix the constants $A^i, A'^i$, \textit{i.e.}, we specify the model we wish to study. However, in the special case where $A^i = A\,, A'^i = A' \; \forall i$ (which holds for several contributions in~\cref{tab:TFtab1}), there is some simplification for Class~1 MOMs. \cref{eq:GFeq4}, then, gives,
\begin{align}
    \sigma = \sum_i\left(\frac{\left|x^i\right|}{\left|A\right|} + \frac{\left|x'^i\right|}{\left|A'\right|}\right)\,.
\label{eq:GFeq5}
\end{align}
Now, the structure of Class~1 MOMs allows us the freedom to choose
$x^i, x'^i$ suitably such that $A'=A$ without losing any generality. 
Then, 
\begin{align}
    \sigma = \frac{1}{\left|A\right|}\sum_i\left(\left|x^i\right| + \left|x'^i\right|\right)\,.
\label{eq:GFeq6}
\end{align}
Thus, with the above choice of the $x^i, x'^i$ variables, the RPV 
amount is directly proportional to $\sum_i\left(\left|x^i\right| + 
\left|x'^i\right|\right)$ -- a model-independent quantity. This allows us to find the point maximizing or minimizing the RPV amount without specifying the details of the model; determining the absolute scale, though, still requires the constant $\left|A\right|$ to be specified. 

\begin{figure}[t]
\centering
\includegraphics[width=0.4\textwidth]{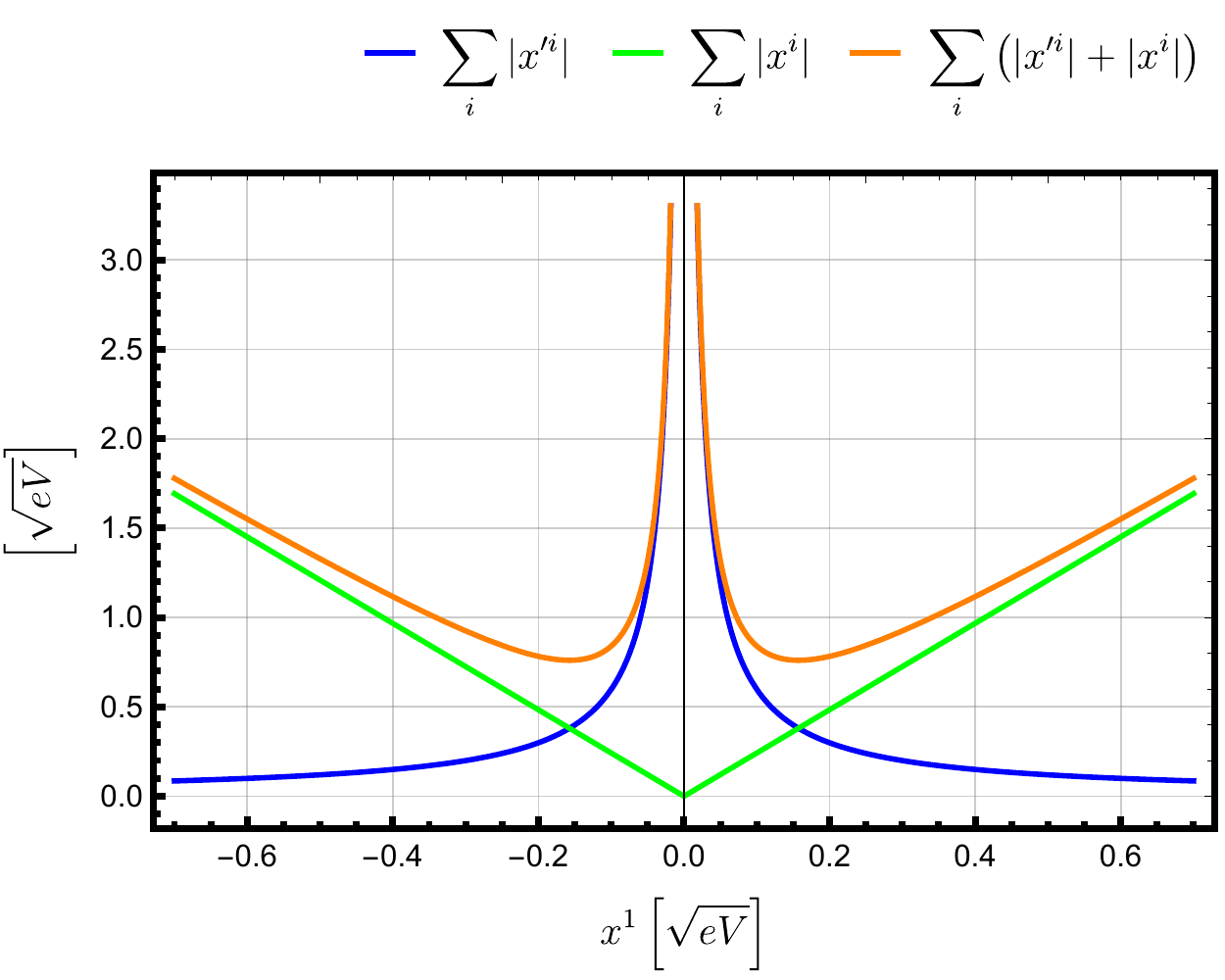}%
\caption{A measure of the amount of RPV required by each point in the 
solution space for Class~1 models. The plot corresponds to the IO limit of 
 the TBM case.}
\label{fig:GFplot2}%
\end{figure}

In~\cref{fig:GFplot2}, we plot the sum of the magnitudes of the $x^i$ 
and $x'^i$ for the IO limit of the TBM case, as well as the overall sum. 
We see that the latter varies from a clear minimum to an unbounded value 
for $|x^1|\to0$. Thus, the neutrino data can be described by relatively 
small or large 
amounts of RPV, depending on the point one chooses. The minimum is 
situated precisely at the point where the individual sums of the $x^i$ 
and $x'^i$ sets are equal. The general expression for this point is 
lengthy. However, for the $CP$-conserving case -- and if the conditions 
$\left(\mathbf{M}_{\nu}^{12}\right)^2 < \mathbf{M}_{\nu}^{11}\times 
\mathbf{M}_{\nu}^{22}$ and an analogous one with the generation index 
$2$ replaced by $3$ are satisfied -- the point is given by, 
\begin{equation}
|x^1|=\sqrt{\frac
{\left|\mathbf{M}_{\nu}^{11}\right|}{2}}\,,
\end{equation}
and the magnitude of the minimum is $\sum_i \sqrt{2\left|\mathbf{M}
_{\nu}^{ii}\right|}$. This holds for a general complex 
$x^1$. The condition we mention above is satisfied by the TBM matrix 
as well as the experimental data we use in our numerical fits.

\subsection{Class~2: \texorpdfstring{$x^ix^j + \left(x^ix'^j + 
x'^ix^j\right)$}{}}
There are four distinct solution sets  related in the same way as 
in the previous case. Consulting~\cref{eq:App2} in~\cref{sec:A}, we see
that $x^2 ,x^3$ satisfy the same relations as for the Class~1 case. The 
behavior of the $x'^i$ is different, however. For $|x^1| \ll 
\sqrt{\left|\mathbf{M}_{\nu}^{11}\right|}$, it is 
as before. However, for $|x^1| \gg 
\sqrt{\left|\mathbf{M}_{\nu}^{11}\right|}$, they grow
 linearly with $|x^1|$. 
In particular, 
$x'^1$ vanishes precisely at $x^1 = \pm 
\sqrt{\mathbf{M}_{\nu}^{11}}$ without any of the other couplings diverging. 
 $x^2, x^3$ can not vanish without other 
couplings diverging. $|x'^2|, |x'^3|$ can also vanish but we 
skip the long general expressions.

\begin{figure}[t]
\centering
\includegraphics[width=0.4\textwidth]{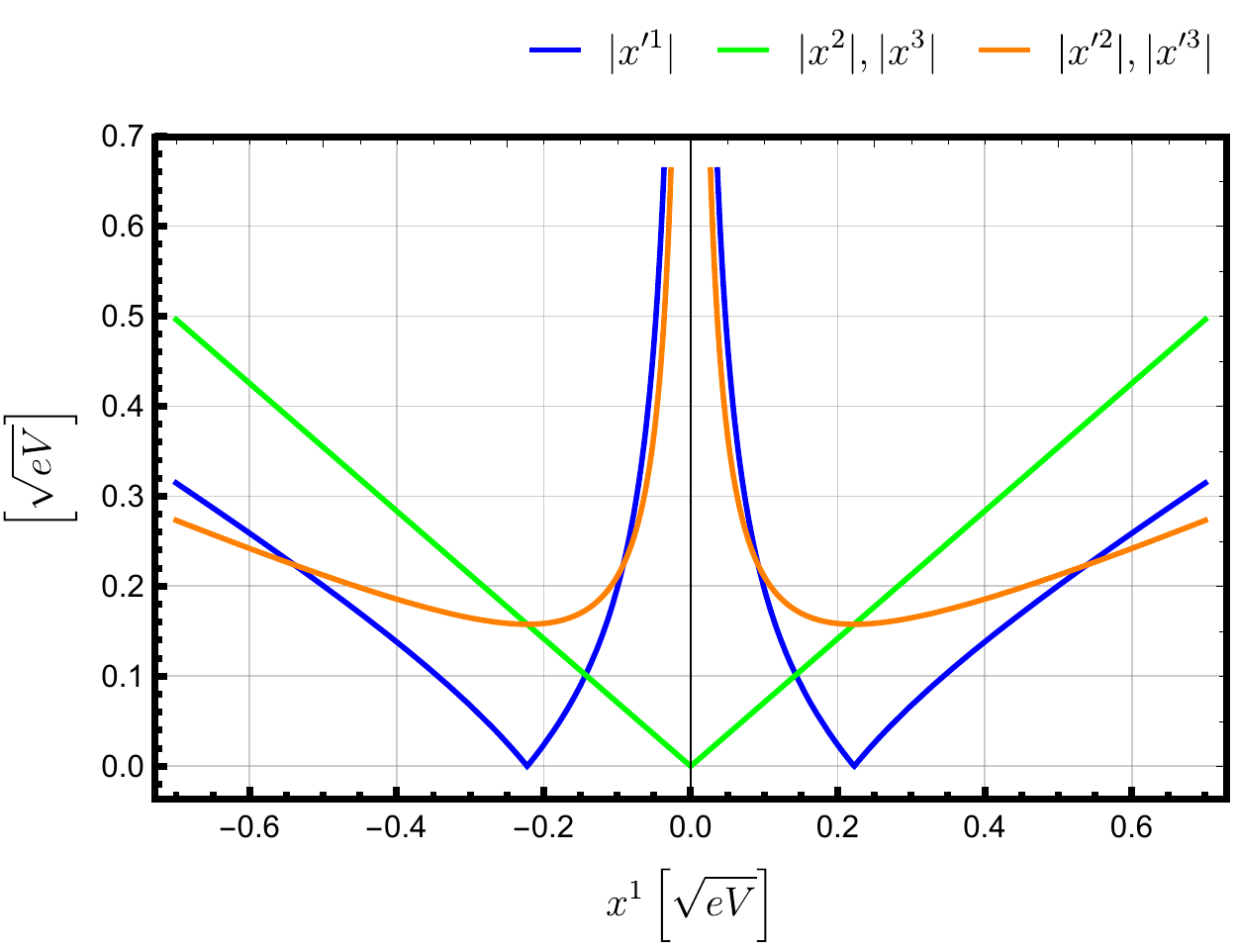}%
\caption{Absolute values of the couplings required to fit the IO limit of the TBM scenario in models with Class~2 structure.}
\label{fig:GFplot3}%
\end{figure}

We plot one of the solution sets corresponding to the TBM-IO limit for 
this class in~\cref{fig:GFplot3}, for real $x^1$. The
symmetry under $x^1\leftrightarrow-x^1$  is evident 
and again intrinsic. The relation between $x^2 
(x'^2)$ and $x^3 (x'^3)$ is TBM-specific. We see the behavior described above.
Indeed $x'^1=0$ at $|x^1| = \sqrt{|\mathbf{M}_{\nu}^{11}|}$; $|x'^2|, |x'^3|$ have their minima at $|x^1| = \sqrt{|\mathbf 
{M}_{\nu}^{11}|}$ too. This is not a general feature but holds in the $CP$-conserving case if, as before,
$\left(\mathbf{M}_{\nu}^{12}\right)^2 < \mathbf{M}_{\nu}^{11}\times \mathbf{M}_{\nu}^{22}$ and the analogous condition with the index $2$ replaced by $3$ are satisfied. $x'^2, 
x'^3=0$ in general requires a non-zero phase for $x^1$.

\begin{figure}[t]
\centering
\includegraphics[width=0.4\textwidth]{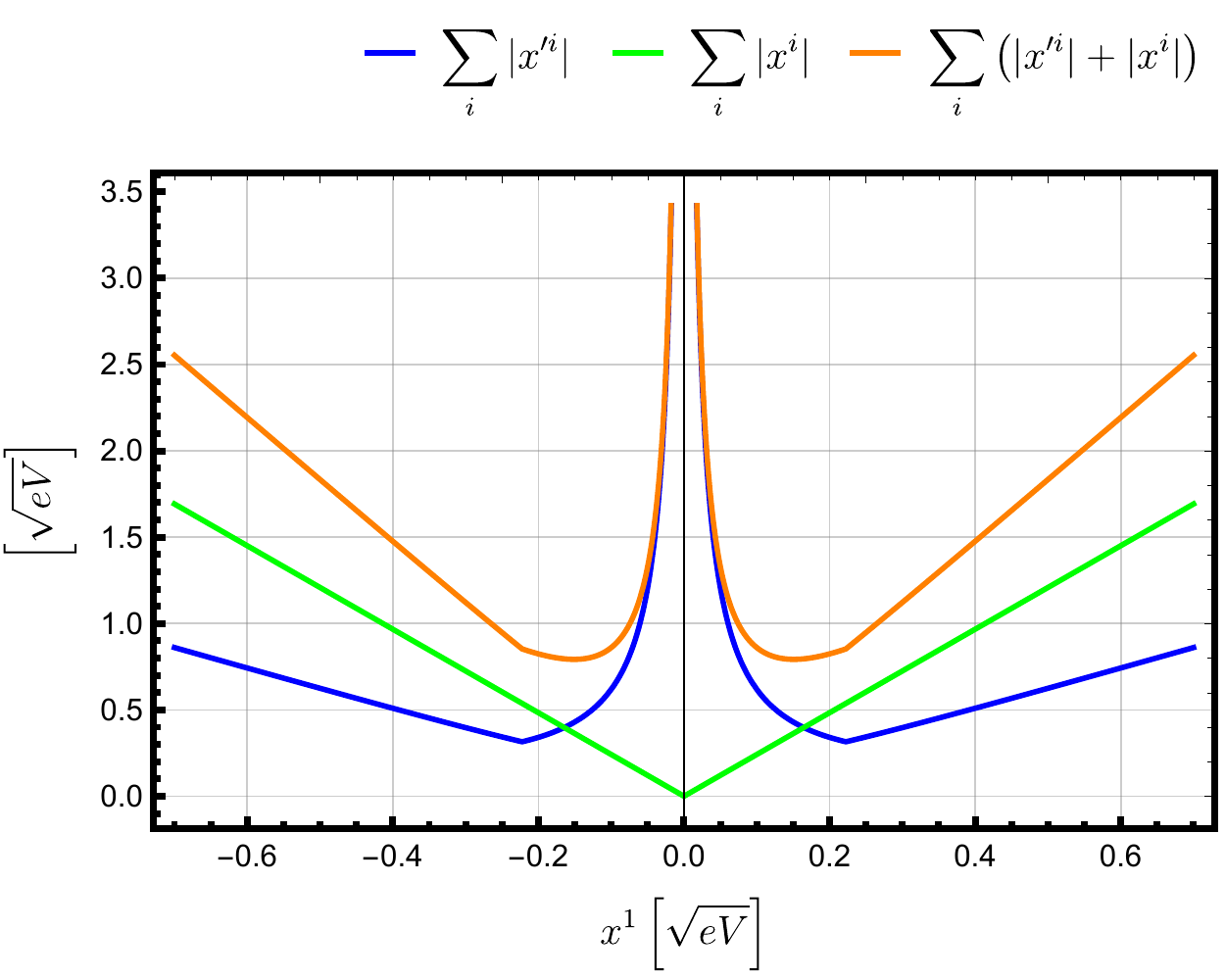}%
\caption{A measure of the amount of RPV required by each point in the solution 
space for Class~2 models. The plot corresponds to the IO limit of the TBM case.}
\label{fig:GFplot4}%
\end{figure}

We plot the sum of magnitudes for $|x^i|$ and $|x'^i|$ for the IO limit in~\cref{fig:GFplot4}. The individual sums 
are directly proportional to the RPV amount for each set and can be
interpreted as before. However, the overall sum is no longer 
directly related to the total RPV amount. Unlike the case of Class~1 MOMs, we do not always have the freedom to choose $A=A'$ in~\cref{eq:GFeq2} for Class~2 MOMs. We still plot the quantity;
however, it should only be used for models where $A=A'$ holds. 

\subsection{Class~3: \texorpdfstring{$x^ix^j + x'^ix'^j$}{}}
Class~3 MOMs have eight distinct solution sets. Four can be obtained using the same arguments as before; this time the invariance is under the simultaneous transformations,
\begin{align}
    x^2\left(x^1\right) &\mapsto -x^2\left(-x^1\right)\,, \qquad  x'^2 \mapsto x'^2\left(-x^1\right)\,.
\end{align}
and the analogous ones for $x^3$ and $x'^3$. In addition, the whole
system of equations is invariant under the simultaneous transformations,
\begin{equation}
    x'^i \mapsto -x'^i\,.
\end{equation}
Thus, for each of the four solution sets, we can obtain one more by 
changing the signs of all the $x'^i$ couplings.

In general, the solution space is more complicated than for the other two classes. Consulting~\cref{eq:App3} in~\cref{sec:A}, in the limit $|x^1|\gg\sqrt{\left|\mathbf{M}_{\nu}^{11} 
\right|}$, all the coupling magnitudes increase roughly linearly with $|x^1|$. This class is somewhat special: It allows solutions where all the couplings are simultaneously real; this occurs when $|x^1| \leq 
\sqrt{|\mathbf{M}_{\nu}^{11}|}$, with $x^1$ real. This also requires $\delta_{{CP}}=0$, $\left(\mathbf{M}_{\nu}^{12}\right)^2<\mathbf{M}_{\nu}^{11}\times 
\mathbf{M}_{\nu}^{22}$ and the analogous condition with the index $2$ replaced by $3$ to hold.  

\begin{figure}
\centering
\begin{subfigure}[b]{0.4\textwidth}
    \includegraphics[width=\textwidth]{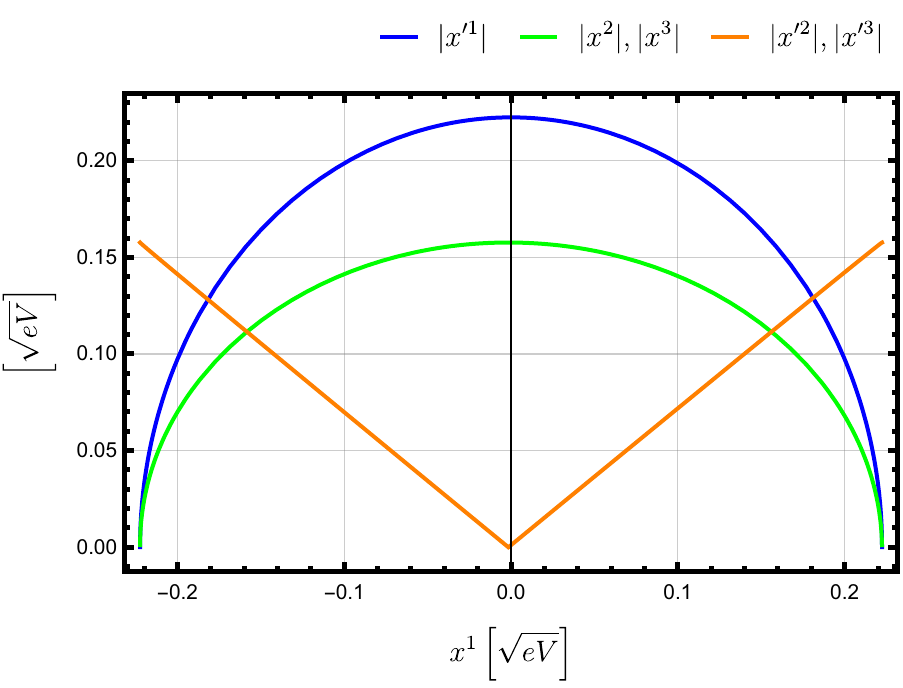}
    \caption{Inverted Ordering}
\end{subfigure}
\begin{subfigure}[b]{0.4\textwidth}
    \includegraphics[width=\textwidth]{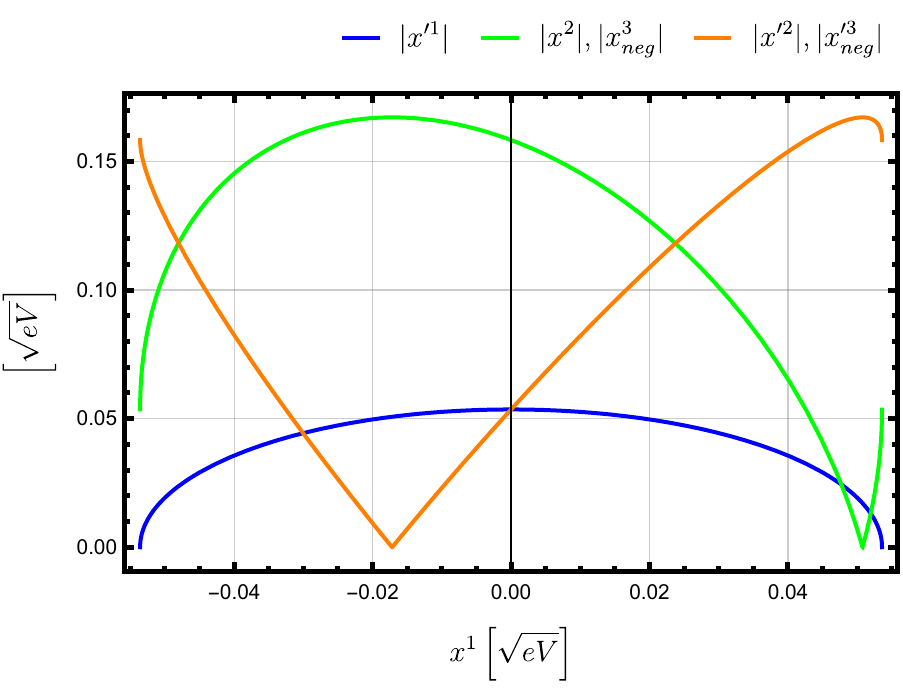}
    \caption{Normal Ordering}
\end{subfigure}
\caption{Absolute values of the couplings required to fit the TBM 
scenario in models with Class~3 structure. The notation $f_{neg}$ 
means $f\left(-x^1\right)$ has been plotted instead of $f\left( x^1 
\right)$. 
}
\label{fig:GFplot5}
\end{figure}

We plot one solution set for the TBM-IO and TBM-NO limits in~\cref{fig:GFplot5}, 
restricted to the above region. The symmetry of $x'^1$ under $x^1\leftrightarrow-x^1$ is an 
intrinsic feature of the model structure. Although the TBM-IO limit numbers conspire to make 
it look otherwise in our plot, the other couplings do not generally possess such a symmetry -- 
this is clear after looking at the NO limit. As before, the $x^2\left(x'^2\right)$ and $x^3 
\left(x'^3\right)$ relation is TBM-specific.

Within our region of interest, we see that the magnitude of $x'^1$ always falls as that of 
$x^1$ increases and it is zero precisely at the point $x^1=\pm\sqrt{\mathbf{M}_{\nu}^{11}}$. 
There are no such universal trends for the other couplings. Unlike before, however, they can 
each be made to vanish in appropriate regions of the solution space. The $x^2\left(x^3 \right 
)$ vanish at $x^1 = \pm\sqrt{\mathbf{M}_{\nu}^{11}-\frac{\mathbf{M}_{\nu}^{12(13)}\times
\mathbf{M}_{\nu}^{12(13)}}{\mathbf{M}_{\nu}^{22(33)}}}$ while $x'^2\left(x'^3\right)$ vanish 
at $\pm{\mathbf{M}_{\nu}^{12(13)}}/{\sqrt{\mathbf{M}_{\nu}^{22(33)}}}$. 

\begin{figure}
\centering
\begin{subfigure}[b]{0.4\textwidth}
    \includegraphics[width=\textwidth]{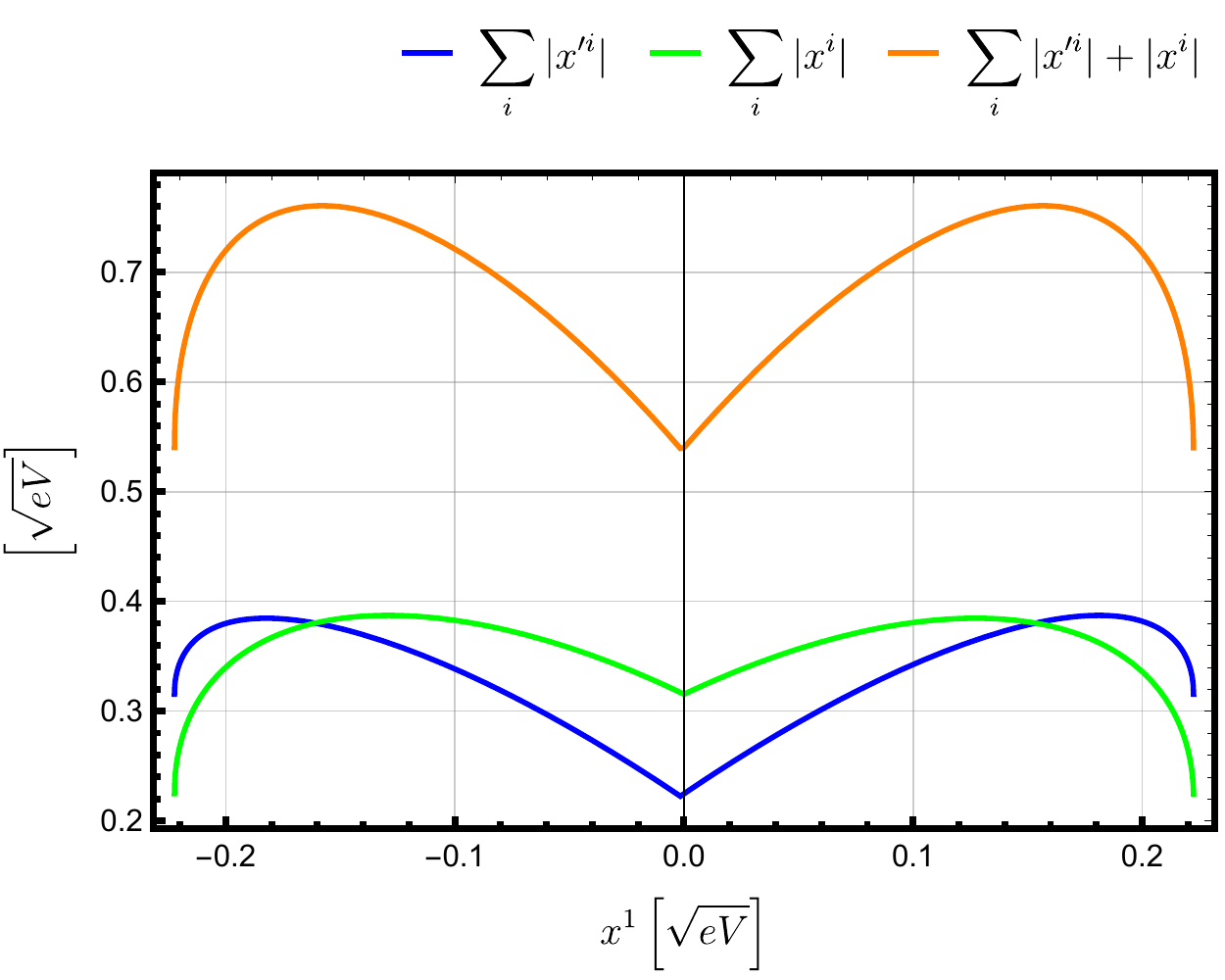}
    \caption{Inverted Ordering}
\end{subfigure}
\begin{subfigure}[b]{0.4\textwidth}
    \includegraphics[width=\textwidth]{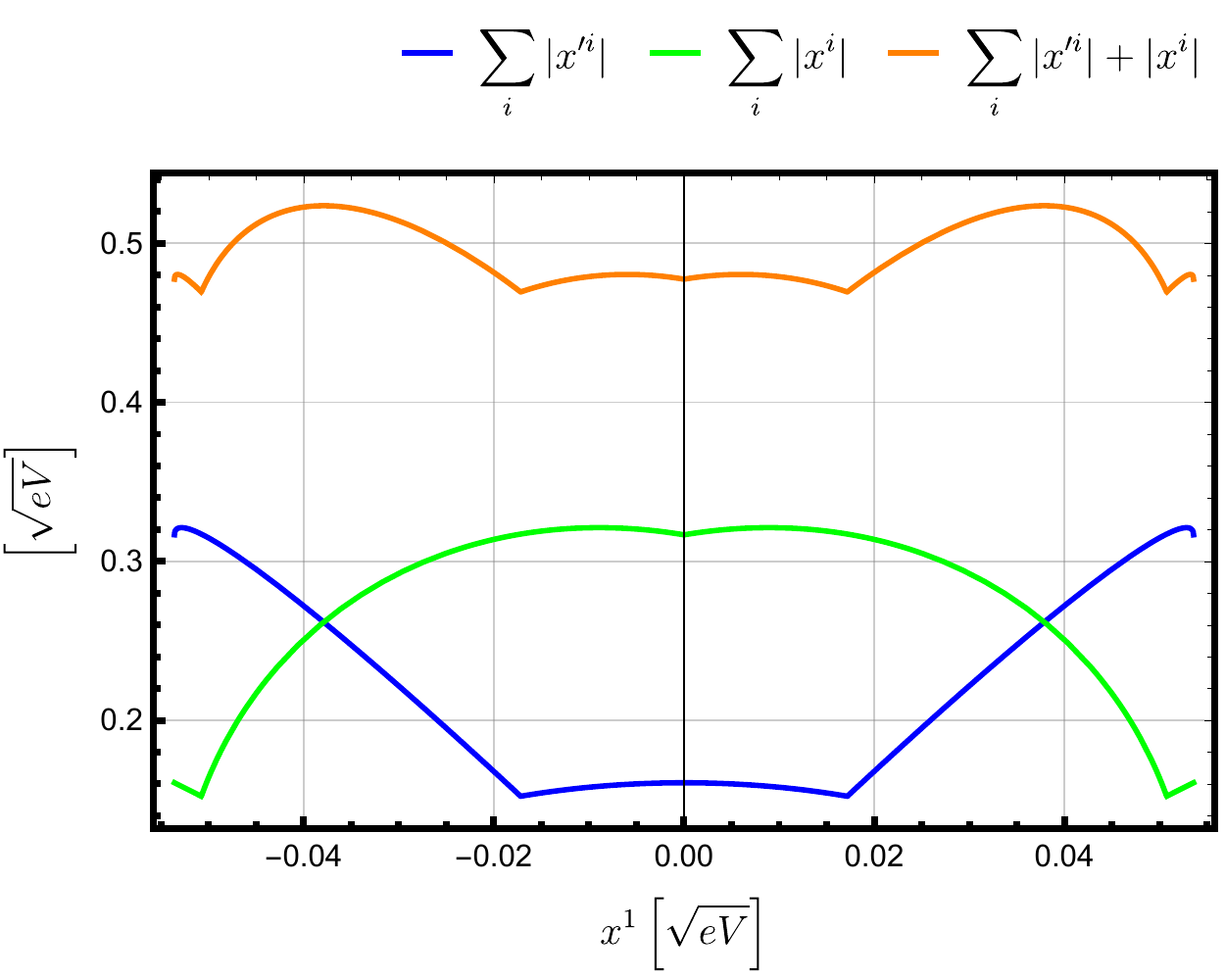}
    \caption{Normal Ordering}
\end{subfigure}
\caption{A measure of the amount of RPV required by each point in the solution space for Class~3 models. The plots correspond to the TBM scenario.}
\label{fig:GFplot7}
\end{figure}

We study the relative RPV amount in~\cref{fig:GFplot7}. As for Class~2 MOMs, the overall sum 
may only be interpreted as the total RPV amount if $A=A'$ in~\cref{eq:GFeq2}. Here, the amount 
of RPV is dominated by the $x^i$ for vanishing $x^1$, with the $x'^i$ share growing as $|x^1|$
grows. The amount of RPV is minimal near the two $|x^1|$ extremes.

\section{Numerical Fits}
\label{sec:7}
We now present the solution space for the experimental data. We numerically solve
the first three MOM classes for each of the dependent parameters with $x^1$ as the free 
variable. We estimate the couplings by means of an error-weighted least-squares fit. We use 
the neutrino data of~\cref{tab:neudata} (with $\delta_{CP} = 0$) at the $1\sigma$ level. 
In order to extract predictions for the couplings, we define a $\chi^2$ function:
\begin{align}
\label{eq:chisquare}
\chi^2 \equiv \frac{1}{N_\text{obs}} \sum_{i=1}^{3} \sum_{j=i}^{3}  \left(
\frac{ x^{ij} - M^{ij}}{\delta^{ij}} \right)^2\,,
\end{align}
where $M^{ij}$ are the central values of the $N_\text{obs}$ experimentally determined
parameters of the mass matrix defined in~\cref{eq:MOMeq1}, $x^{ij}$ are the parameters to be 
determined, and $\delta^{ij}$ are the $1\sigma$ experimental uncertainties. 

We initiate the fit using the TBM approximation for the $x^{ij}$. We minimize the $\chi^2$ of~\cref{eq:chisquare} by using the program package \textsf{MINUIT2}~\cite{James:1975dr}. We 
consider both the NO and IO limits. We accept the minimization result as a success if the 
routine yields  $\chi^2<\mathcal{O}\left(10^{-5}\right)$.

To handle complex couplings, we fit the real and imaginary parts of each parameter separately. This extends the definition of our $\chi^2$ function:
\begin{align}
\label{eq:chisquare2}
\chi^2 =& \frac{1}{N_\text{obs}} \sum_{i=1}^{3} \sum_{j=i}^{3}  \left[\left(\frac{ \text{Re}\left(x^{ij}\right) - \text{Re}\left(M^{ij}\right)}{\delta^{ij} } \right)^2\right. 
\notag \\
&+ \left.\left(\frac{ \text{Im}\left(x^{ij}\right) - \text{Im}\left(M^{ij}\right)}{\delta^{ij}} \right)^2\right]\,,
\end{align}
where we demand that the imaginary components of the neutrino mass matrix 
vanish, since we are working in the $CP$-conserving limit. 

\begin{figure}
\centering
\begin{subfigure}[b]{0.4\textwidth}
   {\includegraphics[scale=0.5]{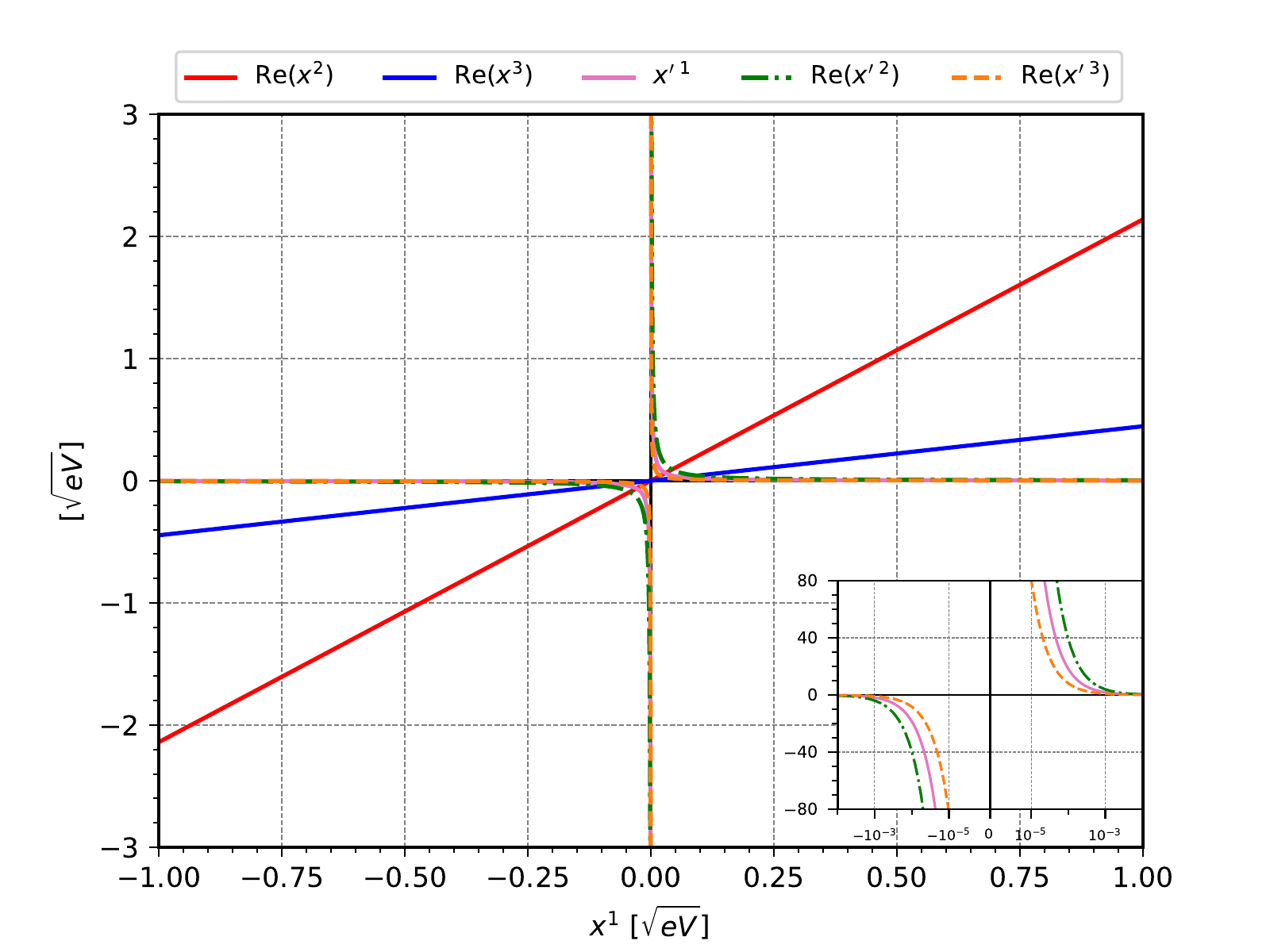}}
    \end{subfigure}
    \begin{subfigure}[b]{0.4\textwidth}
    {\includegraphics[scale=0.5]{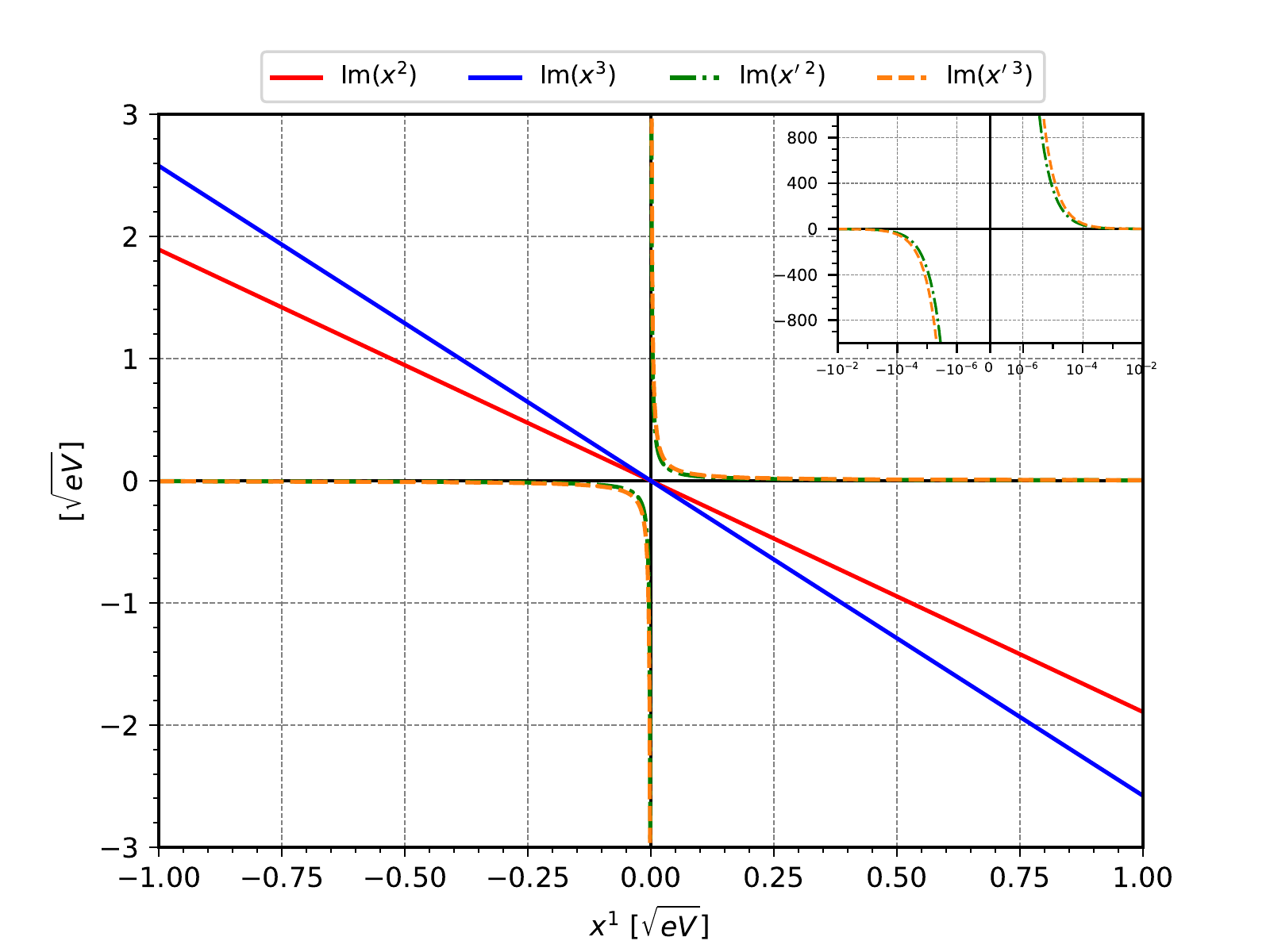}}
    \end{subfigure}
    \caption{Real (top) and imaginary (bottom) values of the couplings required to fit the actual neutrino data for the NO limit in models with Class~1 structure.}
    \label{fig::Model1}%
   \end{figure}

~\cref{fig::Model1} shows the numerical result using the neutrino data, assuming the NO limit, for Class~1~MOMs. We restrict ourselves to real $x^1$. This automatically implies that $x'^1$ has to be real. As before, we depict only one of the multiple
solution sets. We see that the solution space reproduces the 
general features discussed in~\cref{subsec:VIA}.  The analogous 
results for the IO limit for Class~1 MOMs (~\cref{fig::Model1IM}), as well as the plots corresponding to Class~2 MOMs (~\cref{fig::Model2}), and Class~3 MOMs (~\cref{fig::Model3}) can be found in~\cref{sec:C}. A corresponding solution including a non-zero $\delta_ 
{CP}$ can be found in~\cref{fig:CPV} with more details in~\cref{sec:C}.

\begin{figure}[htp]
\centering
\includegraphics[width=0.4\textwidth]{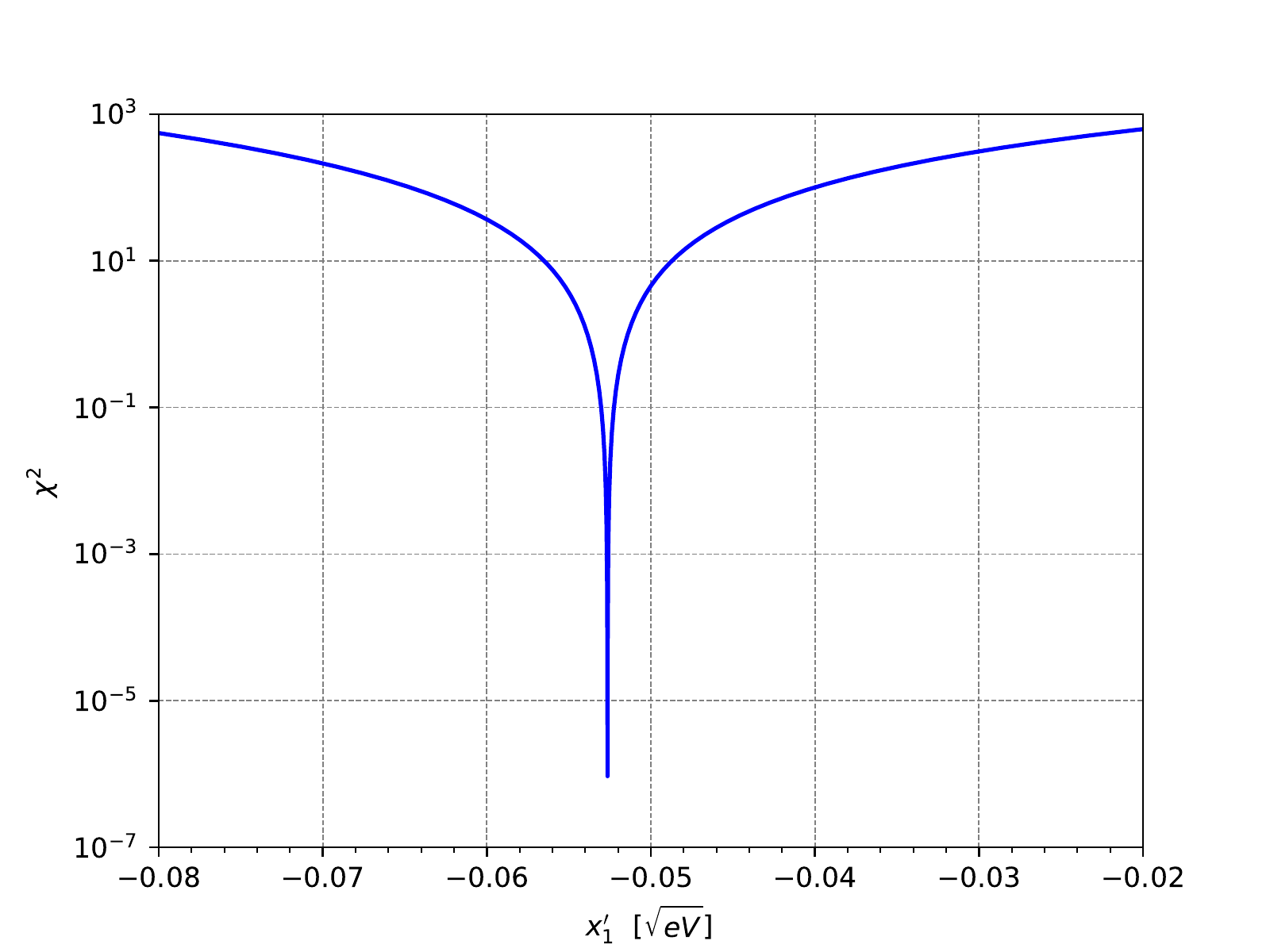}%
\caption{Variation of $\chi^2$ as a function of the fitted parameter $x'_1$ for the Class~3 NO scenario around the best-fit point as determined by \textsf{MINUIT2}. The other couplings are held fixed.}
\label{fig::chisquare}%
\end{figure}

To depict the robustness of our procedure, we show, in~\cref{fig::chisquare}, the 
variation of $\chi^2$ by varying one of the fitted couplings -- $x'^1$ -- about the
best-fit point. 
The other couplings are held fixed. The minimum is extremely well-defined, indicating
excellent convergence.

\section{Example Applications}
\label{sec:8}

As long as a model has a MOM structure, our general results can be directly translated into 
model-specific numbers. We now demonstrate this by considering several examples of RPV models. The statement that only certain RPV couplings are 
non-vanishing in a given model is $\mathrm{U}(4)$-basis dependent; our statements in this section apply to the vanishing-sneutrino-vev basis. 

\subsection{\texorpdfstring{$\kappa$}{}-only Models}
\label{subsec:A}
In a model where the only RPV sources are the $\delta_{\kappa}$ invariants, the effective neutrino mass matrix has contributions at tree level, and of types 7, 8 and 13 in~\cref{tab:TFtab1} at one-loop level. The expression for the mass matrix is~\cite{Davidson:2000ne, Allanach:2003eb},
    \begin{align}
    \mathbf{M}_{\nu}^{ij} &= m_{0}\delta_{\kappa}^i\delta_{\kappa}^j + \frac{g_2\left[\left(m_{e_i}\right)^2 + \left(m_{e_j}\right)^2\right]}{16\pi^2v}\times\notag\\
    \qquad &\times \left(1 + \sin^2\beta + \tan\beta\sin^2\beta\right)\delta_{\kappa}^i\delta_{\kappa}^j + \ldots\,,
    \label{eq:EAeq1}
    \end{align}
where $m_0 = -\frac{M^2_Z\cos^2\beta m_\text{SUSY}}{m^2_\text{SUSY} - M^2_Z\sin2\beta}$ is the
tree-level mass scale of~\cref{eq:TFeq5}, $v$ is the electroweak vev, $g_2$ is the $\mathrm{SU}(2)_{L}$ gauge coupling, and the other notation is as in~\cref{tab:TFtab1}. There are three 
separate diagrams of type 13 that lead to the second term~\cite{Davidson:2000ne}. The ellipsis
indicates all the terms of higher (fourth) order in the lepton Yukawas, due to contributions of
types $7, 8$. We have set all SUSY mass scales to $m_\text{SUSY}$.

\cref{eq:EAeq1} does not have a MOM structure. However, we can neglect the terms in the ellipsis to a first approximation, given their suppression by two extra powers of the small Yukawas. Then, making the identifications,
\begin{align}
    x^i &= \sqrt{m_0}\delta^i_{\kappa}\,, \quad \text{and}\notag\\
    x'^i &= \frac{g_2\left(m_{e_i}\right)^2}{16\pi^2v\sqrt{m_0}}\left(1 + \sin^2\beta + \tan\beta\sin^2\beta\right)\delta^i_{\kappa}\,,
\end{align}
we see that the model reduces to a Class~2 MOM structure, and our framework can be applied. One can easily show that such a model cannot solve the neutrino pattern. The above equations imply the following relations involving the lepton masses:
\begin{equation}
\frac{1}{m^2_e}\frac{x'^1}{x^1} = \frac{1}{m^2_{\mu}}\frac{x'^2}{x^2} = \frac{1}{m^2_{\tau}}\frac{x'^3}{x^3}\,.
\end{equation}
Consulting~\cref{eq:App2} in~\cref{sec:A}, there is no point in the solution space of Class~2 models satisfying this.

\subsection{\texorpdfstring{$\kappa - B$}{} Models}
\label{subsec:B}
We next consider a model also including the soft-breaking bilinear terms, \textit{i.e.}, $\delta_{\kappa},\delta_B
\not=0$ with all other RPV couplings zero (see also Ref.~\cite{Hirsch:2000ef}). We have the contributions, \textit{cf.}~\cref{tab:TFtab1}: Tree-level, and of types 
3, 7, 8, 9, 11, 12, 13, 16 and 17. The complete expression is~\cite{Davidson:2000ne},
\begin{align}
    \mathbf{M}_{\nu}^{ij} &= m_{0}\delta_{\kappa}^i\delta_{\kappa}^j + \frac{g^2_2 m_\text{SUSY}}{64\pi^2 \cos^2\beta}\delta_{B}^i\delta_{B}^j\notag\\
    \qquad &+ \frac{g^2_2 m_\text{SUSY}}{64\pi^2 \cos\beta}\left(\delta_{\kappa}^i\delta_{B}^j + \delta_{B}^i\delta_{\kappa}^j\right) +  \ldots\,.
    \label{eq:EA1prime}
\end{align}
The ellipsis again proxies contributions of higher (second and above) 
order in the Yukawas. As before, the full model does not have a MOM 
structure but neglecting the Yukawa-suppressed terms\footnote{Some of
these Yukawa-suppressed terms have $\tan\beta$ factors which may enhance them for large $\tan\beta$; however, even in this case the
second and third terms in~\cref{eq:EA1prime} dominate due to the 
$\cos\beta$ factors.}, and making the identifications,
\begin{align}
    x^i &= \sqrt{m_0}\delta^i_{\kappa}\,, \notag\\
    x'^i &= \frac{g_2\sqrt{m_\text{SUSY}}}{8\pi \cos\beta}\delta^i_{B}\,, \notag\\
    A &= \frac{g_2\sqrt{m_\text{SUSY}}}{8\pi \sqrt{m_0}}\,,
\end{align}
the model reduces to a Class~4 MOM.

As a numerical illustration, we set $m_\text{SUSY} = \SI{1}{\tera\electronvolt}$, $\tan\beta=10$, and substitute the other known parameters. This gives,
\begin{align}
    x^i &\approx \left(\SI{9.081e+3}{\sqrt{\electronvolt}}\right)i\delta^i_{\kappa}\,, \notag\\
    x'^i &= \left(\SI{2.607e+5}{\sqrt{\electronvolt}}\right)\delta^i_{B}\,, \notag\\
    A &\approx - 2.857i\,.
    \label{eq:EAeq1a}
\end{align}
Numerically solving this for the TBM-IO limit -- for instance, at the point with $x^1=0$ -- yields,
\begin{align}
    x^2 &= -x^3 \approx 0.052\;\SI{}{\sqrt{\electronvolt}}\,,\notag\\
    x'^1 &\approx \ -0.222\;\SI{}{\sqrt{\electronvolt}}\,,\notag\\
    x'^2 &= -x'^3 \approx \left(-0.001+0.149i\right)\;\SI{}{\sqrt{\electronvolt}}\,,
\end{align}
or, using~\cref{eq:EAeq1a},
\begin{align}
    \kappa^1 &= 0 \,,\notag\\
    \kappa^2 &= -\kappa^3 \approx -|\kappa|\left(\SI{5.73e-6}{}i\right)\,,\notag\\
    B^1 &\approx-|B|\left(\SI{8.52e-7}{}i\right)\,,\notag\\
    \quad B^2 &= -B^3 \approx -|B|\left(\SI{3.84e-9}{}-\SI{5.72e-7}{}i\right)\,.
\end{align}
In the above, we have made use of the forms of the $\delta$ invariants in the vanishing-sneutrino-vev basis, \textit{cf.}~\cref{eq:TFeq5a}. Similarly, one could also numerically solve at the point corresponding to the minimal RPV amount, and use that in order to derive a minimal bound on the couplings.

\subsection{Diagonal Trilinear Models I}
\label{subsec:C}
We now consider models with the trilinear sector contributing, and assume the other contributions are negligible. The effective neutrino mass matrix is zero at tree level but receives contributions at loop-level of types $1$ and $2$ in~\cref{tab:TFtab1}. The expression for the matrix is,
\begin{align}
    \mathbf{M}_{\nu}^{ij} &= \frac{1}{8\pi^2 m_\text{SUSY}}\delta_{\lambda}^{ink}\delta_{\lambda}^{jkn}m_{e_n}m_{e_k}\notag\\
    & + \frac{3}{8\pi^2 m_\text{SUSY}}\delta_{\lambda'}^{ink}\delta_{\lambda'}^{jkn}m_{d_n}m_{d_k}\,,
    \label{eq:EAeq2}
\end{align}
with a summation implied over repeated indices. The equation has too many parameters to have a MOM structure, or any predictivity in general.

In a minimal model where only the diagonal (in the last two indices) trilinear couplings 
contribute, the above expression simplifies:{\small
\begin{align}
&    \mathbf{M}_{\nu}^{ij} =  \label{eq:EAeq3}\\
& \frac{1}{8\pi^2 m_\text{SUSY}}\left(\lambda^{i11}\lambda^{j11}m^2_{e} + \lambda^{i22}\lambda^{j22}m^2_{\mu} + \lambda^{i33}\lambda^{j33}m^2_{\tau}\right)\notag\\
    &+ \frac{3}{8\pi^2 m_\text{SUSY}}\left(\lambda'^{i11}\lambda'^{j11}m^2_{d} + \lambda'^{i22}\lambda'^{j22}m^2_{s} + \lambda'^{i33}\lambda'^{j33}m^2_{b}\right),
\notag    
\end{align}
}

\vspace{-0.3cm}

\noindent where we have used the fact that the $\delta$ invariants can be simply replaced by the $\lambda$ couplings in the vanishing-sneutrino-vev basis. 

~\cref{eq:EAeq3} still has too many terms for a MOM structure. We can further reduce the 
system, by assuming the couplings have a comparable magnitude. The terms then have a 
natural hierarchy due to the fermion masses. Considering only the contributions of the two 
heaviest particles -- the $b$ quark and the $\tau$ lepton -- the largest neglected term is a 
factor $\frac{ m_{\tau}^2}{m_{\mu}^2} \sim 300$ smaller. The model reduces to:
\begin{align}
    \mathbf{M}_{\nu}^{ij} &= \frac{1}{8\pi^2 m_\text{SUSY}}\lambda^{i33}\lambda^{j33}m^2_{\tau} \notag \\
    &+ \frac{3}{8\pi^2 m_\text{SUSY}}\lambda'^{i33}\lambda'^{j33}m^2_{b}\,,
\end{align}
which has a Class~3 MOM structure with the identifications,
\begin{align}
    x^i &= \sqrt{\frac{1} {8\pi^2 m_\text{SUSY}}}m_\tau \lambda^{i33}\,, \notag\\
    x'^i &= \sqrt{\frac{3} {8\pi^2 m_\text{SUSY}}}m_b \lambda'^{i33}\,.
\end{align}
Due to the antisymmetry of the LLE couplings in the first two indices, $x^3 \sim\lam^{333}= 
0$. This uniquely determines the solution to the point where $x^3=0$ vanishes. We had discussed the location of this point earlier. Plugging in the numbers for the TBM-IO limit 
gives:
\begin{align}
    x^1 &\approx \SI{-0.2224}{\sqrt{\electronvolt}}\,, \quad x^2 = x^3 \approx \SI{0}{\sqrt{\electronvolt}}\,,\notag\\
    x'^1 &\approx \SI{-0.0016}{\sqrt{\electronvolt}}\,, \quad x^2 = -x^3 \approx \SI{-0.1577}{\sqrt{\electronvolt}}\,.
\end{align}
One can plug in the values of $m_{\text{SUSY}}$ and the lepton masses to see what 
this implies for the $\lam$ couplings.

\subsection{Diagonal Trilinear Models II}
\label{subsec:D}
To discuss a slightly more complex application, we consider a cMSSM-like scenario, called the 
$B_3$ cMSSM in Ref.~\cite{Dreiner:2011ft}. At the GUT scale, the five cMSSM parameters are appended 
by one (or two) RPV trilinear coupling(s). All other RPV couplings are assumed to be 
zero. Through the renormalization group equations (RGEs), further couplings are generated at 
the electroweak scale. Thus, we end up with multiple contributions to the neutrino mass matrix. The most relevant are the bilinear terms since these contribute at 
tree-level; the RGE-generated trilinear couplings are suppressed and only contribute at one-loop 
level. The neutrino mass matrix has the structure,
\begin{align}
    \mathbf{M}_{\nu}^{ij} \sim \delta^i_{\kappa}\delta^j_{\kappa} + \delta^i_{B}\delta^j_{B} + \delta^i_{\lambda'}\delta^j_{\lambda'} + \delta^i_{\lambda}\delta^j_{\lambda} + \ldots\,,
    \label{eq:EAeq4}
\end{align}
where we assume two non-zero GUT-scale couplings $\lambda, \lambda'$ and only symbolically 
depict the type of terms contributing. The
ellipsis indicates potential cross-terms. The above model again has too many terms. To a good 
approximation, the generated bilinear parameters are of the form,
\begin{align}
    \delta^i_{\kappa} &\approx a_1\delta^i_{\lambda'} + b_1\delta^i_{\lambda}\,,\notag\\
    \delta^i_{B} &\approx a_2\delta^i_{\lambda'} + b_2\delta^i_{\lambda}\,,
    \label{eq:EAeq4a}
\end{align}
where the $a_i, b_i$ are numerical constants. Thus, the RGEs ensure that the generated 
couplings are approximately linearly dependent on the original $\delta_\lam^i,\,\delta^i_ 
{\lam'}$.\footnote{The exact forms of the RGEs can be found in Ref.~\cite{Dreiner:2011ft}, 
where this approximation is also discussed.} The model has only two linearly independent 
structures appearing and the MOM framework applies. Substituting~\cref{eq:EAeq4a} in~\cref{eq:EAeq4}, the matrix reduces to the form,
\begin{align}
    \mathbf{M}_{\nu}^{ij} \sim \delta^i_{\lambda'}\delta^j_{\lambda'} + \delta^i_{\lambda}\delta^j_{\lambda} + \left(\delta^i_{\lambda'}\delta^j_{\lambda} + \delta^i_{\lambda'}\delta^j_{\lambda}\right)\,,
    \label{eq:EAeq4b}
\end{align}
which is a Class~4 MOM.

\subsection{Non-diagonal Trilinear Models}
\label{subsec:E}
Next, we consider the dominant contributions to arise from the non-diagonal (in the last two indices) 
trilinear couplings. The effective neutrino mass matrix has the form of~\cref{eq:EAeq2}, 
except now the $n, k$ indices are not equal. Again, we exploit the natural hierarchy of the 
structures to reduce the model to a MOM. Performing the expansion in the vanishing-sneutrino-vev
basis, we have,
\begin{align}
    \mathbf{M}_{\nu}^{ij} =& \frac{3}{8\pi^2 m_\text{SUSY}}\left(\lambda'^{i32}\lambda'^{j23} + \lambda'^{j32}\lambda'^{i23}\right)m_{b}m_{s} + \ldots\,,     
\end{align}
where the ellipsis hides the other terms. For instance, assuming similar magnitudes of
couplings, the next highest contribution is the one proportional to $m_{\tau}m_{\mu}$ and is 
smaller by a factor $\frac{3 m_{b} m_{s}}{m_{\tau} m_{\mu}} \sim 8$ compared to the first 
term. The factor of $3$ here is due to the quark colors. Thus, we only consider the first 
term. This reduces the model to a MOM of Class~1 structure, with the identifications,
\begin{align}
    x^i &= \sqrt{\frac{3 m_b m_s}{8 \pi^2 m_\text{SUSY}}}\lambda'^{i23}\,,\notag\\
    x'^i &= \sqrt{\frac{3 m_b m_s}{8 \pi^2 m_\text{SUSY}}}\lambda'^{i32}\,.
    \label{eq:EAeq5}
\end{align}

\begin{figure*}[htp]
   \centering
    \subfloat[Inverted Ordering]{\includegraphics[scale=0.48]{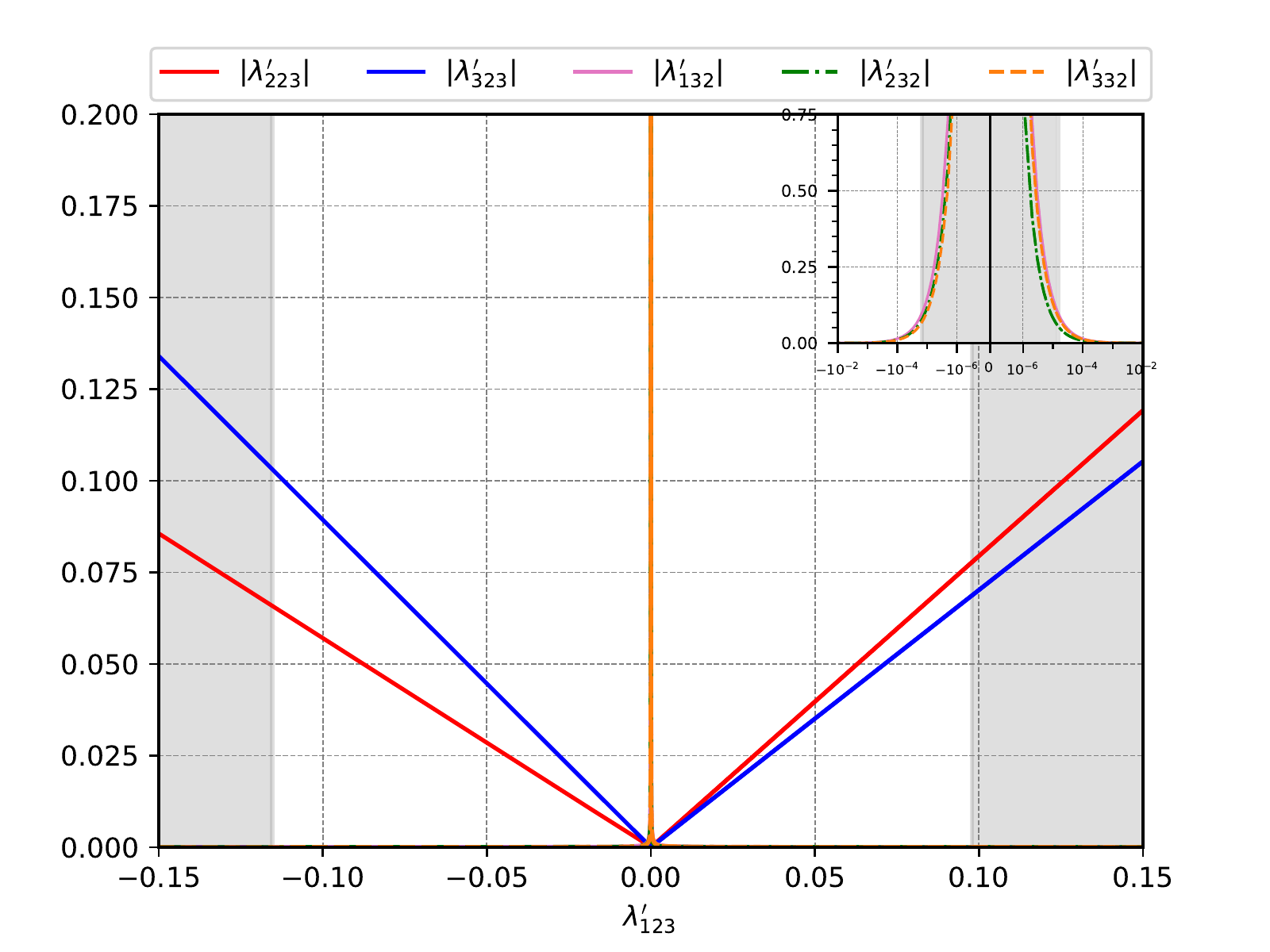}}
    \subfloat[Normal Ordering]{\includegraphics[scale=0.48]{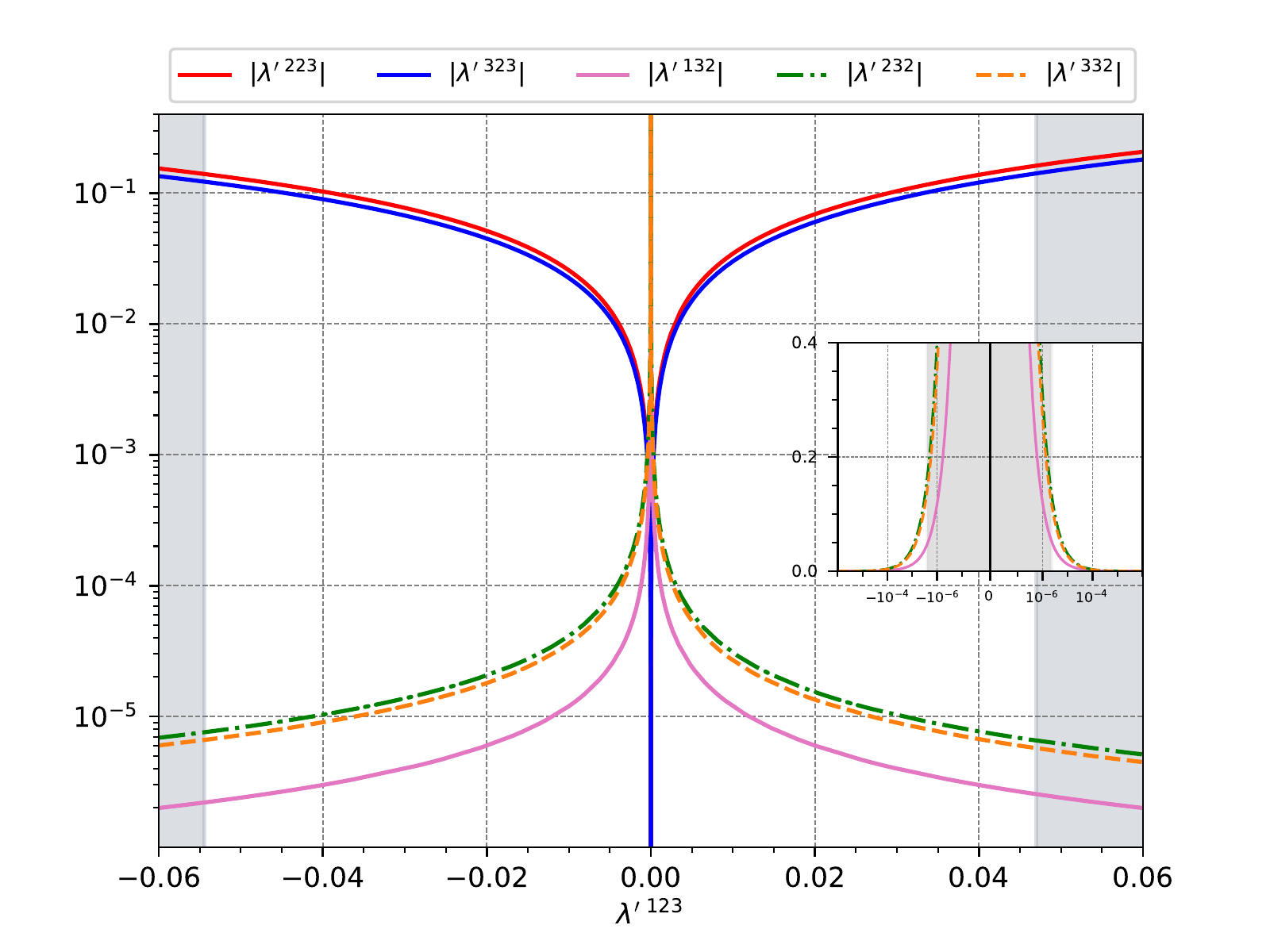}}%
    \caption{The IO (left) and NO (right) solution spaces for the non-diagonal trilinear model of~\cref{subsec:E} including $\delta_{CP}$. The grey regions are the ones ruled out by the bounds of~\cref{eq:EAeq7}. The bounds of~\cref{eq:EAeq6} are beyond the scale of the plots.}
\label{fig:EAplot1}
\end{figure*}

We had mentioned earlier how a broader phenomenological perspective can sometimes make it 
relevant to know which couplings can be made smaller by trading for others. We can see an 
example of that here. The $\lambda'$ couplings above need to satisfy certain single bounds, \textit{cf.} Ref.~\cite{Allanach:1999ic}:
\begin{align}
    |\lambda'^{123}| &\leq 0.43, \qquad |\lambda'^{132}| \leq 1.04\,,\notag\\
    |\lambda'^{223}| &\leq 1.12, \qquad |\lambda'^{232}| \leq 1.04\,,\notag\\
    |\lambda'^{323}| &\leq 1.12, \qquad |\lambda'^{332}| \leq 1.04\,.
\label{eq:EAeq6}
\end{align}
In addition, there are also product bounds:
\begin{align}
    |\lambda'^{123*}\lambda'^{223}| &\leq 0.0076\,,\notag\\
     |\lambda'^{132*}\lambda'^{232}| &\leq 0.0076\,.
\label{eq:EAeq7}
\end{align}
In the above, we have assumed all sfermion masses to be \SI{1}{\tera\electronvolt}, 
or, if stricter, we have used the perturbativity constraint. Thus, for model building, 
solutions with, for instance, small $\lam'_{123}$ are preferable. We recast the solution space 
for the two limits of~\cref{fig::Model1}, and depict it in terms of the RPV couplings using~\cref{eq:EAeq5} in~\cref{fig:EAplot1}. The plot also depicts the regions ruled out by the 
above bounds as shaded grey regions.

\subsection{Bilinear-Trilinear Models}
\label{subsec:F}
\begin{figure}[htp]
\centering
\includegraphics[width=0.4\textwidth]{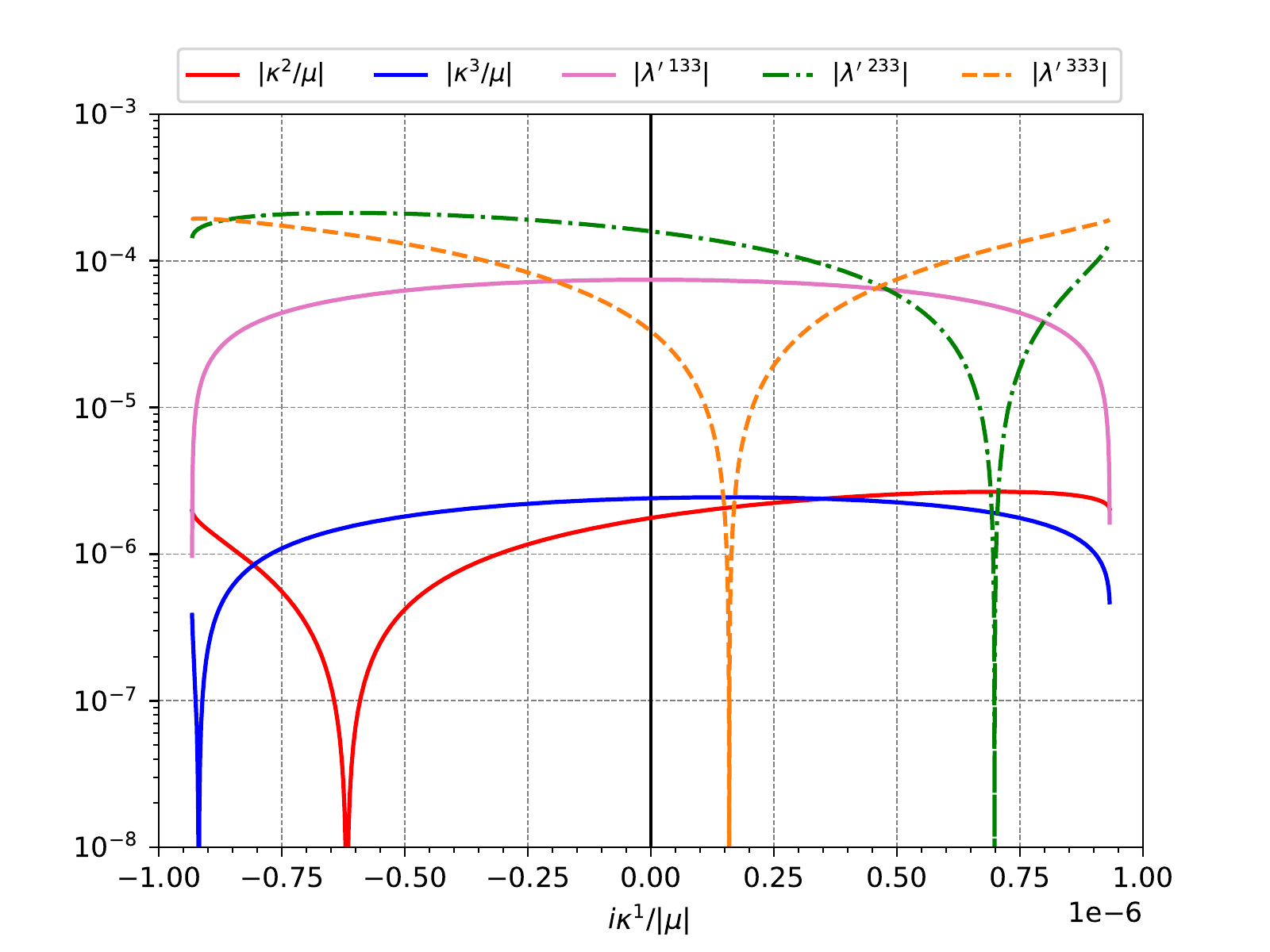}%
\caption{The NO limit solution space for the bilinear-trilinear mixed model of~\cref{subsec:F} including $\delta_{CP}$.}
\label{fig:EAplot2}%
\end{figure}

The final model we consider has contributions from both the bilinear and trilinear sectors. To have predictivity, we consider a scenario where all the $\delta_{\kappa}$ bilinears and the \textit{diagonal} trilinears contribute. The effective neutrino mass matrix is,
\begin{align}
    \mathbf{M}_{\nu}^{ij} &= m_{0}\delta_{\kappa}^i\delta_{\kappa}^j + \frac{g_2\left[\left(m^i_e\right)^2 + \left(m^j_e\right)^2\right]}{16\pi^2v}\notag\\
    \qquad &\times \left(1 + \sin^2\beta + \tan\beta\sin^2\beta\right)\delta_{\kappa}^i\delta_{\kappa}^j\notag\\
    &+ \frac{3}{8\pi^2 m_\text{SUSY}}\delta_{\lambda'}^{i33}\delta_{\lambda'}^{j33}m^2_{b} +  \ldots\,,     \end{align}
where, the ellipsis indicates terms that are suppressed by extra powers of the Yukawas. The above does not have a MOM structure. However, as long as $\tan\beta$ is not too large, the second term is expected to be suppressed compared to the first and third. The former is due to the extra Yukawas, while the latter follows from the fact that the bilinear invariants typically have to satisfy bounds at least a couple of orders of magnitude more stringent than the trilinear ones in order to fit the neutrino data -- for instance, \textit{cf.} the numbers in the previous applications. Ignoring the second term, the model reduces to a MOM with Class~3 structure, as can be seen by making the identifications,
\begin{align}
    x^i &= \sqrt{m_0}\delta^i_{\kappa}\,, \notag\\
    x'^i &= \sqrt{\frac{3} {8\pi^2 m_\text{SUSY}}}m_b \delta_\lambda'^{i33}\,.
\end{align}
Once again, we show what the solution space looks like for the above model by recasting the plot corresponding to the NO limit of~\cref{fig::Model3} in~\cref{fig:EAplot2}.

\section{Conclusions}
\label{sec:9}

In this paper, we have explored neutrino-mass generation in the $B_3$-conserving, but $R$-parity-violating MSSM. The main obstacle to a systematic phenomonelogical study in general RPV models is the large number of undetermined parameters. Typically, to deal with this, one specializes to specific models; this, however, restricts the applicability of the study. Here, we have taken a different route. By analyzing the structures of the neutrino mass matrix, we have identified four classes of minimal models -- the Minimal Oscillation Models (MOMs) -- that are consistent with the neutrino oscillation data for the case of two massive neutrinos. This allows for a model-independent study, at least for all models that satisfy the MOM criteria. Our study can be generalized to the case of three massive neutrinos.

We have analyzed each MOM class individually, and shown that it is possible to obtain solution-points consistent with the observed neutrino masses and mixings; for each class there is actually an infinite space of solutions. We have explored the general features of these solution spaces. Finally, we have presented numerical fits that can be adapted to any (MOM-like) specific RPV model without the need for re-performing the least-squares fit. As a demonstration, we have studied several examples that show the wide range of applicability of MOMs. This includes bilinear-only models, trilinear-only models (diagonal and non-diagonal), as well as mixed models.

MOMs do not solve the most general RPV case; we have described the limitations of the framework in the main text. However, given its simplicity, predictivity, and range of applicability, we believe the MOM framework is a useful way to think about neutrino masses in general RPV settings. 
\section*{Acknowledgments}
We thank Philip Bechtle for useful discussions. We acknowledge partial 
ﬁnancial support by the Deutsche Forschungsgemeinschaft (DFG, German 
Research Foundation) through the funds provided to the Sino-German 
Collaborative Research Center TRR110 “Symmetries and the Emergence of 
Structure in QCD” (DFG Project ID 196253076 - TRR 110).
\appendix
\section{Analytical Expressions for the MOM Solution Spaces}
\label{sec:A}
We write the explicit analytical solutions for the first three MOM classes here. As mentioned in the main text, the expressions for Class~4 MOMs are lengthy; we skip presenting them. Throughout, we treat $x^1$ as our free variable and solve~\cref{eq:MOMeq1} for the other variables. For short, we use the notation $M^{ij}\equiv\mathbf{M}_{\nu}^{ij}$.
\subsection*{Class~1: \texorpdfstring{$x^ix'^j + x'^ix^j$}{}}
\begin{align}
    x'^1 &= \frac{M^{11}}{2x^1}\,, \quad x'^2 = \frac{M^{22}}{2x^2}\,, \quad x'^3 = \frac{M^{33}}{2x^3}\,, \quad \text{with}\notag\\
    x^2 &= \frac{M^{12}}{M^{11}}x^1 \pm \frac{\sqrt{\left(M^{12}\right)^2\left(x^1\right)^2-M^{11}M^{22}\left(x^1\right)^2}}{{M^{11}}}\,,\notag\\
    x^3 &= \frac{M^{13}}{M^{11}}x^1 \pm \frac{\sqrt{\left(M^{13}\right)^2\left(x^1\right)^2-M^{11}M^{33}\left(x^1\right)^2}}{{M^{11}}}\,.
    \label{eq:App1}
\end{align}
This represents four distinct solutions corresponding to the various sign choices. The above expressions are general as long as $M^{ii} \neq 0$ for any $i$, which is true for the experimental neutrino mass matrix.

\subsection*{Class~2: \texorpdfstring{$x^ix^j + \left(x^ix'^j + x'^ix^j\right)$}{}}
\begin{align}
    x'^1 &= \frac{M^{11}-\left(x^1\right)^2}{2x^1}\,, \quad x'^2 = \frac{M^{22}-\left(x^2\right)^2}{2x^2}\,,\notag\\
    x'^3 &= \frac{M^{33}-\left(x^3\right)^2}{2x^3}\,,\notag 
\end{align}
with
\begin{align}
x^2 &= \frac{M^{12}}{M^{11}}x^1 \pm \frac{\sqrt{\left(M^{12}\right)^2\left(x^1\right)^2-M^{11}M^{22}\left(x^1\right)^2}}{{M^{11}}}\,,\notag\\
    x^3 &= \frac{M^{13}}{M^{11}}x^1 \pm \frac{\sqrt{\left(M^{13}\right)^2\left(x^1\right)^2-M^{11}M^{33}\left(x^1\right)^2}}{{M^{11}}}\,.
    \label{eq:App2}
\end{align}
This represents four distinct solutions. Once again, the expressions are valid as long as $M^{ii} \neq 0$ for any $i$.

\subsection*{Class~3: \texorpdfstring{$x^ix^j + x'^ix'^j$}{}}
\begin{align}
    \quad x'^2 &= \frac{M^{12}-x^1x^2}{x'^1}\,, \quad x'^3 = \frac{M^{13}-x^1x^3}{x'^1}\,, \quad \text{with}\notag\\
    x'^1 &= \pm\sqrt{M^{11}-\left(x^1\right)^2}\,,\notag\\
    x^2 &= \frac{M^{12}}{M^{11}}x^1\notag\\
    \quad &\pm\sqrt{\frac{\left(M^{12}\right)^2}{\left(M^{11}\right)^2}\left(x^1\right)^2-\frac{M^{22}}{M^{11}}\left(x^1\right)^2 - \frac{\left(M^{12}\right)^2}{M^{11}} + M^{22}}\,,\notag
\end{align} 
\begin{align}
    x^3 &= \frac{M^{13}}{M^{11}}x^1\notag\\
    \quad &\pm\sqrt{\frac{\left(M^{13}\right)^2}{\left(M^{11}\right)^2}\left(x^1\right)^2-\frac{M^{33}}{M^{11}}\left(x^1\right)^2 - \frac{\left(M^{13}\right)^2}{M^{11}} + M^{33}}\,.
    \label{eq:App3}
\end{align}
This represents eight distinct solutions. The above expressions are valid for $\left(x^1\right)^2 \neq M^{11}$, and $M^{11} \neq 0$. For the case $\left(x^1\right)^2 = M^{11}$, the solution sets are:
\begin{align}
    \quad x'^1 &= 0\,, \quad x'^2 = \pm \frac{\sqrt{M^{11}M^{22}-\left(M^{12}\right)^2}}{{\sqrt{M^{11}}}}\,,\notag\\
    x'^2 &= \pm \frac{\sqrt{M^{11}M^{33}-\left(M^{13}\right)^2}}{{\sqrt{M^{11}}}}\,,\notag\\
    x^2 &= \frac{M^{12}}{x^1}\,, \quad x^3 = \frac{M^{13}}{x^1}\,.
\end{align}

\section{A Mini-guide to MOMs}
\label{sec:B}
In this appendix, we expand upon certain points related to MOMs that were only briefly mentioned in the main text.

\subsection*{RPV parameters and Linearly Independent \texorpdfstring{$x^i$}{}}
Recall from the main text that the variables $x^i$ are directly proportional to the RPV couplings. Nevertheless, there is no simple relation between the number of contributing RPV parameters in a model and the number of linearly independent $x^i$ needed to describe their contributions. We demonstrate this through an explicit example.

Consider a $\kappa$-only model with all other RPV parameters zero in some basis. Consulting~\cref{tab:TFtab1}, our neutrino mass matrix receives contributions at tree level, as well as of types 7, 8 and 13 at one-loop level:
\begin{align}
    \mathbf{M}_{\nu}^{ij} &\sim m_{0}\delta_{\kappa}^{i}\delta_{\kappa}^{j} + \frac{\delta^i_\kappa \delta^j_\kappa m_{e_i} m_{e_j} h^i_e h^j_e}{16\pi^2m_{\text{SUSY}}}\notag\\ 
     &+ \frac{\delta^i_\kappa \delta^j_\kappa \left[ \left(m_{e_i} h^i_e \right)^2+ \left(m_{e_j} h^j_e\right)^2 \right]}{16\pi^2m_{\text{SUSY}}}\notag\\ &+ \frac{g\, \delta^i_\kappa \delta^j_\kappa \left(m^2_{e_i}+ m^2_{e_j} \right)}{16\pi^2m_{\text{SUSY}}}\notag\\
    &\sim x^ix^j + x'^ix'^j\notag\\
    &+ \left(x^ix''^j+x''^ix^j\right)\notag\\
    &+ \left(x^ix'''^j +x'''^ix^j\right)\,,
\end{align}
where, we have defined,
\begin{align}
x^i&\equiv \sqrt{m_{0}}\delta_{\kappa}^{i} \,, &&x'^i\equiv\frac{m_{e_i}h^i_e\delta^i_\kappa}{4\pi\sqrt{m_{\text
{SUSY}}}}\,,\notag\\
x''^i&\equiv \frac{\left(m_{e_i} h^i_e
\right)^2 \delta^i_\kappa}{16\pi^2m_{\text{SUSY}}\sqrt{m_{0}}} \,,
&&x'''^i\equiv \frac{2g\, 
\delta^i_\kappa m^2_{e_i}}{16\pi^2m_{\text{SUSY}}\sqrt{m_{0}}} \,.
\end{align}
It can easily be checked that any three of these four sets are linearly independent. Even though all the 
contributions come from only one RPV parameter set -- $\kappa^i$ -- we need three linearly independent sets to describe the structure.

We can also have situations where the opposite is true, \textit{i.e.}, where
several RPV parameter sets lead to fewer linearly independent sets $x^i$. We 
already saw an example of this in~\cref{subsec:D}.

\subsection*{Deriving the Class~4 MOM Structure}
Recall our observation that the neutrino mass matrix only has contributions of two forms: $x^ix^j$, and $x^ix'^j + x'^ix^j$. Given this fact, and restricting ourselves to the case of two linearly independent sets, it is immediately clear how the first three classes of MOMs arise. Here, we describe how we get the fourth one.

With only the linearly independent sets, $x^i$ and $x'^i$, contributing, the most general form of the neutrino mass matrix is,
\begin{align}
    \mathbf{M}_{\nu}^{ij} = x^ix^j + x'^ix'^j + \left(x^ix'^j + x'^ix^j\right)\,. 
\end{align}
Now, consider an additional contributing set, $x''^i$. The most general form the matrix can then take is,
\begin{align}
    \mathbf{M}_{\nu}^{ij} &= x^ix^j + x'^ix'^j + \left(x^ix'^j + x'^ix^j\right)\notag\\ 
    \qquad &+ x''^ix''^j + \left(x^ix''^j + x''^ix^j\right) + \left(x'^ix''^j + x''^ix'^j\right)\,. 
\end{align}
Since we assume there are only two linearly independent sets, there have to be some $a, b$ (not both zero) such that $x''^i = ax^i + bx'^i$. Substituting this in the above expression, we get, after some algebra,
\begin{align}
    \mathbf{M}_{\nu}^{ij} &= \left(1+a^2+2a\right)x^ix^j + \left(1+b^2+2b\right)x'^ix'^j \notag\\
    \qquad &+ \left(1+ab+2a+2b\right)\left(x^ix'^j + x'^ix^j\right)\,. 
\end{align}
Finally, defining,
\begin{align}
\tilde{x}^i &\equiv \sqrt{\left(1+a^2+2a\right)}x^i \,, \quad \tilde{x}'^i \equiv \sqrt{\left(1+b^2+2b\right)}x'^i\,,\notag \\
A &\equiv \frac{\left(1+ab+2a+2b\right)}{\sqrt{\left(1+a^2+2a\right)}\sqrt{\left(1+b^2+2b\right)}}\,,
\end{align}
 we get,
\begin{align}
    \mathbf{M}_{\nu}^{ij} &= \tilde{x}^i\tilde{x}^j + \tilde{x}'^i\tilde{x}'^j + A\left(\tilde{x}^i\tilde{x}'^j + \tilde{x}'^i\tilde{x}^j\right)\,,
\end{align}
which is the Class~4 MOM structure. Note that this is not the most general form of $A$ since further couplings could contribute; the arguments remain the same.

\subsection*{Most General Solution and MOMs}
We stressed in the main text that the MOM approach does not solve the most general case since there can always be conspiring cancellations. We demonstrate this here with an example.

Consider a scenario where we have three linearly independent sets -- $x^i, x'^i, x''^i$ -- with the explicit form,
\begin{align}
    x^i &= \left(1, 0, 0\right)\,,\notag\\
    x'^i &= \left(0, 1, 0\right)\,,\notag\\
    x''^i &=\left(0, 0, 1\right)\,.
\end{align}
Now, consider a fourth contribution $x'''^i$. This can always be written in the form,
\begin{align}
    x'''^i = ax^i + bx'^i + cx''^i\,,
\end{align}
for some $a, b, c$. Finally, consider the matrix with the structure,
\begin{align}
    \mathbf{M}_{\nu}^{ij} = x^ix^j + x'^ix'^j + \left(x''^ix'''^j + x'''^ix''^j\right)\,.
\end{align}
The above matrix clearly does not have a MOM form since it has three linearly
independent sets. However, the matrix is rank two as long as the condition $a^2+b^2 = 2c$ is satisfied. Thus, a matrix being rank two does
not imply that the matrix has a MOM-form. There 
can always be additional hidden structure -- for instance through specific 
cancellations/relations as in the above 
case.

\section{Numerical Plots}
\label{sec:C}
\subsection*{\texorpdfstring{$CP$}{}-Conserving Solutions of MOMs}
\begin{figure}[!h]
\centering
\begin{subfigure}[b]{0.45\textwidth}
    \includegraphics[width=\textwidth]{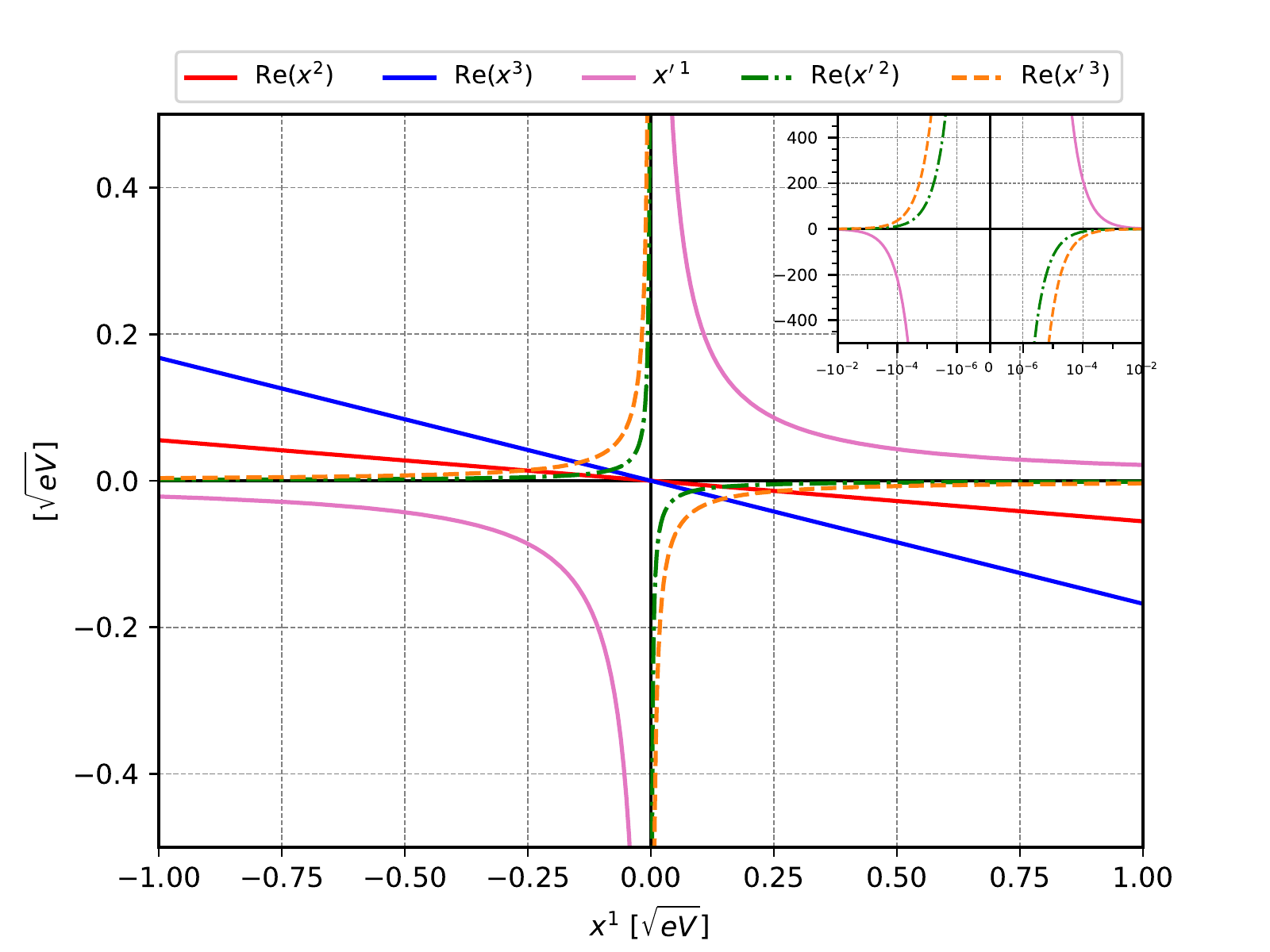}
\end{subfigure}
\begin{subfigure}[b]{0.45\textwidth}
    \includegraphics[width=\textwidth]{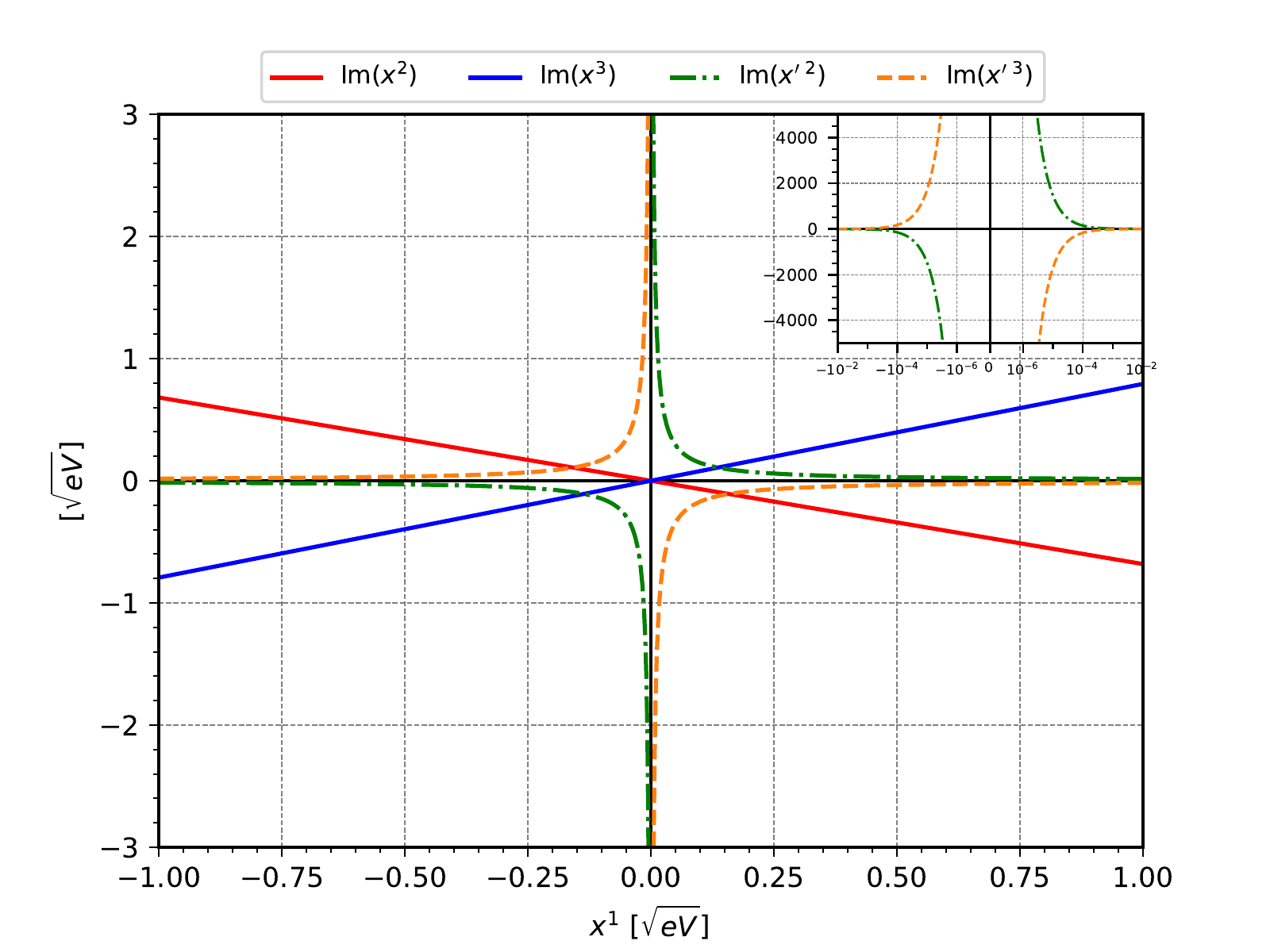}
\end{subfigure}
 \caption{Real (top) and imaginary (bottom) values of the couplings required to fit the actual neutrino data for the Inverted Ordering limit in models with Class~1 structure.}
    \label{fig::Model1IM}%
\end{figure}

\begin{figure*}[!h]
    \centering\begin{tabular}[b]{c c}
    \subfloat[Inverted Ordering]{\includegraphics[scale=0.5]{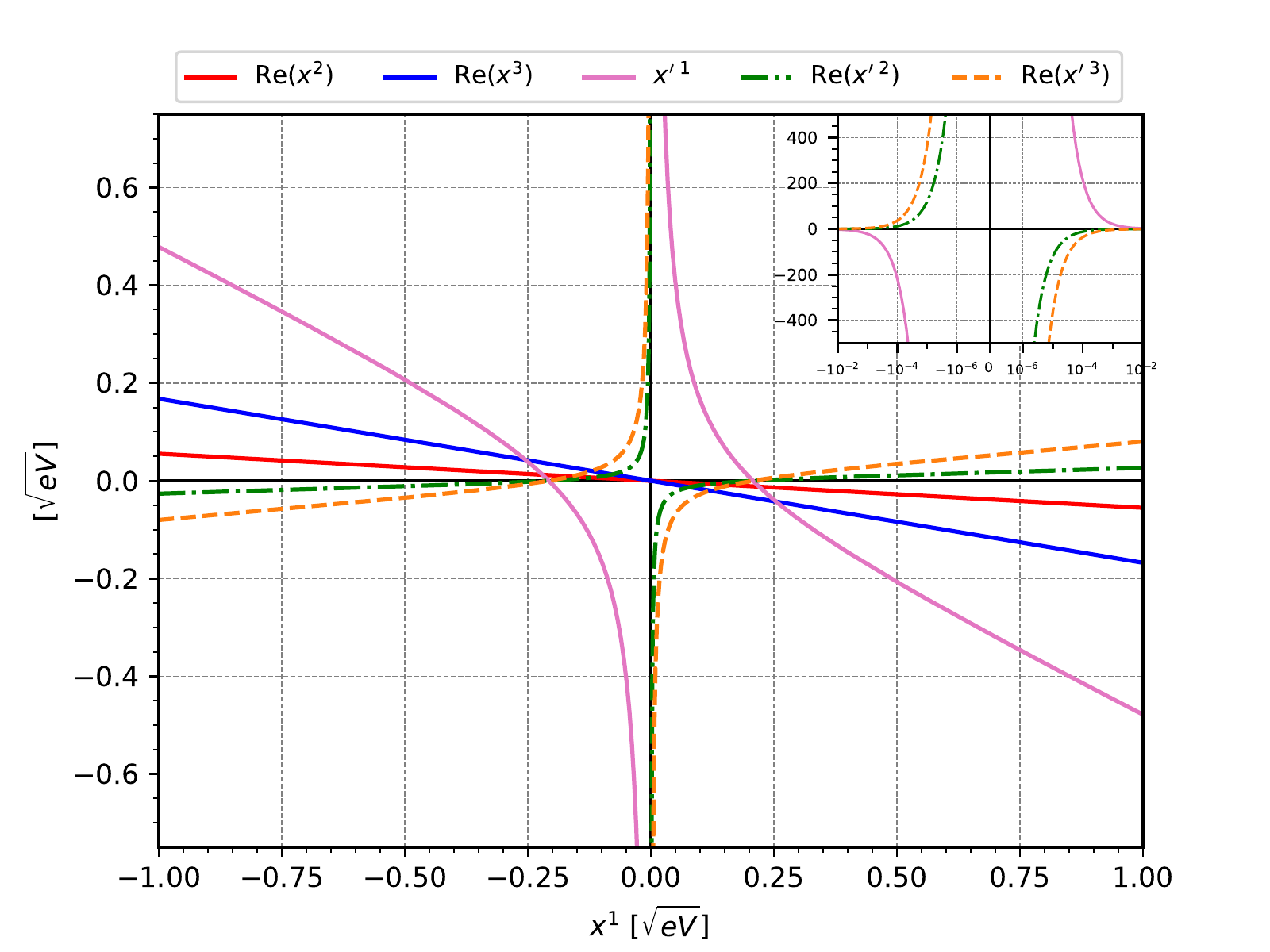}}%
   &
    \subfloat[Inverted Ordering]{\includegraphics[scale=0.5]{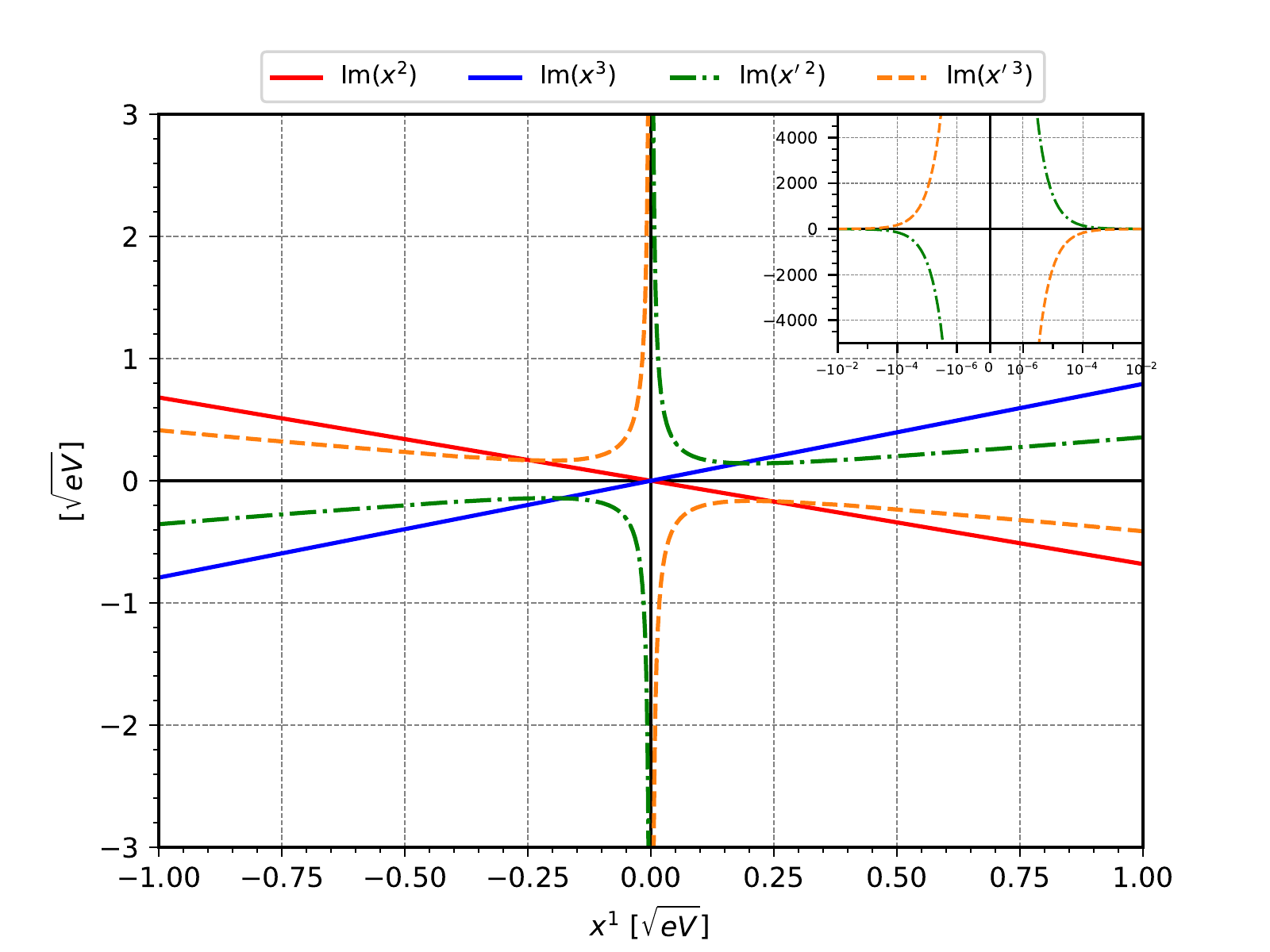}}\\
    \subfloat[Normal Ordering]{\includegraphics[scale=0.5]{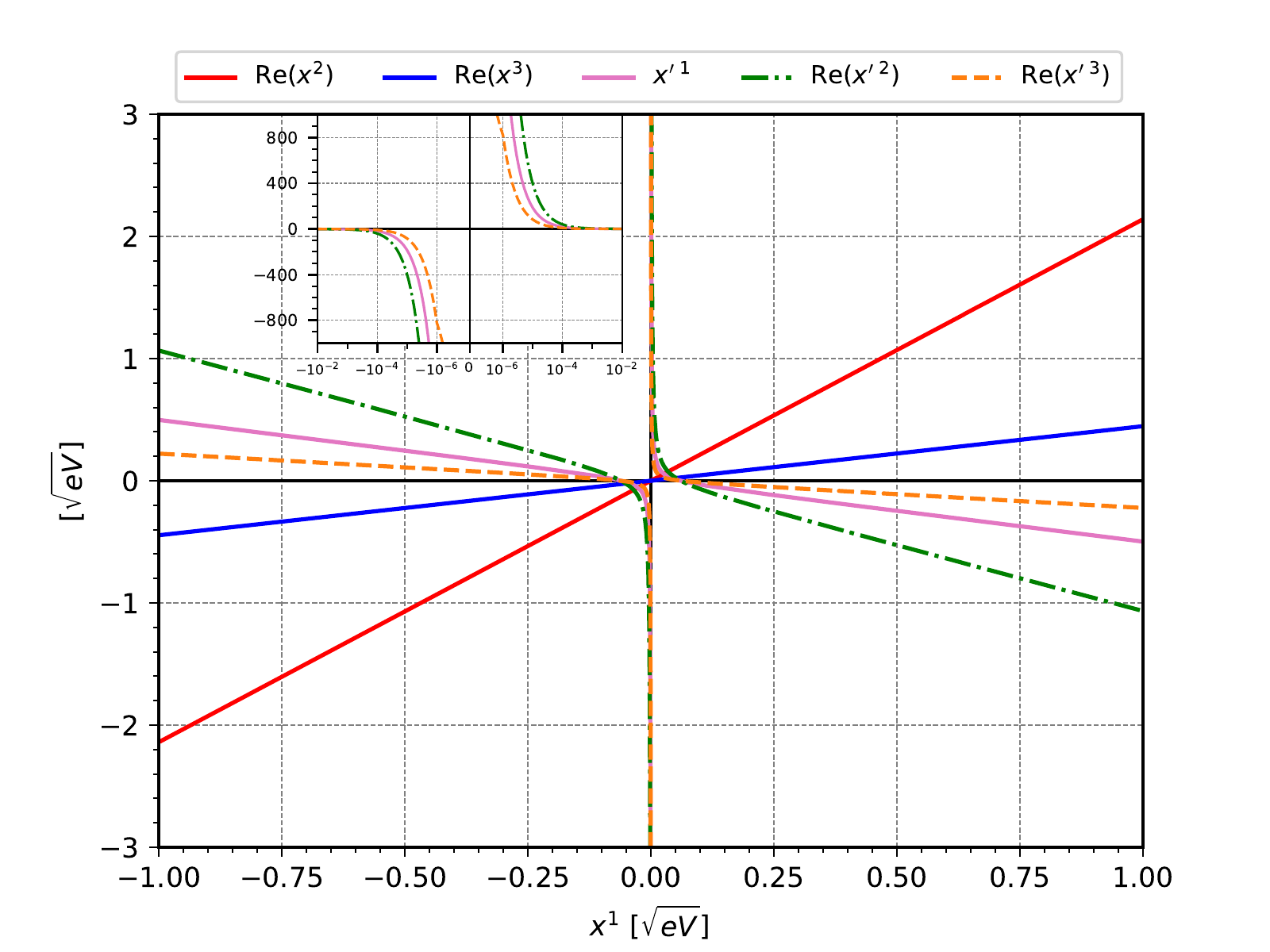}}%
   &
    \subfloat[Normal Ordering]{\includegraphics[scale=0.5]{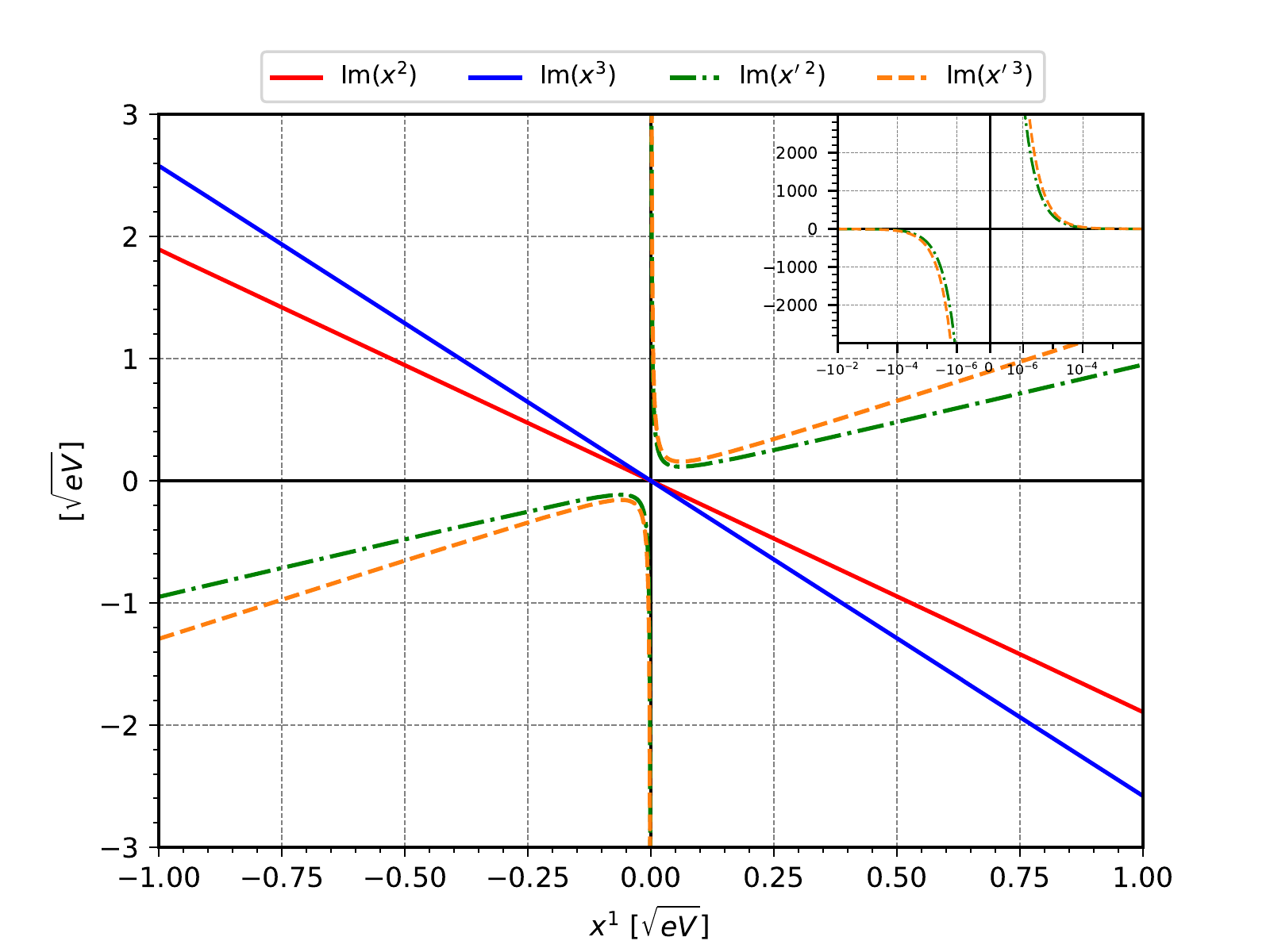}}
     \end{tabular}
    \caption{Real (left) and imaginary (right) values of the couplings required in models with Class~2 structure to fit the actual neutrino data for the two mass orderings: IO (top) and NO (bottom).}
    \label{fig::Model2}%
   \end{figure*}

\begin{figure*}[!h]
\centering
\begin{subfigure}[b]{0.4\textwidth}
    \includegraphics[width=\textwidth]{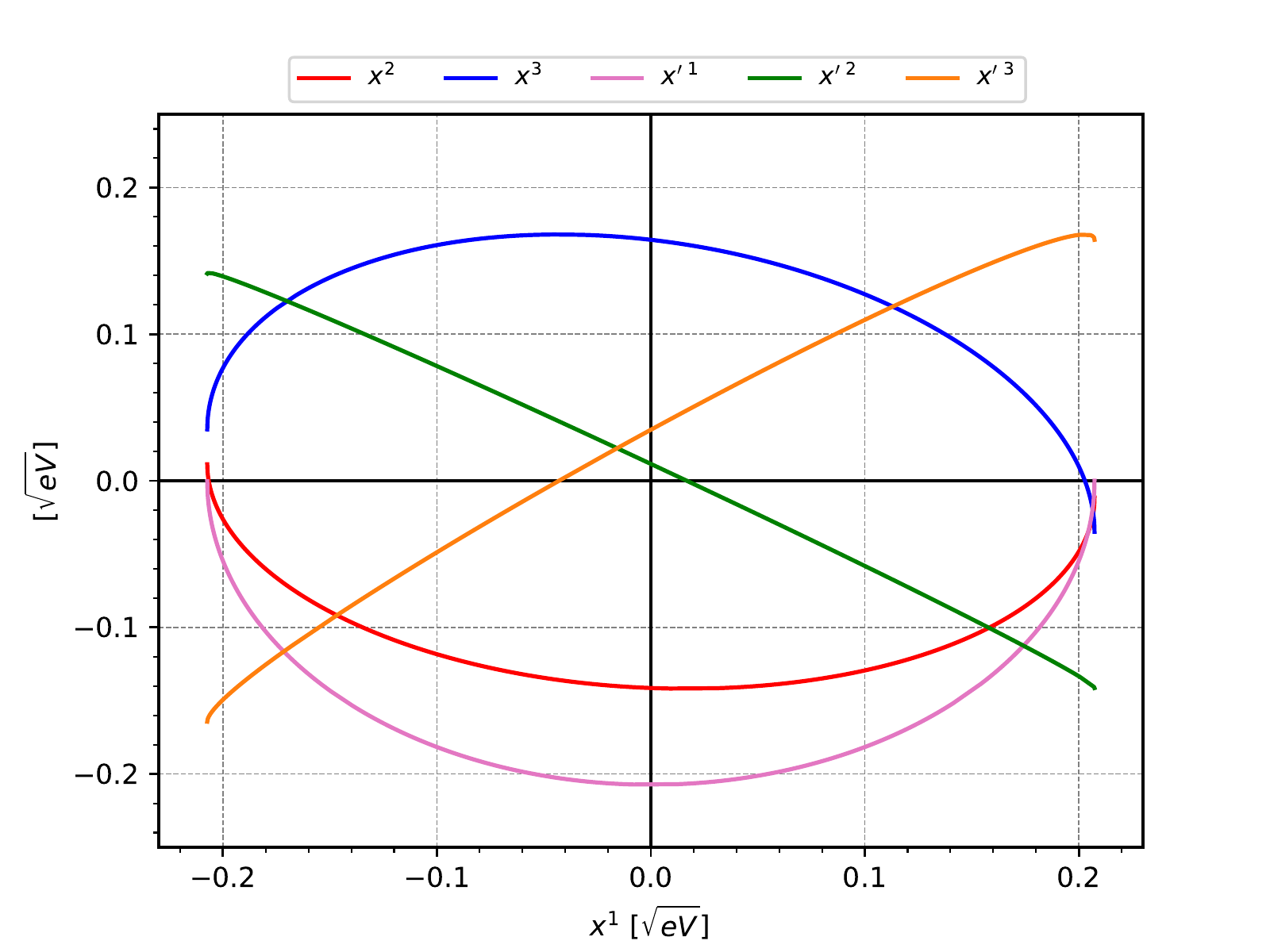}
\end{subfigure}
\begin{subfigure}[b]{0.4\textwidth}
    \includegraphics[width=\textwidth]{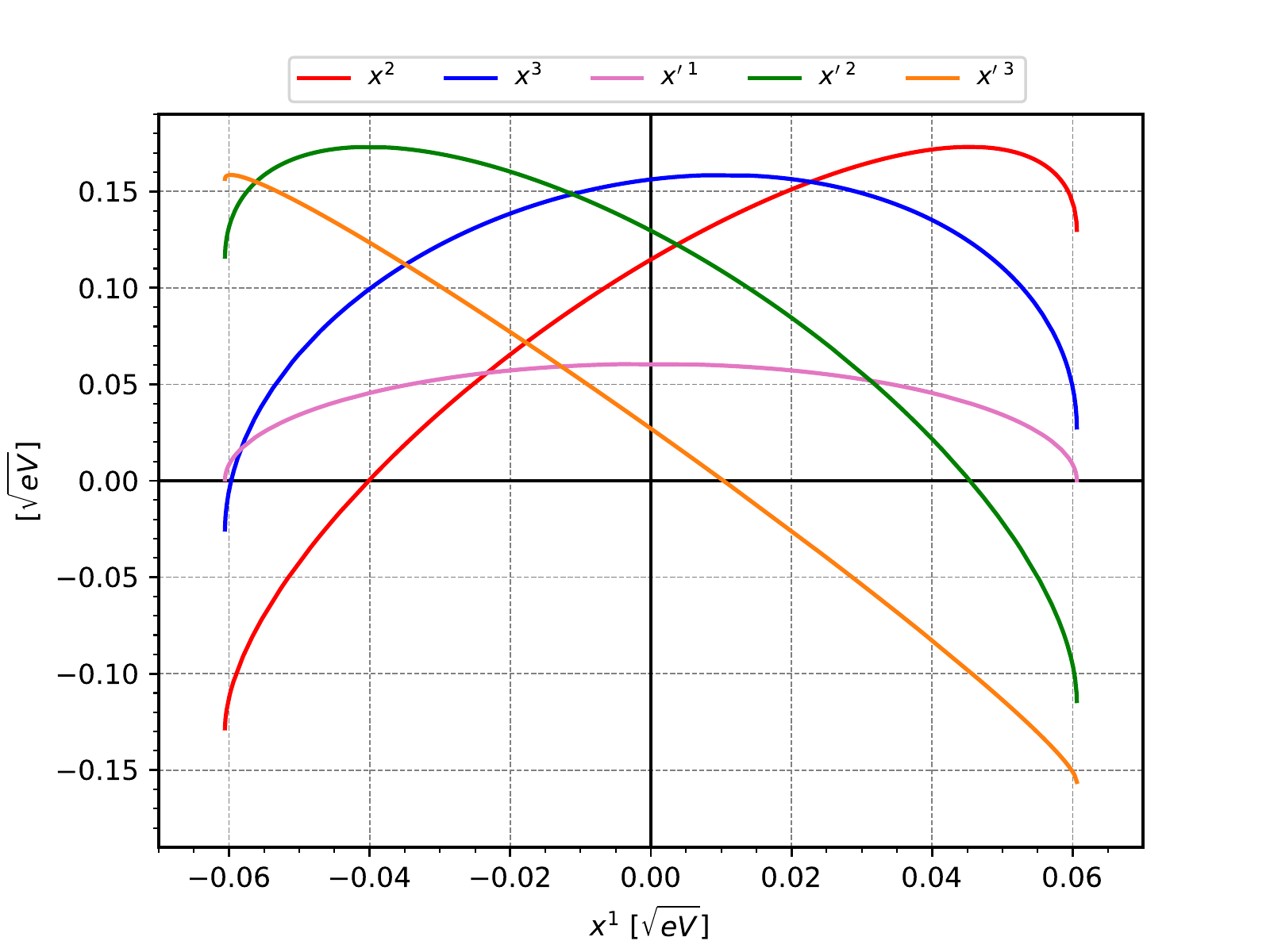}
\end{subfigure}
\caption{Values of the couplings required in models with Class~3 structure to fit the actual neutrino data for the two mass orderings: IO (left) and NO (right). For this class, all couplings are real.}
\label{fig::Model3}
\end{figure*}

This appendix contains the numerical fits to the experimental data. ~\cref{fig::Model1IM} shows the solution for Class~1 structures, assuming Inverted Ordering. In~\cref{fig::Model2} and~\cref{fig::Model3}, we display the solution for Class~2 and Class~3 structures correspondingly. The fits have been performed using the neutrino data of~\cref{tab:neudata} (with $\delta_{CP} = 0$) as described in~\cref{sec:7}.
\vspace{1cm}


\clearpage


\subsection*{MOMs and a Non-zero \texorpdfstring{$\delta_{CP}$}{}}
\label{app:CP}
\begin{widetext}

\begin{figure*}[h!]
  \centering
    \subfloat[Inverted Ordering]{\includegraphics[scale=0.48]{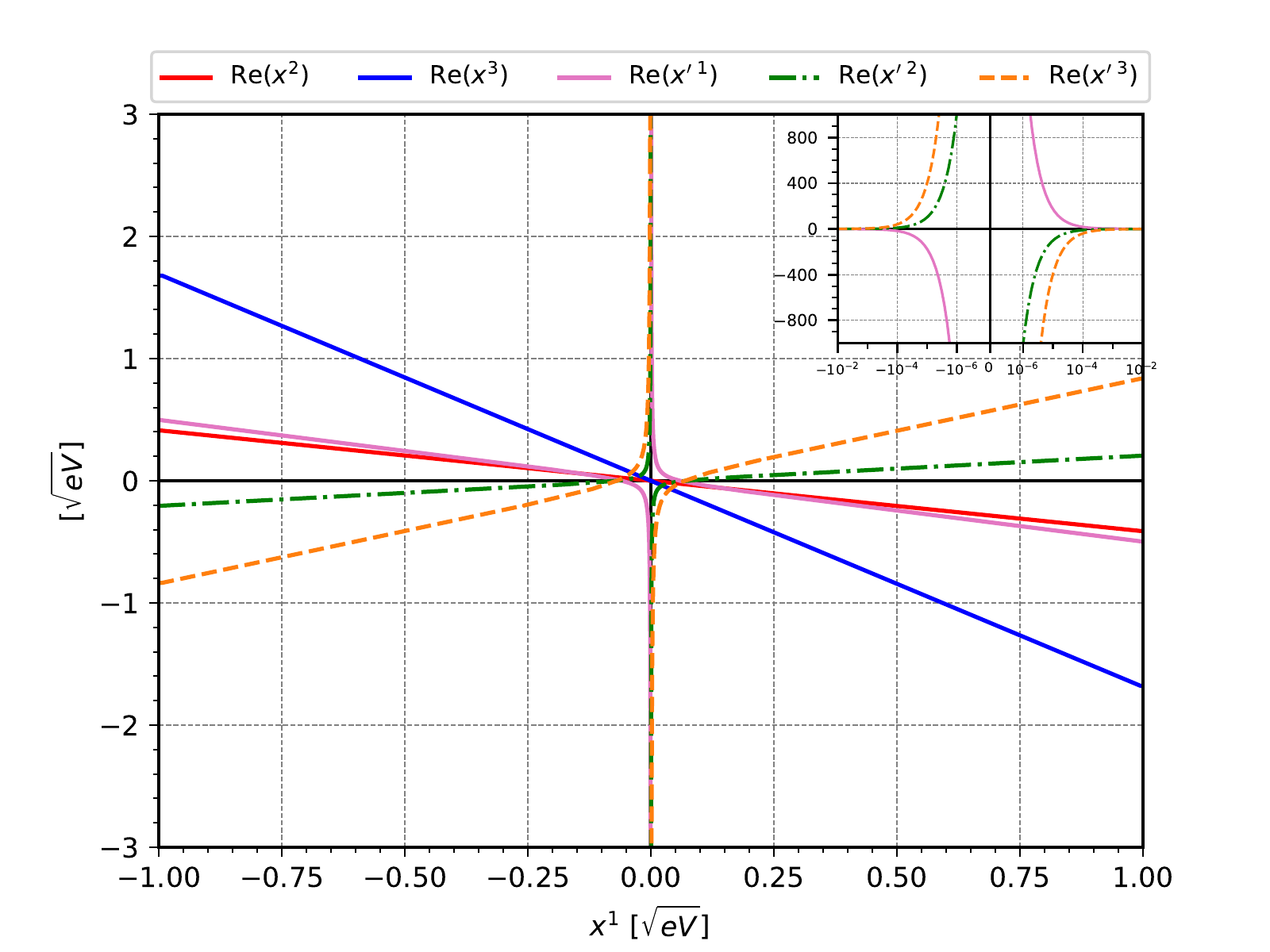}}
    \subfloat[Normal Ordering]{\includegraphics[scale=0.48]{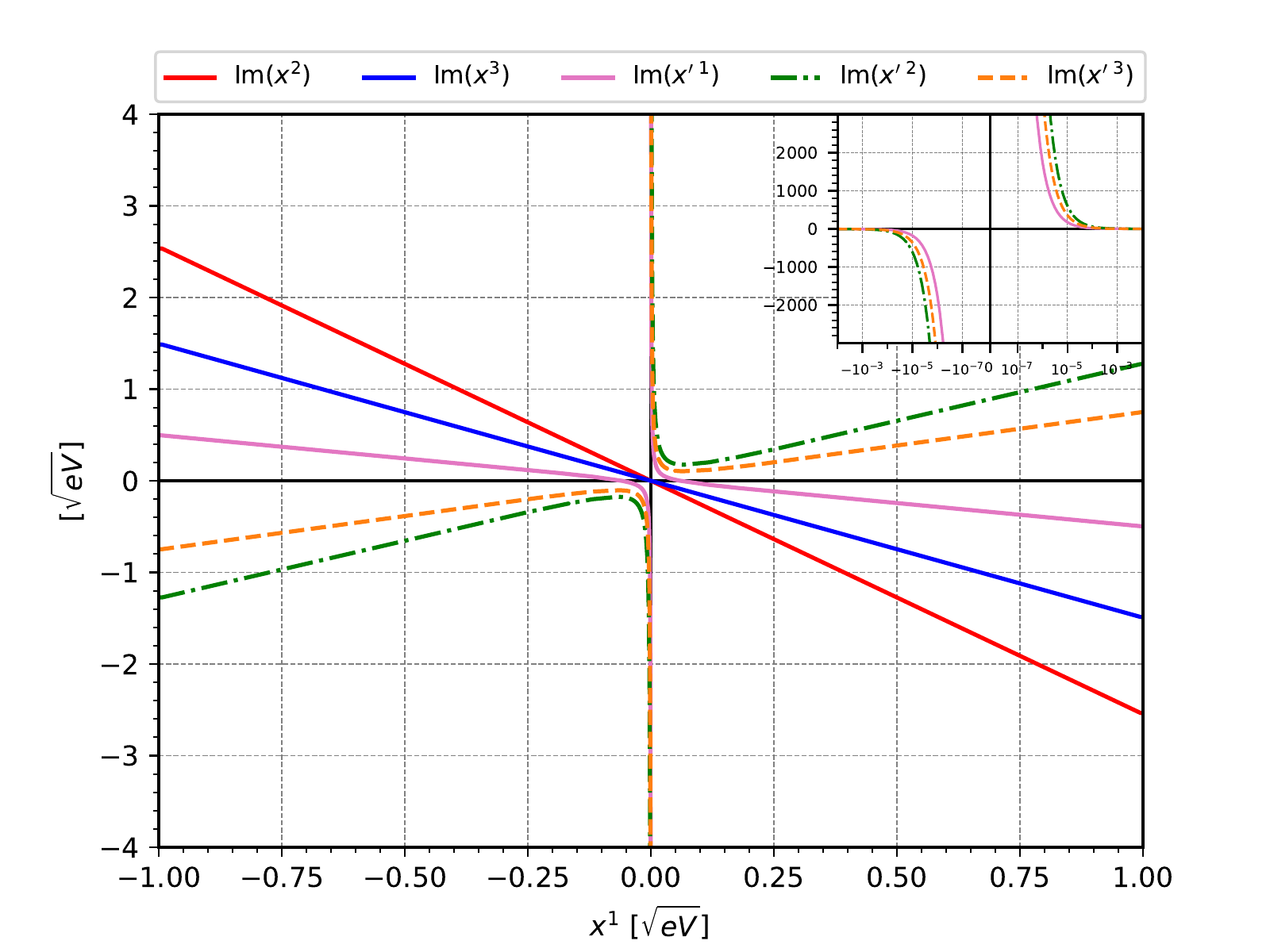}}%
  \caption{Values of the couplings required to fit the actual neutrino data including $\delta_{CP}$ in models with Class~2 structures for normal ordering.}
\label{fig:CPV}
\end{figure*} 

\end{widetext}

We include, here, a fit with non-zero $\delta_{{CP}}$ to show that it is 
possible to accommodate $CP$ violation in our framework. We show, in~\cref{fig:CPV}, the fit corresponding to the Normal Ordering data of~\cref{tab:neudata} for Class~2 MOMs, including the best-fit value for $\delta_{CP}$. The qualitative features are as before. The overall goodness of the fit is also stable, indicating that the minimum of the $\chi^2$ is determined as robustly as
before.
\bibliographystyle{JHEP}
\bibliography{refs}

\providecommand{\href}[2]{#2}\begingroup\raggedright\begin{thebibliography}{10}

\bibitem{Workman:2022ynf}
{\bf Particle Data Group} Collaboration, R.~L. Workman, {\it {Review of
  Particle Physics}},  {\em PTEP} {\bf 2022} (2022) 083C01.

\bibitem{KATRIN:2021uub}
{\bf KATRIN} Collaboration, M.~Aker et~al., {\it {Direct neutrino-mass
  measurement with sub-electronvolt sensitivity}},  {\em Nature Phys.} {\bf 18}
  (2022), no.~2 160--166, [\href{http://arxiv.org/abs/2105.08533}{{\tt
  arXiv:2105.08533}}].

\bibitem{Nilles:1983ge}
H.~P. Nilles, {\it {Supersymmetry, Supergravity and Particle Physics}},  {\em
  Phys. Rept.} {\bf 110} (1984) 1--162.

\bibitem{Martin:1997ns}
S.~P. Martin, {\em {A Supersymmetry primer}}, vol.~21, pp.~1--153.
\newblock Kane, Gordon L., 2010.
\newblock \href{http://arxiv.org/abs/hep-ph/9709356}{{\tt hep-ph/9709356}}.

\bibitem{Dreiner:1997uz}
H.~K. Dreiner, {\it {An Introduction to explicit R-parity violation}},  {\em
  Adv. Ser. Direct. High Energy Phys.} {\bf 21} (2010) 565--583,
  [\href{http://arxiv.org/abs/hep-ph/9707435}{{\tt hep-ph/9707435}}].

\bibitem{Barbier:2004ez}
R.~Barbier et~al., {\it {R-parity violating supersymmetry}},  {\em Phys. Rept.}
  {\bf 420} (2005) 1--202, [\href{http://arxiv.org/abs/hep-ph/0406039}{{\tt
  hep-ph/0406039}}].

\bibitem{Allanach:2003eb}
B.~C. Allanach, A.~Dedes, and H.~K. Dreiner, {\it {R parity violating minimal
  supergravity model}},  {\em Phys. Rev. D} {\bf 69} (2004) 115002,
  [\href{http://arxiv.org/abs/hep-ph/0309196}{{\tt hep-ph/0309196}}]. [Erratum:
  Phys.Rev.D 72, 079902 (2005)].

\bibitem{Hall:1983id}
L.~J. Hall and M.~Suzuki, {\it {Explicit R-Parity Breaking in Supersymmetric
  Models}},  {\em Nucl. Phys. B} {\bf 231} (1984) 419--444.

\bibitem{Hirsch:2000ef}
M.~Hirsch, M.~A. Diaz, W.~Porod, J.~C. Romao, and J.~W.~F. Valle, {\it
  {Neutrino masses and mixings from supersymmetry with bilinear R parity
  violation: A Theory for solar and atmospheric neutrino oscillations}},  {\em
  Phys. Rev. D} {\bf 62} (2000) 113008,
  [\href{http://arxiv.org/abs/hep-ph/0004115}{{\tt hep-ph/0004115}}]. [Erratum:
  Phys.Rev.D 65, 119901 (2002)].

\bibitem{Joshipura:1994ib}
A.~S. Joshipura and M.~Nowakowski, {\it {'Just so' oscillations in
  supersymmetric standard model}},  {\em Phys. Rev. D} {\bf 51} (1995)
  2421--2427, [\href{http://arxiv.org/abs/hep-ph/9408224}{{\tt
  hep-ph/9408224}}].

\bibitem{Nowakowski:1995dx}
M.~Nowakowski and A.~Pilaftsis, {\it {W and Z boson interactions in
  supersymmetric models with explicit R-parity violation}},  {\em Nucl. Phys.
  B} {\bf 461} (1996) 19--49, [\href{http://arxiv.org/abs/hep-ph/9508271}{{\tt
  hep-ph/9508271}}].

\bibitem{Banks:1995by}
T.~Banks, Y.~Grossman, E.~Nardi, and Y.~Nir, {\it {Supersymmetry without
  R-parity and without lepton number}},  {\em Phys. Rev. D} {\bf 52} (1995)
  5319--5325, [\href{http://arxiv.org/abs/hep-ph/9505248}{{\tt
  hep-ph/9505248}}].

\bibitem{Hempfling:1995wj}
R.~Hempfling, {\it {Neutrino masses and mixing angles in SUSY GUT theories with
  explicit R-parity breaking}},  {\em Nucl. Phys. B} {\bf 478} (1996) 3--30,
  [\href{http://arxiv.org/abs/hep-ph/9511288}{{\tt hep-ph/9511288}}].

\bibitem{Chun:1999bq}
E.~J. Chun and S.~K. Kang, {\it {One loop corrected neutrino masses and mixing
  in supersymmetric standard model without R-parity}},  {\em Phys. Rev. D} {\bf
  61} (2000) 075012, [\href{http://arxiv.org/abs/hep-ph/9909429}{{\tt
  hep-ph/9909429}}].

\bibitem{Kaplan:1999ds}
D.~Kaplan and A.~E. Nelson, {\it {Solar and atmospheric neutrino oscillations
  from bilinear R parity violation}},  {\em JHEP} {\bf 01} (2000) 033,
  [\href{http://arxiv.org/abs/hep-ph/9901254}{{\tt hep-ph/9901254}}].

\bibitem{Grossman:1997is}
Y.~Grossman and H.~E. Haber, {\it {Sneutrino mixing phenomena}},  {\em Phys.
  Rev. Lett.} {\bf 78} (1997) 3438--3441,
  [\href{http://arxiv.org/abs/hep-ph/9702421}{{\tt hep-ph/9702421}}].

\bibitem{Grossman:1998py}
Y.~Grossman and H.~E. Haber, {\it {(S)neutrino properties in R-parity violating
  supersymmetry. 1. CP conserving phenomena}},  {\em Phys. Rev. D} {\bf 59}
  (1999) 093008, [\href{http://arxiv.org/abs/hep-ph/9810536}{{\tt
  hep-ph/9810536}}].

\bibitem{Grossman:2003gq}
Y.~Grossman and S.~Rakshit, {\it {Neutrino masses in R-parity violating
  supersymmetric models}},  {\em Phys. Rev. D} {\bf 69} (2004) 093002,
  [\href{http://arxiv.org/abs/hep-ph/0311310}{{\tt hep-ph/0311310}}].

\bibitem{Diaz:2003as}
M.~Diaz, M.~Hirsch, W.~Porod, J.~Romao, and J.~Valle, {\it {Solar neutrino
  masses and mixing from bilinear R parity broken supersymmetry: Analytical
  versus numerical results}},  {\em Phys. Rev. D} {\bf 68} (2003) 013009,
  [\href{http://arxiv.org/abs/hep-ph/0302021}{{\tt hep-ph/0302021}}]. [Erratum:
  Phys.Rev.D 71, 059904 (2005)].

\bibitem{Davidson:2000uc}
S.~Davidson and M.~Losada, {\it {Neutrino masses in the R(p) violating MSSM}},
  {\em JHEP} {\bf 05} (2000) 021,
  [\href{http://arxiv.org/abs/hep-ph/0005080}{{\tt hep-ph/0005080}}].

\bibitem{Abada:2001zh}
A.~Abada, S.~Davidson, and M.~Losada, {\it {Neutrino masses and mixings in the
  MSSM with soft bilinear R(p) violation}},  {\em Phys. Rev. D} {\bf 65} (2002)
  075010, [\href{http://arxiv.org/abs/hep-ph/0111332}{{\tt hep-ph/0111332}}].

\bibitem{Davidson:2000ne}
S.~Davidson and M.~Losada, {\it {Basis independent neutrino masses in the R(p)
  violating MSSM}},  {\em Phys. Rev. D} {\bf 65} (2002) 075025,
  [\href{http://arxiv.org/abs/hep-ph/0010325}{{\tt hep-ph/0010325}}].

\bibitem{Dedes:2006ni}
A.~Dedes, S.~Rimmer, and J.~Rosiek, {\it {Neutrino masses in the lepton number
  violating MSSM}},  {\em JHEP} {\bf 08} (2006) 005,
  [\href{http://arxiv.org/abs/hep-ph/0603225}{{\tt hep-ph/0603225}}].

\bibitem{Allanach:2007qc}
B.~Allanach and C.~Kom, {\it {Lepton number violating mSUGRA and neutrino
  masses}},  {\em JHEP} {\bf 04} (2008) 081,
  [\href{http://arxiv.org/abs/0712.0852}{{\tt arXiv:0712.0852}}].

\bibitem{Borzumati:1996hd}
F.~Borzumati, Y.~Grossman, E.~Nardi, and Y.~Nir, {\it {Neutrino masses and
  mixing in supersymmetric models without R parity}},  {\em Phys. Lett. B} {\bf
  384} (1996) 123--130, [\href{http://arxiv.org/abs/hep-ph/9606251}{{\tt
  hep-ph/9606251}}].

\bibitem{Drees:1997id}
M.~Drees, S.~Pakvasa, X.~Tata, and T.~ter Veldhuis, {\it {A Supersymmetric
  resolution of solar and atmospheric neutrino puzzles}},  {\em Phys. Rev. D}
  {\bf 57} (1998) 5335--5339, [\href{http://arxiv.org/abs/hep-ph/9712392}{{\tt
  hep-ph/9712392}}].

\bibitem{Chun:1998gp}
E.~Chun, S.~Kang, C.~Kim, and U.~Lee, {\it {Supersymmetric neutrino masses and
  mixing with R-parity violation}},  {\em Nucl. Phys. B} {\bf 544} (1999)
  89--103, [\href{http://arxiv.org/abs/hep-ph/9807327}{{\tt hep-ph/9807327}}].

\bibitem{Joshipura:1999hr}
A.~S. Joshipura and S.~K. Vempati, {\it {Sneutrino vacuum expectation values
  and neutrino anomalies through trilinear R-parity violation}},  {\em Phys.
  Rev. D} {\bf 60} (1999) 111303,
  [\href{http://arxiv.org/abs/hep-ph/9903435}{{\tt hep-ph/9903435}}].

\bibitem{Choi:1998wc}
K.~Choi, K.~Hwang, and E.~J. Chun, {\it {Atmospheric and solar neutrino masses
  from horizontal U(1) symmetry}},  {\em Phys. Rev. D} {\bf 60} (1999) 031301,
  [\href{http://arxiv.org/abs/hep-ph/9811363}{{\tt hep-ph/9811363}}].

\bibitem{Kong:1998bs}
O.~C. Kong, {\it {Neutrino oscillations and flavor structure of supersymmetry
  without R-parity}},  {\em Mod. Phys. Lett. A} {\bf 14} (1999) 903--912,
  [\href{http://arxiv.org/abs/hep-ph/9808304}{{\tt hep-ph/9808304}}].

\bibitem{Rakshit:1998kd}
S.~Rakshit, G.~Bhattacharyya, and A.~Raychaudhuri, {\it {R-parity violating
  trilinear couplings and recent neutrino data}},  {\em Phys. Rev. D} {\bf 59}
  (1999) 091701, [\href{http://arxiv.org/abs/hep-ph/9811500}{{\tt
  hep-ph/9811500}}].

\bibitem{Adhikari:1999pa}
R.~Adhikari and G.~Omanovic, {\it {LSND, solar and atmospheric neutrino
  oscillation experiments, and R-parity violating supersymmetry}},  {\em Phys.
  Rev. D} {\bf 59} (1999) 073003.

\bibitem{Abada:2000xr}
A.~Abada and M.~Losada, {\it {Constraints on both bilinear and trilinear
  R-parity violating couplings from neutrino laboratories and astrophysics
  data}},  {\em Phys. Lett. B} {\bf 492} (2000) 310--320,
  [\href{http://arxiv.org/abs/hep-ph/0007041}{{\tt hep-ph/0007041}}].

\bibitem{Rakshit:2004rj}
S.~Rakshit, {\it {Neutrino masses and R-parity violation}},  {\em Mod. Phys.
  Lett. A} {\bf 19} (2004) 2239--2258,
  [\href{http://arxiv.org/abs/hep-ph/0406168}{{\tt hep-ph/0406168}}].

\bibitem{Dreiner:2011ft}
H.~K. Dreiner, M.~Hanussek, J.-S. Kim, and C.~Kom, {\it {Neutrino masses and
  mixings in the baryon triality constrained minimal supersymmetric standard
  model}},  {\em Phys. Rev. D} {\bf 84} (2011) 113005,
  [\href{http://arxiv.org/abs/1106.4338}{{\tt arXiv:1106.4338}}].

\bibitem{Romao:1999up}
J.~Romao, M.~Diaz, M.~Hirsch, W.~Porod, and J.~Valle, {\it {A Supersymmetric
  solution to the solar and atmospheric neutrino problems}},  {\em Phys. Rev.
  D} {\bf 61} (2000) 071703, [\href{http://arxiv.org/abs/hep-ph/9907499}{{\tt
  hep-ph/9907499}}].

\bibitem{Cheung:1999az}
K.-m. Cheung and O.~C. Kong, {\it {Zee neutrino mass model in SUSY framework}},
   {\em Phys. Rev. D} {\bf 61} (2000) 113012,
  [\href{http://arxiv.org/abs/hep-ph/9912238}{{\tt hep-ph/9912238}}].

\bibitem{Farrar:1978xj}
G.~R. Farrar and P.~Fayet, {\it {Phenomenology of the Production, Decay, and
  Detection of New Hadronic States Associated with Supersymmetry}},  {\em Phys.
  Lett. B} {\bf 76} (1978) 575--579.

\bibitem{Dreiner:2009ic}
H.~K. Dreiner, S.~Heinemeyer, O.~Kittel, U.~Langenfeld, A.~M. Weber, and
  G.~Weiglein, {\it {Mass Bounds on a Very Light Neutralino}},  {\em Eur. Phys.
  J. C} {\bf 62} (2009) 547--572, [\href{http://arxiv.org/abs/0901.3485}{{\tt
  arXiv:0901.3485}}].

\bibitem{Chamoun:2020aft}
N.~Chamoun, F.~Domingo, and H.~K. Dreiner, {\it {Nucleon decay in the R-parity
  violating MSSM}},  \href{http://arxiv.org/abs/2012.11623}{{\tt
  arXiv:2012.11623}}.

\bibitem{Dreiner:2006xw}
H.~K. Dreiner, C.~Luhn, H.~Murayama, and M.~Thormeier, {\it {Baryon triality
  and neutrino masses from an anomalous flavor U(1)}},  {\em Nucl. Phys. B}
  {\bf 774} (2007) 127--167, [\href{http://arxiv.org/abs/hep-ph/0610026}{{\tt
  hep-ph/0610026}}].

\bibitem{Ibanez:1991hv}
L.~E. Ibanez and G.~G. Ross, {\it {Discrete gauge symmetry anomalies}},  {\em
  Phys. Lett. B} {\bf 260} (1991) 291--295.

\bibitem{Ibanez:1991pr}
L.~E. Ibanez and G.~G. Ross, {\it {Discrete gauge symmetries and the origin of
  baryon and lepton number conservation in supersymmetric versions of the
  standard model}},  {\em Nucl. Phys. B} {\bf 368} (1992) 3--37.

\bibitem{Dreiner:2005rd}
H.~K. Dreiner, C.~Luhn, and M.~Thormeier, {\it {What is the discrete gauge
  symmetry of the MSSM?}},  {\em Phys. Rev. D} {\bf 73} (2006) 075007,
  [\href{http://arxiv.org/abs/hep-ph/0512163}{{\tt hep-ph/0512163}}].

\bibitem{Dreiner:1991pe}
H.~K. Dreiner and G.~G. Ross, {\it {R-parity violation at hadron colliders}},
  {\em Nucl. Phys. B} {\bf 365} (1991) 597--613.

\bibitem{Dercks:2017lfq}
D.~Dercks, H.~Dreiner, M.~E. Krauss, T.~Opferkuch, and A.~Reinert, {\it
  {R-Parity Violation at the LHC}},  {\em Eur. Phys. J. C} {\bf 77} (2017),
  no.~12 856, [\href{http://arxiv.org/abs/1706.09418}{{\tt arXiv:1706.09418}}].

\bibitem{Dreiner:2003hw}
H.~K. Dreiner and M.~Thormeier, {\it {Supersymmetric Froggatt-Nielsen models
  with baryon and lepton number violation}},  {\em Phys. Rev. D} {\bf 69}
  (2004) 053002, [\href{http://arxiv.org/abs/hep-ph/0305270}{{\tt
  hep-ph/0305270}}].

\bibitem{Dreiner:2007yz}
H.~K. Dreiner, J.~S. Kim, O.~Lebedev, and M.~Thormeier, {\it {Supersymmetric
  Jarlskog invariants: The Neutrino sector}},  {\em Phys. Rev. D} {\bf 76}
  (2007) 015006, [\href{http://arxiv.org/abs/hep-ph/0703074}{{\tt
  hep-ph/0703074}}].

\bibitem{Dreiner:2011fp}
H.~K. Dreiner, M.~Hanussek, J.~S. Kim, and S.~Sarkar, {\it {Gravitino cosmology
  with a very light neutralino}},  {\em Phys. Rev. D} {\bf 85} (2012) 065027,
  [\href{http://arxiv.org/abs/1111.5715}{{\tt arXiv:1111.5715}}].

\bibitem{Esteban:2020cvm}
I.~Esteban, M.~Gonzalez-Garcia, M.~Maltoni, T.~Schwetz, and A.~Zhou, {\it {The
  fate of hints: updated global analysis of three-flavor neutrino
  oscillations}},  {\em JHEP} {\bf 09} (2020) 178,
  [\href{http://arxiv.org/abs/2007.14792}{{\tt arXiv:2007.14792}}].

\bibitem{Hosaka:2005um}
{\bf Super-Kamiokande} Collaboration, J.~Hosaka et~al., {\it {Solar neutrino
  measurements in super-Kamiokande-I}},  {\em Phys. Rev. D} {\bf 73} (2006)
  112001, [\href{http://arxiv.org/abs/hep-ex/0508053}{{\tt hep-ex/0508053}}].

\bibitem{Ashie:2004mr}
{\bf Super-Kamiokande} Collaboration, Y.~Ashie et~al., {\it {Evidence for an
  oscillatory signature in atmospheric neutrino oscillation}},  {\em Phys. Rev.
  Lett.} {\bf 93} (2004) 101801,
  [\href{http://arxiv.org/abs/hep-ex/0404034}{{\tt hep-ex/0404034}}].

\bibitem{Kelly:2020fkv}
K.~J. Kelly, P.~A.~N. Machado, S.~J. Parke, Y.~F. Perez-Gonzalez, and R.~Z.
  Funchal, {\it {Neutrino mass ordering in light of recent data}},  {\em Phys.
  Rev. D} {\bf 103} (2021), no.~1 013004,
  [\href{http://arxiv.org/abs/2007.08526}{{\tt arXiv:2007.08526}}].

\bibitem{Harrison:2002er}
P.~F. Harrison, D.~H. Perkins, and W.~G. Scott, {\it {Tri-bimaximal mixing and
  the neutrino oscillation data}},  {\em Phys. Lett. B} {\bf 530} (2002) 167,
  [\href{http://arxiv.org/abs/hep-ph/0202074}{{\tt hep-ph/0202074}}].

\bibitem{Allanach:1999ic}
B.~Allanach, A.~Dedes, and H.~K. Dreiner, {\it {Bounds on R-parity violating
  couplings at the weak scale and at the GUT scale}},  {\em Phys. Rev. D} {\bf
  60} (1999) 075014, [\href{http://arxiv.org/abs/hep-ph/9906209}{{\tt
  hep-ph/9906209}}].

\bibitem{James:1975dr}
F.~James and M.~Roos, {\it {Minuit: A System for Function Minimization and
  Analysis of the Parameter Errors and Correlations}},  {\em Comput. Phys.
  Commun.} {\bf 10} (1975) 343--367.

\end{thebibliography}\endgroup

\end{document}